\documentclass[article]{elsarticle}

\usepackage[hmarginratio=1:1,top=32mm,left=20mm,columnsep=1cm]{geometry} % 
\usepackage{bm}
\usepackage{amsmath}
\usepackage{relsize} % e.g. used for \mathsmaller
\usepackage[svgnames]{xcolor} 
\usepackage{grffile} % allows to have dots in the name of graphic files
\usepackage{lineno}
\usepackage{graphicx}
\newcommand{\Dist}{{\bm{A}}}
\newcommand{\GG}{{\bm{G}}}
\newcommand{\II}{{\bm{I}}}
\newcommand{\A}{\mathbf{A}} 
\newcommand{\dev}[1]{#1\bm{'}}%{\overset{\bullet}{\GG}}

\newcommand{\G}{\mathbf{G}} 
\newcommand{\Q}{{\mathbf{Q}}}
\newcommand{\B}{{\mathbf{B}}}
\newcommand{\F}{{\bm{F}}}
\newcommand{\FF}{{\mathbf{F}}}
\newcommand{\ff}{{\mathbf{f}}}
\renewcommand{\S}{{\mathbf{S}}}

\newcommand{\csh}{c_\textrm{sh}}
\newcommand{\taus}{\tau_\textrm{s}}
\newcommand{\tauf}{\tau_\textrm{f}}
\newcommand{\vv}{{\bm{v}}}
\newcommand{\nn}{{\bm{n}}}
\newcommand{\xx}{{\bm{x}}}
\newcommand{\pd}{{\partial}}
\newcommand{\Efluid}{{E_\textrm{f}}}
\newcommand{\Esolid}{{E_\textrm{s}}}
\newcommand{\tr}{\textnormal{tr}}
\newcommand{\transpose}{{\mathsf{\mathsmaller T}}}
\newcommand{\dotgam}{\dot{\tensor{\gamma}}}
\newcommand{\doteps}{\dot{\tensor{\varepsilon}}}
\newcommand{\etaPL}{\eta_{\ms{P}\hspace{-0.4mm}\ms{L}}}
\newcommand{\etaHB}{\eta_{\ms{H}\hspace{-0.4mm}\ms{B}}}

\newcommand{\etaeff}{\eta_{\ms{e\hspace{-0.4mm}f\hspace{-0.5mm}f}}}
\newcommand{\sigmaY}{\sigma_{\mathsmaller{Y}}}
\newcommand{\Bi}{\textrm{Bi}}
\renewcommand{\Re}{\textrm{Re}}
\newcommand{\halb}{{\frac{1}{2}}}

\newcommand{\ms}[1]{\mathsmaller{#1}}

\usepackage{hyperref}
\hypersetup{
	colorlinks=true,                          
	linkcolor=DarkRed,
	citecolor=DarkRed,
	urlcolor=DarkRed  }

\DeclareMathAlphabet{\mathsfbi}{OT1}{\sfdefault}{bx}{sl}
\DeclareMathVersion{sfletters}
\SetSymbolFont{letters}{sfletters}{OML}{ntxsfmi}{b}{it}

\makeatletter
\newcommand{\mathbfsbilow}[1]{%
	\text{\mathversion{sfletters}$\m@th#1$}%
}
\DeclareRobustCommand{\tensor}[1]{%
	\begingroup
	\ifcat\noexpand #1\relax
	\edef\greek@test{\detokenize{#1}}%
	\edef\greek@test{\expandafter\@cdr\greek@test\@nil}%
	\edef\greek@test{\expandafter\@car\greek@test\@nil}%
	\edef\x{\the\lccode\expandafter`\greek@test}%
	\edef\y{\number\expandafter`\greek@test}%
	\ifnum\x=\y\relax
	\mathbfsbilow{#1}%
	\else
	\mathsfbi{#1}%
	\fi
	\else
	\mathsfbi{#1}%
	\fi
	\endgroup
}
\makeatother
\makeatletter
\newcommand{\sbullet}{%
	\hbox{\fontfamily{lmr}\fontsize{.4\dimexpr(\f@size pt)}{0}\selectfont\textbullet}}

\makeatother

\usepackage[numbers]{natbib}
\journal{Computers \& Fluids}

\begin{document}
	
\begin{frontmatter}
		
\title{\textbf{Simulation of non-Newtonian viscoplastic flows with a unified first order 
hyperbolic 
model and a structure-preserving semi-implicit~scheme}}
%\tnotetext[mytitlenote]{Fully documented templates are available in the elsarticle package on 
%\href{http://www.ctan.org/tex-archive/macros/latex/contrib/elsarticle}{CTAN}.}

%\fntext[myfootnote]{On leave from Sobolev Institute of Mathematics, Novosibirsk, Russia}
\cortext[mycorrespondingauthor]{Corresponding author}
%\ead[pesh]{ilya.peshkov@unitn.it}
%\author[mymainaddress,mysecondaryaddress]{Ilya Peshkov}

\address[UniTn]{Laboratory of Applied Mathematics, University of Trento,
	Via Mesiano 77, 38123 Trento, Italy}
\address[UniFe]{Department of Mathematics and Computer Science, University of 
	Ferrara, via Machiavelli 30, I-44121 Ferrara, Italy}
\address[Sobolev]{Sobolev Institute of Mathematics, 4 Acad. Koptyug 
	Avenue,	Novosibirsk, Russia}

\author[UniTn]{Ilya Peshkov\corref{mycorrespondingauthor}} 
\ead{ilya.peshkov@unitn.it}
%\fnref{myfootnote}
\author[UniTn]{Michael Dumbser}
\author[UniFe]{Walter Boscheri}
\author[Sobolev]{Evgeniy Romenski}
\author[UniTn]{Simone Chiocchetti}
\author[UniTn]{Matteo Ioriatti}

% ABSTRACT
\begin{abstract}
	We discuss the applicability of a unified hyperbolic model for continuum fluid and solid 
	mechanics to modeling non-Newtonian flows and in particular to modeling the 
	stress-driven solid-fluid transformations in flows of viscoplastic fluids, also called 
	yield-stress 
	fluids. 
	In contrast to the conventional approaches relying on the non-linear viscosity concept of the 
	Navier-Stokes theory and 
	representation of the solid state as an infinitely rigid non-deformable solid, the solid state 
	in our theory is deformable and the fluid state is considered rather as a ``\emph{melted}'' 
	solid via a certain procedure of relaxation of tangential stresses similar to Maxwell's 
	visco-elasticity theory. The model is formulated as a system of first-order hyperbolic partial 
	differential equations with possibly stiff non-linear relaxation source terms. The 
	computational strategy is based on a staggered semi-implicit scheme which can be applied in 
	particular to low-Mach number flows as usually required for flows of non-Newtonian fluids. The 
	applicability of the model and numerical scheme is demonstrated on a few standard benchmark 
	test cases such as Couette, Hagen-Poiseuille, and lid-driven cavity flows. The numerical 
	solution is compared with analytical or numerical solutions of the Navier-Stokes theory with 
	the Herschel-Bulkley constitutive model for nonlinear viscosity. 
\end{abstract}


\end{frontmatter}

%\linenumbers

\section{Introduction} \label{sec:introduction}

In \cite{HPR2016,DPRZ2016,HYP2016}, a unified first-order hyperbolic formulation for continuum 
fluid and solid mechanics was proposed. In a single system of partial differential equations 
(PDEs), such a model can describe various material responses including inviscid\footnote{Of course, 
	in the inviscid case, one can use directly the Euler equations for ideal fluids. But formally
	speaking, the Euler equations can be recovered in the stiff relaxation limit of the unified 
	theory, 
	e.g. see the asymptotic analysis in \cite{DPRZ2016}.} and viscous fluids, 
elastic and elastoplastic solids, see also 
\cite{BartonRom2010,GodPesh2010,Hyper-Hypo2019,Busto2020}. Due to its geometric,
\cite{PRD-Torsion2019} 
and thus intrinsically covariant nature, this model was also formulated in the general 
relativistic setting \cite{PTRSA2020}. Another useful extension of the model is based on the 
coupling with the equations for continuous modeling of damage and fracture which allows to describe 
both ductile and brittle fractures, e.g. see \cite{Cracks2020,Cracks2021}. The model originated from 
the works by Godunov and Romenski on Eulerian non-linear elastoplasticity theory 
\cite{GodRom1972,God1978,Romenski1979,GodRom2003} in 1970s. Later it was suggested by Peshkov and 
Romenski 
\cite{HPR2016} that this theory can be applied to describe viscous Newtonian flows. Therefore, 
sometimes we shall refer to this model as the Godunov-Peshkov-Romenski model or GPR model.

In this paper, we focus on investigating the capabilities of the unified model to describe flows 
exhibiting
solid-fluid-type transformations such as in flows of viscoplastic fluids \cite{FrigaardReview2014}, 
also called yield-stress fluids. Indeed, being able to model both fluid an solid states, 
this unified formulation is a good candidate for a monolithic continuous modeling of the 
solid-fluid transition such as thermally-driven melting and solidification\footnote{For example, 
	the importance of developing such an approach as ours is dictated by the necessity to predict 
	the 
	distribution of residual stresses in solidified materials in additive manufacturing. Recall 
	that 
	residual stresses can not appear in solids if they are treated as an infinitely viscous 
	fluids.} 
process in additive 
manufacturing \cite{Francois2017,King2017,Khairallah2016} as well as stress-driven fluidization and 
solidification process in industrial and environmental flows (landslides, avalanches, lava flows) 
of 
viscoplastic fluids and of dense granular flows \cite{Forterre2013}. This 
paper presents a preliminary step towards this ultimate goal. In particular, we shall assume that 
the solid-fluid transition occurs in a time-independent manner, i.e. infinitely fast. 

Under these assumptions, we found out that the unified model is able to describe a quite general 
theoretical class of fluid flows, called Herschel-Bulkley fluids, which unites the nonlinear 
power-law-type viscosity with the yield-stress behavior of Bingham-type materials. The ability 
of the unified model to describe flows of non-Newtonian fluids with a power-law viscosity was 
already 
demonstrated by Jackson and Nikiforakis \cite{Jackson2019a}. Also, the model was used by Hank et al 
\cite{Hank2016a,Hank2017} in the elastoviscoplastic regime to model high-strain-rate impact 
problems of 
clay-type materials. In this paper, the ability of the unified model to 
describe the yield-stress behavior of viscoplastic fluids in low-Mach number regime is discussed 
for the first time.

The most dramatic feature of viscoplastic fluids 
\cite{FrigaardReview2014}, is the \textit{yield stress}, 
from both the physical and mathematical perspective, because below this threshold such 
a material behaves as an elastic solid (i.e. it is able to recover its original 
shape after the loading has been removed) while it flows irreversibly like a viscous fluid 
if the stresses exceed a threshold. Thus, such materials exhibit drastically different behavior 
above and below the yield stress.

While it is apparent that real viscoplastic fluids are essentially two-phase media 
with a stress-induced solid-fluid phase transformation, such materials have been 
considered as a two-phase media only relatively recently by Putz 
and Burghelea in~\cite{Putz2009} and Grmela \cite{Grmela2003}. In 
particular, it was demonstrated in~\cite{Putz2009} that some 
experimental data 
for a typical viscoplastic fluid such as shear thinning physical gel (Carbopol 
940) can be fitted quite well if the chemical kinetics of the solid-fluid transition (i.e. 
breaking and restoration of bonds between molecules) are taken into account via an additional 
equation for the mass fraction of the solid state. Nevertheless, from the mathematical modeling 
viewpoint, to apply a two-phase 
approach to modeling viscoplastic flows is very challenging because fluids and solids are 
traditionally modeled by 
very different mathematical equations. Namely, fluid mechanics relies on the second-order 
Navier-Stokes equations of \emph{parabolic} type, while solid mechanics relies on the first-order 
elastodynamic equations which are of \emph{hyperbolic} type. Hence, a description of the 
solid-fluid transition would require 
changing the type of the PDEs which is a non-trivial problem from the mathematical and 
computational 
standpoints. 
Therefore, viscoplastic fluids are traditionally described by means of phenomenological 
Bingham-type models in which a material is considered 
as a \emph{nondeformable} solid below the yield stress (infinitely rigid solid represented by the 
Navier-Stokes equations 
with an infinitely large viscosity), while the flow starts 
abruptly once the applied stress exceeds a threshold. Nevertheless, in the last couple of decades, 
mixed elastoviscoplastic
approaches which combine the Navier-Stokes and elasticity theory have been proposed by several 
authors \cite{Saramito2007,Saramito2009,Benito2008,Fusi2007}.  The main idea of this type of models 
is 
to 
represent the Cauchy stress tensor $ \tensor{\sigma} $ as the sum of two parts $ \tensor{\sigma} = 
\tensor{\sigma}_{\ms{NS}}(\doteps) 
+ \tensor{\sigma}_{\ms{E}}(\tensor{\varepsilon}) $ with $ 
\tensor{\sigma}_{\ms{NS}}(\doteps) $ being the 
conventional 
Navier-Stokes stress defined as the function of the strain rate $ \doteps $, and $ 
\tensor{\sigma}_{\ms{E}} $ being the  
elastic stress which is a function of the strain $ \tensor{\varepsilon} $. Yet, in these 
approaches, the fluid and solid states are 
characterized by the state variables of opposite nature, i.e. $ \doteps $ being non-local (needs 
spatial gradients of velocity) and 
dissipative and  $ 
\tensor{\varepsilon} $ being local and non-dissipative. It is thus, not clear at all how one can 
build a 
thermodynamically consistent 
theory describing transformation of one into another.

In contrast to the traditional way of modeling viscoplastic flows and the new aforementioned 
approaches, our approach does not rely on the Navier-Stokes theory and does not consider viscous 
fluids as a strain-rate-type constitutive theory, i.e. the strain rate $ \doteps $ is excluded from 
being the state variable. Instead, both solids and 
fluids are considered from a deformation, and thus unified, geometrical standpoint\footnote{In the 
language of differential geometry, this theory can be considered as a Riemann-Cartan geometry with 
non-zero torsion \cite{PRD-Torsion2019}.}. This, in particular, allows to use exactly the same 
state variables for fluids and solids and to compute the Cauchy stress tensor in a unified manner. 
Furthermore, instead of the traditional viscosity, the parameter that characterizes the ability of 
the material to flow in our theory is the strain relaxation time $ \tau $ which is a continuous 
analog of Frenkel's concept of relaxation time $ \tau_{\ms{F}} $ in liquids 
\cite{Frenkel1955,brazhkin2012two,bolmatov2013thermodynamic,bolmatov2015revealing} which was 
proposed by Frenkel \cite{Frenkel1955} and essentially provides a solid-like view on the kinetic 
description of liquids, e.g. see the discussion in 
\cite{HYP2016} on Frenkel's relaxation time and its continuous interpretation in our theory.
We 
believe that the characterization of fluids and solids by the same set of state variables, the same 
form of the stress tensor, and the same ``fluidity'' characteristic (relaxation time) have a 
great potential for modeling of solidification/melting processes as it provides a unified 
thermodynamically consistent framework 
for continuous tracking of material properties across\footnote{See also further developments of 
Frenkel's seminal idea in unified descriptions \cite{Heo2014,Caplan2014,Bolmatov2015a,Frank2018} of 
thermodynamical properties of matter across the 
solid-fluid interfaces.} the solid-fluid transformation fronts.

Concerning our numerical strategy, in contrast to our previous papers 
\cite{DPRZ2016,DPRZ2017,Busto2020,Hyper-Hypo2019} in which the 
model was studied using explicit Finite Volume and Discontinuous Galerkin methods belonging to the 
family of high-order ADER-Finite-Volumes and ADER-Discontinuous-Galerking schemes 
\cite{toro1,toro3,toro4,titarevtoro,Toro:2006a,BTVC2016,Fambri2019,DumbserEnauxToro,AMR3DCL,ADERDGVisc} and 
ADER schemes in the Arbitrary 
Lagrangian Eulerian framework
\cite{Boscheri2013,Boscheri2014,Boscheri2015,Gaburro2020a,Gaburro2020},
in this 
paper, we rely 
on our recent Semi-Implicit Structure Preserving Finite Volume (SISPFV) scheme \cite{SIGPR2021} 
which allows to run the model efficiently in the low-Mach number regime typical for problems 
involving non-Newtonian fluids.

%We thus consider a 
%viscoplatic fluid as a two-phase media which behaves as a deformable elastic solid if the 
%stresses are essentially below the yield stress and as a power-law non-Newtonian fluid if the 
%stresses are above the yield stress with a smooth transition between these 
%two states which means that there might be zones where two phases coexist. 
%According to our 
%unified approach, these two states are represented by the strain dissipation 
%times $ \tau=\infty $ and $ 0<\tau=const<\infty $ for the solid and fluid 
%state respectively.

%Recent advances in rheometry offer new insight into this issue. Important 
%results are that (a) flows observed at low stresses are not always steady, and 
%(b) flow complexity (shear bands, wall slip) can lead to erroneous 
%interpretation of macroscopic data~\cite{FrigaardReview2014}

\section{Mathematical model}\label{sec.math.model}

\subsection{System of governing PDEs}\label{sec.gov.PDE}

A quite general mixture model for modeling stress-induced solid-fluid phase 
transition was proposed in~\cite{PeshGrmRom2015}. 
In this paper, however, we consider a much simpler model. In particular, we assume that the 
mobility of the solid and fluid phases in the 
transition zone is negligible, i.e. the relative velocity of the phases is zero.  Furthermore, we 
shall 
assume that the fluid and solid states are only differ by the so-called strain relaxation time $ 
\tau $ (to be introduced later) while all other material parameters such as longitudinal 
and shear sound speed\footnote{Recall that in our theory, 
	a viscous fluid is also characterized by a shear sound speed.} are equal. 
 
The state parameters of interest are
\begin{equation}\label{eqn.state.par}
	(\rho,s,\vv,\Dist),
\end{equation}
where $ \rho $ is the mass density, $ s $ is the specific entropy, $ \vv $ is 
the material velocity, and $ \Dist $ is the distortion matrix which characterizes the morphology of 
the material elements. The distortion can be seen as a local basis triad (three vectors). For 
example, these vectors are denoted as $ \bm{e}^\beta $, $ \beta=1,2,3 $ in papers 
\cite{Gavrilyuk2008} 
by Gavrilyuk and 
Favrie. By 
measuring the changes in angles between these three vectors and coefficients of 
elongation/contraction, 
one can use $ \Dist $ as a deformation measure of the material elements\footnote{In the language of 
differential geoemtry, the distortion field can be viewed as a non-holonomic frame field (local 
basis triad), or 
Cartan's moving frame, e.g. see \cite{Hehl2007,PRD-Torsion2019}.}. Also, the word 
``\emph{local}'' means that it is impossible to recover the global deformation of the whole 
continuum by integrating the distortion field $ \Dist $. 
The distortion matrix $ \Dist $ can be related to the elastic part $ \F^e $ of the deformation 
gradient $ \F = \F^e \F^p$ in the  conventional finite-strain inelasticity as $ 
\Dist =(\F^e)^{-1} $, where $ \F^p $ is the plastic component of the full deformation gradient $ \F 
$. Remark that if one is 
working in the Eulerian frame, the plastic component $ \F^p $ is not required as discussed for 
example in \cite{Hyper-Hypo2019,PeshGrmRom2015}. 

The system of governing PDEs in a Cartesian coordinate system can be written as 
\cite{HPR2016,DPRZ2016}
\begin{subequations}\label{eqn.GPR}
	\begin{align}
		& \frac{\partial \rho}{\partial t}+\frac{\partial (\rho v_k)}{\partial 
			x_k}=0,\label{eqn.conti}\\[2mm]
		&\frac{\partial (\rho v_i)}{\partial t}+\frac{\partial 
			\left(\rho v_i v_k + p \delta_{ki} - \sigma_{ki} \right)}{\partial x_k}=0, 
		\label{eqn.momentum}\\[2mm]
		&\frac{\partial A_{i k}}{\partial t}+\frac{\partial (A_{im} 
			v_m)}{\partial x_k}+v_j\left(\frac{\partial A_{ik}}{\partial 
			x_j}-\frac{\partial A_{ij}}{\partial x_k}\right)
		=-\dfrac{ E_{A_{ik}} }{\theta},\label{eqn.dist}\\[2mm]
		%	&\frac{\partial (\rho c)}{\partial t}+\frac{\partial 
		%	\left(\rho c v_k\right)}{\partial x_k}=-\chi, \label{eqn.heatflux}\\[2mm]
		&\frac{\partial (\rho s)}{\partial t}+\frac{\partial \left(\rho 
			s v_k \right)}{\partial x_k}=\dfrac{\rho}{\theta T} 
		E_{A_{ik}} E_{A_{ik}}, \label{eqn.entropy}\\[2mm]
		&\frac{\partial (\rho  E)}{\partial t}+\frac{\partial \left( \rho E v_k  + v_i (p \, 
			\delta_{ki} 
			- \sigma_{ki}) \right)}{\partial x_k}=0. \label{eqn.energy} 
	\end{align}
\end{subequations}
where the pressure $ p $ and the tangential stresses $ \sigma_{ik} $ should be defined via the total
energy potential  $ E(\rho,s,\vv,\Dist) $ as $ p = \rho^2 E_\rho $ and $ \sigma_{ki} = - \rho  
A_{ji} E_{A_{jk}}$, where, in 
turn, $ E_{\rho} =\frac{\pd E}{\pd \rho}$ and $ E_{A_{jk}} = \frac{\pd E}{\pd A_{jk}} $, $ 
\delta_{ki} $ is the Kronecker delta, $ T=E_s $ is the temperature, and $ \theta > 0 $ is one of 
the 
main constitutive functions of our theory which controls the rate of strain relaxation and will be 
specified in Sec.\,\ref{sec.inelast}. 

System \eqref{eqn.GPR} is an overdetemined  system which has one more equations than unknowns.
In fact, the total energy $ E $ as mentioned above, is not an unknown but a potential $ 
E=E(\rho,s,\vv,\Dist) $ and the last equation (the total energy conservation law) is, in fact, the 
consequence of the others, i.e. it can be obtained as a linear combination of the other equations, 
e.g. see \cite{SHTC-GENERIC-CMAT,GodRom2003}. Yet, in order to guarantee the energy conservation 
exactly at the 
discrete level, it is equation \eqref{eqn.energy} that is used in the discretization of system 
\eqref{eqn.GPR} while the entropy equation is obtained as a consequence, see \cite{SIGPR2021}. 

One can see that in order to close system \eqref{eqn.GPR} (to specify the pressure, stress, etc) 
one needs to specify the energy potential $ E $ which is done in the following section.

\subsection{Closure of the reversible part via energy potential}

To close the system of governing PDEs \eqref{eqn.GPR}, one needs to propose a model for the total 
energy $ E $ and the strain relaxation function $ \theta $. 
Here, we specify the total energy 
potential as the sum of three contributions\footnote{In a general situation, the 
		fluid and solid state of the material are governed by their own equations of state 
		(energies) $ 
		\Efluid $ and $ \Esolid $ so that the total energy of the material is 
		$
		E = c \Esolid + (1-c) \Efluid
		$
		with $ c $ being the solid mass fraction governed by its own time evolution equation.}
\begin{equation}\label{eqn.EOS}
	E(\rho,s,\vv,\Dist) = E_{\rm i}(\rho,s) + E_{\rm e}(\Dist) + E_{\rm k}(\vv) \equiv E_{\rm 
	i}(\rho,s) + 
	\dfrac{\csh^2}{2} \Vert\dev{\GG}\Vert^2 + 
	\frac{1}{2}\Vert \vv \Vert^2,
\end{equation}
where $ E_{\rm i}(\rho,s) $ is the internal energy of volumetric deformation, $ E_{\rm e}(\Dist) $ is the elastic energy of shear deformation, and 
$ E_{\rm k}(\vv) $ is the kinetic energy, $ \dev{\GG} = \GG - \frac{1}{3}\tr(\GG) \II $ is the 
deviatoric (or trace-less) part of the metric tensor $ \GG = 
\Dist^\transpose\Dist $, $ \II $ is the identity matrix, and we use the following matrix and vector 
norms
\begin{equation}\label{eqn.norms}
	\Vert\dev{\GG}\Vert  = \sqrt{\frac12 
	\dev{G}_{ij}\dev{G}_{ij}}, \qquad  \Vert\vv\Vert = 
\sqrt{v_i v_i}.
\end{equation}Also, $ \csh $ is the shear sound speed, which, as was mentioned earlier, is 
assumed to be the same in the solid and fluid states. It characterizes the rigidity of the 
infinitesimal frames represented by the basis 
triads $ \Dist $. 

The internal energy $ E_{\rm i}(\rho,s) $, in the case of solids and liquids, can 
be taken 
as 
a stiffened gas equations of state, e.g. see \cite{DPRZ2016,SIGPR2021}. However, in this paper, we 
only consider test cases that are considered in the literature for incompressible isothermal flows 
in confined geometries. This means that the choice of the internal energy $ E_{\rm 
i}(\rho,s) $ plays no role in this study. Nevertheless, in order to stay in the hyperbolic region 
of model \eqref{eqn.GPR}, we always solve the full \emph{compressible} model 
\eqref{eqn.GPR} for the full vector of state variables \eqref{eqn.state.par} and hence, an 
equations of state for the internal energy $ E_{\rm i}(\rho,s) $ has to be always provided. 
In particular, in all the test cases, we shall use the ideal gas equation of state for $ E_{\rm 
i}(\rho,s) $, i.e. 
$ E_{\rm i}(\rho,s) = \frac{\rho^{\gamma-1}}{\gamma-1} e^{s/c_v} $ or $ E_{\rm i}(\rho,p) = 
\frac{p}{\rho(\gamma-1)} $ with $ c_v $ being the heat capacity at constant volume and $ \gamma $ 
the ratio of specific heats.

Remark that, when dealing with incompressible flows, we do not impose the incompressibity 
constraint $ \nabla \cdot \vv = 0 $, which would change the type of the governing PDEs from 
hyperbolic to a mixed hyperbolic-elliptic type. Instead, we solve the full compressible formulation 
\eqref{eqn.GPR} with the new semi-implicit SISPFV scheme \cite{SIGPR2021} in the low-Mach number 
limit $ {\rm Ma}\to 0 $.
Thus, as shown in \cite{SIGPR2021}, with this type of schemes, the density and velocity divergence 
fluctuations scale with $ {\rm Ma}^2 $ in the low-Mach number limit.

Using \eqref{eqn.EOS}, one can explicitly compute the derivatives $ E_{A_{ik}} $ used in the right 
hand side of \eqref{eqn.GPR} and in the computation of the stress $ \sigma_{ki} $. 
Thus, we have
\begin{equation}\label{eqn.stress}
	E_{\Dist} = \csh^2 \Dist\, \dev{\GG}, \qquad \tensor{\sigma} = -\rho \Dist^\transpose E_\Dist = 
	-2\rho 
	\GG E_\GG =  
	-\rho \csh^2 
	\GG\, 
	\dev{\GG}.
\end{equation}

We remark that in contrast to the Navier-Stokes theory where the tangential stresses are of a 
dissipative nature, the tangential stresses of our theory are due to elastic forces which are 
intrinsically non-dissipative. The dissipation is introduced only via the right-hand side of system 
\eqref{eqn.GPR}. In other words, in the context of viscous fluids, viscous solution of the 
parabolic Navier-Stokes equations is approximated by a solution of hyperbolic system 
\eqref{eqn.GPR} of visco-elastic medium, see the following section for the details.

\subsection{Closure of the irreversible part}\label{sec.inelast}

Recall that if the source term in \eqref{eqn.dist} vanishes then system \eqref{eqn.GPR} is simply 
the 
system of nonlinear elasticity written in the Eulerian coordinates \cite{GodRom2003,DPRZ2016}. We 
thus say that the left hand-side of \eqref{eqn.GPR} is the reversible part of the time evolution 
\cite{SHTC-GENERIC-CMAT,PKG-Book2018}. However, the most drastic change in the material behavior 
is controlled by the right hand-side of \eqref{eqn.dist} which, together with the entropy 
production 
source term in \eqref{eqn.entropy}, constitutes the irreversible (dissipative) part of the time 
evolution. Note that the entropy production source term is obviously positive and hence, the second 
law of thermodynamics is always respected \cite{SHTC-GENERIC-CMAT}.  

In particular, denoting by $ \tau $ the strain relaxation time discussed in the introduction and
taking the energy potential in the form \eqref{eqn.EOS} and $ \theta $ 
in the form
\begin{equation}\label{eqn.theta}
	\theta = \frac13\tau \csh^2 |\Dist|^{-5/3},
\end{equation}
it has been shown in \cite{HPR2016,DPRZ2016} that, for small $ 0 < \tau \ll 1 $, the stress tensor 
of 
our theory $ \tensor{\sigma} 
= -\rho \Dist^\transpose E_{\Dist}$ (which is computed fully locally, i.e. as an algebraic function 
of state variables) approximates the Navier-Stokes stress (which is non-local in space, i.e. its 
computation involves spatial derivatives)
\begin{equation}\label{eqn.NS.stress}
	\tensor{\sigma}_{\ms{NS}} = 
	\eta \, \dotgam, 
	\qquad 
	\dotgam = \doteps  - \frac{1}{3} \tr(\doteps) \II, 
	\qquad \doteps = \nabla \vv + \nabla \vv^\transpose,
\end{equation}
at first order in $ \tau $, i.e. one can expand $ \tensor{\sigma} $ in series in $ \tau $
\begin{equation}\label{eqn.series}
	\tensor{\sigma} = \tensor{\sigma}_0 + \tau \tensor{\sigma}_1 + \tau^2 \tensor{\sigma}_2 + \ldots
\end{equation}
such that $ \tensor{\sigma}_0 = \tensor{0} $ and
\begin{equation}\label{eqn.NS.GPR}
		\tensor{\sigma}_1 = \frac16 \rho \tau \csh^2
	 \, \dotgam.
\end{equation}
Here, $ \frac12\doteps $ is the strain rate tensor and $ \dotgam $ is the rate-of-shear tensor.

Thus, by comparing \eqref{eqn.NS.stress} and \eqref{eqn.NS.GPR}, one can conclude that 
the \emph{effective}
shear viscosity of our theory is expressed as\footnote{This formula for $ \etaeff $ is valid only 
for the energy given by \eqref{eqn.EOS} and $ \theta $ given by \eqref{eqn.theta}. For a different 
energy potential and $ \theta $, the coefficient $ \frac{1}{6} $ may change or may depend on 
the density 
$ \rho $ and entropy $ s $. Nevertheless, in general, one has that $ \etaeff \sim \rho \tau \csh^2 
$.}
\begin{equation}\label{eqn.viscosity}
	\etaeff = \frac16 \rho \tau \csh^2.
\end{equation}

One can clearly see that the effective viscosity $ \etaeff $ is composed of the two principal 
parameters of our theory: the shear sound speed $ \csh $ which characterizes the elasticity of the 
infinitesimal frames represented by the  
basis triads $ \Dist $ 
and thus, characterizes the reversible part of the model, and $ \tau $ which characterizes the 
rate of dissipation. In other words, even if the model is employed in the diffusive 
regime\footnote{More 
	rigorously, in the viscous regime, the relaxation time $ \tau $ has to be sufficiently smaller 
	than 
	the characteristic time of the macroscopic process.} ($ \tau 
\ll 1 $), the viscous flow is approximated by a \emph{viscoelastic} solution of our theory.
In this 
case, the dissipative dynamics governed by the right hand-side dominates.

Therefore, in the presence of the relaxation term in \eqref{eqn.dist}, the overall dynamics of 
system \eqref{eqn.GPR} is always of viscoelastic type at the time scales $ \tau < 
t < \infty $
because both parts of the time evolution (left and right hand-sides of \eqref{eqn.GPR}) are working 
together. 
Yet, the dynamics is reversible (elastic) at the time scales $ t < \tau $, e.g. see  
\cite{Frenkel1955,brazhkin2012two,bolmatov2013thermodynamic,bolmatov2015revealing,HPR2016}.

Despite the asymptotic 
expansion \eqref{eqn.series} being
only performed under the assumptions of constant viscosity and constant relaxation time, we shall 
assume\footnote{The validity of this assumption will be justified latter via numerical 
	examples.} in this paper that the same procedure can be done even in the case of nonlinear 
viscosity 
which will inevitably result in a nonlinear but still sufficiently small $ \tau $, now being a 
function of state variables $ \tau = \tau(\rho,s,\Dist) $. In 
principle, the ``\emph{sufficient smallness}'' of $ \tau $ can be achieved by taking larger shear 
sound speed  
$ \csh $
that for 
the same viscosity provides a smaller $ \tau $ as can be seen from \eqref{eqn.viscosity}.

\subsubsection{Pure fluids (no yield-stress behavior)}
Based on the aforementioned reasoning, we assume that if the Newton law of viscosity 
\eqref{eqn.NS.stress} is generalized with a nonlinear viscosity
\begin{equation}\label{eqn.Newton.law}
	\tensor{\sigma}_{\ms{NS}} = 
	\eta(\dot{\gamma}) \, \dotgam, 
	\qquad 
	\dot{\gamma} = \Vert\dotgam\Vert = \sqrt{\frac{1}{2}\dot{\gamma}_{ij}\dot{\gamma}_{ij}},
\end{equation}
we still assume that the relation \eqref{eqn.viscosity} holds for some constant shear sound speed $ 
\csh $ and some nonlinear function $ 
\tau(\Dist) $ to be determined.

For example, for incompressible flows of the power-law (PL) fluids with the viscosity 
\begin{equation}\label{eqn.power.law}
	\etaPL(\dot{\gamma}) = \kappa \dot{\gamma}^{n-1},
\end{equation}
with $ \kappa $ and $ n $ being the so-called consistency and power-law indexes,
one should solve \eqref{eqn.Newton.law} with respect to $ \dot{\gamma} $, i.e. $ \dot{\gamma}( 
\sigma )= 
(\sigma/\kappa)^{1/n}$, where $ \sigma = \Vert \tensor{\sigma} 
\Vert $. This is necessary because $ \dotgam $ is not a state variable in our theory. Then, one can 
get the 
power-law viscosity as a function of $ \sigma $ which, in turn, is the function of $ \Dist $:
\begin{equation}\label{eqn.power.law.A}
	\etaPL(\Dist) = \kappa  \left( \frac{\sigma}{\kappa}\right )^{\frac{n-1}{n}} 
\end{equation}
and the corresponding relaxation time $ \tau $ is simply obtained from \eqref{eqn.viscosity} and 
\eqref{eqn.power.law.A} as
$ \tau(\Dist) = 6 \, \etaPL(\Dist)/(\rho \csh^2) $. Remark that the relation  
\eqref{eqn.power.law.A} can be recognized as a Perzyna-type model \cite{Lubliner1990,Perzyna1966a}.
In particular, in this way, the first results 
for 
non-Newtonian power-law fluids with model \eqref{eqn.GPR} were obtained in \cite{Jackson2019a} for 
Hagen-Poiseuille and lid-driven cavity flows. 

Note that not all rheological models can be formulated in simple linear form as 
\eqref{eqn.Newton.law}, e.g. see \cite{Prusa2010}, and here, we do not provide a general recipe for 
finding the proper expression for the relaxation time $ \tau $. For a specific nonlinear 
stress-strain-rate relation of the form $ f(\tensor{\sigma},\dotgam) = 0 $, this question should be 
considered individually.

\subsubsection{Viscoplastic fluids}

In the classical fluid mechanics models such as Bingham, or more general Herschel-Bulkley model 
with 
yield-stress behavior, the material in the solid state is traditionally treated as a 
\emph{rigid solid} (non-deformable) which has an infinite viscosity. 
Thus, if we apply directly the above approach to a constitutive relation with an infinite 
viscosity, 
we obtain an infinite relaxation time $ \tau $ and therefore the smallness assumption under which 
the relation \eqref{eqn.viscosity} was obtained does not hold.

In fact, an infinite relaxation time in our model corresponds to a \emph{deformable} elastic solid 
(the right-hand side in \eqref{eqn.dist} vanishes) 
and there is no need to treat it as a fluid with an infinite viscosity. Therefore, for a 
viscoplastic fluid with a yield-stress $ \sigmaY $, we shall assume smallness of $ \tau(\Dist) 
$ 
only in the fluid region (i.e. $ \sigma > \sigmaY $), while in the solid region ($ \sigma < \sigmaY 
$) the relation \eqref{eqn.viscosity} is not required and thus smallness of $ 
\tau $ is not required as well. This inevitably causes that in a certain vicinity of the yield-stress 
$ \sigmaY $ there is a transition zone where $ \tau $ is not small and the material is neither 
viscous nor elastic but visco-elastic. However, as we shall see 
in Section\,\ref{sec.results}, this zone is negligibly small and does not affect much the overall 
solution.

Therefore, we treat the viscoplastic fluids as elastoviscoplastic with the relaxation time given by
\begin{equation}\label{eqn.tau.general}
	\tau(\Dist) = \left\{
	\begin{array}{ll}
		\taus, & \Vert\tensor{\sigma}\Vert < \sigmaY, \\[2mm]
		\tauf(\Dist),& \Vert\tensor{\sigma}\Vert \geq \sigmaY,
	\end{array} \right.
\end{equation}
where $ \taus = const$ denotes the relaxation time in the solid region which is assumed to be 
elastic. This means that\footnote{In practice, it is sufficient to take $ \taus $ just as a 
very large number (sufficiently larger than the typical time scale of the problem). In the 
simulations, we take $ 
	\taus = 10^{20} $.} $ \taus =\infty $ and the source term in \eqref{eqn.dist} simply 
	vanishes 
$  
\theta^{-1} E_{A_{ik}} = 0$. The relaxation time in the fluid state $ \tauf(\Dist) $ can be 
computed from the viscosity based on the relation \eqref{eqn.viscosity}.

As an example, let us consider the Herschel-Bulkley (HB) constitutive relation for incompressible 
flows 
(e.g. see \cite{Lubliner1990})
which includes the 
Bingham-type behavior
\begin{equation}\label{eqn.HB.const}
	\dotgam=\left\{
	\begin{array}{ll}
		0,& \Vert\tensor{\sigma}\Vert < \sigmaY, \\[2mm]
		\etaHB^{-1} \, \tensor{\sigma},& \Vert\tensor{\sigma}\Vert \geq \sigmaY,
	\end{array} \right.
\end{equation}
where  $ \etaHB = \etaHB(\dotgam) $ is the 
Herschel-Bulkley 
viscosity of the fluid state that reads
\begin{equation}\label{eta.HB}
	\etaHB(\dot{\gamma}) = 
	\kappa \dot{\gamma}^{n-1} + \sigmaY \dot{\gamma}^{-1}, \qquad 
	\dot{\gamma} = \Vert\dotgam\Vert, 
\end{equation}
or
\begin{equation}\label{eta.HB2}
	\etaHB(\sigma) = \left( \frac{\sigma - \sigmaY}{\kappa}\right)^{-\frac{1}{n}} \sigma ,
	\qquad
	\sigma =\Vert \tensor{\sigma} \Vert, 
\end{equation}
where $ \kappa = const $ is the consistency index and $ n $ is the power-law index. Note that 
\eqref{eta.HB} includes the power-law viscosity \eqref{eqn.power.law} as a 
particular case for $ \sigmaY = 0$. Then, the fluid state relaxation time is computed from 
\eqref{eqn.viscosity} as 
\begin{equation}\label{eqn.tauf}
	\tauf(\Dist) = \frac{6 \etaHB(\sigma)}{\rho \csh^2}.
\end{equation}
This strategy is employed in the numerical examples presented in the following section. Note that 
equations \eqref{eqn.GPR} with a 
constitutive relation similar to \eqref{eqn.tau.general}, \eqref{eqn.tauf} was used in  
\cite{Hank2017} for impact dynamics modeling of a bentonite clay suspension which is also 
considered as a elastoviscoplastic-type material. Here, however, we focus on applying these 
constitutive 
relations to modeling \textit{flow phenomena} in viscoplastic materials.

\paragraph{Remark} The solid-fluid transition as it is treated in the Bingham-type models of 
classical fluid mechanics is a time-independent (instantaneous) transition. However, some real 
complex fluids with 
a yield-stress behavior can not be treated in such a simplified way. The solid-fluid transition in 
such fluids 
is time-dependent and requires a chemical kinetics-type time evolution equation for an order 
parameter (e.g. mass fraction of the solid state) \cite{Putz2009}. Therefore, we remark that a more 
general constitutive framework can be adopted in our theory for modeling this and other solid-fluid 
transitions, 
e.g.  thermally driven phase transformations such as melting and solidification. In these settings, 
the material can be considered as a mixture of the solid and fluid states with an 
order-parameter-dependent relaxation time and the elastic modulus, e.g. $ \tau(c) = (c/\taus + 
(1-c)/\tauf)^{-1} $, where $ c $ is the mass fraction of the solid state, and $ \taus $ and $ \tauf 
$ are the relaxation times in the solid and fluid states. In particular, a similar approach has 
been applied to modeling of the brittle and ductile fracture in solids in 
\cite{Cracks2020,Cracks2021} where the damaged material is treated as a fluid or solid with 
degraded elastic modulus and relaxation times.

\section{Numerical scheme}\label{sec.scheme}

Here, we describe the Semi-Implicit Structure Preserving Finite Volume (SISPFV) scheme  
\cite{SIGPR2021} which is used in the numerical simulations in Section\,\ref{sec.results}. The 
SISPFV scheme is first-order accurate in time and second-order accurate in space. It is based on 
the staggered discretization with the pressure and energy stored in 
the cell-centers, velocity field stored on the cell-edges and the distortion field in the 
cell-corners, see Fig.\,\ref{fig.staggered}.  The SISPFV scheme consists of four steps. 
In the first step, see 
Sec.\,\ref{sec.explicit}, the nonlinear 
convective part 
of the fluxes is discretized in an \emph{explicit} manner using a classical 
second order MUSCL-Hancock type TVD finite volume scheme \cite{toro-book}. In the second step, 
the distortion field is 
evolved according to the new \emph{curl-preserving} explicit finite difference discretization.
In the third step, the 
mildly nonlinear pressure 
sub-system (momentum 
and energy fluxes containing the pressure) is solved \emph{implicitly} using the nested Newton 
method of Casulli and Zanolli \cite{CasulliZanolli2010,CasulliZanolli2012,DumbserCasulli2016}.
And in the fourth step, the stiff source term in \eqref{eqn.dist} is treated using the implicit 
Euler method. Thanks to the implicit treatment of the pressure sub-system, the CFL condition for 
the time step is formulated based on the flow velocity $ \vv $ and shear sound speed $ \csh $ and 
not on the adiabatic sound speed. For semi-implicit schemes on unstructured staggered meshes, the reader 
is referred to \cite{Tavelli2015,TavelliDumbser2017,PODFVFE,Hybrid1,BFTVC2018,SIDGConv} and references 
therein.   

The scheme is consistent with the low Mach number limit of the GPR equations, it is exactly 
curl-free for the homogeneous part of the PDE system in the absence of source terms and is 
consistent with the Navier-Stokes limit of the model in the stiff relaxation limit when $ \tau \to 
0 $ and with the large-strain hyperelasticity limit of the equations when $ \tau \to \infty $. Yet, 
due to the strong non-linearity of the relaxation source term, the current version of the scheme is 
not truly asymptotic-preserving in the limit $ \tau\to0 $ for the stress tensor $ \tensor{\sigma} $ 
but only \emph{quasi} 
asymptotic-preserving, i.e. up to the second order error terms. In particular, this imposes certain 
limitations when we shall consider very stiff examples in Section\,\ref{sec.results}, e.g. 
lid-driven cavity flows of HB fluid with $ n=0.5 $.

System \eqref{eqn.GPR} can be written more compactly in the following matrix-vector notation   
\begin{equation}
	\label{eqn.pde} 
	\frac{\pd \Q}{\pd t} + \nabla \cdot \FF(\Q) + \B(\Q) \cdot \nabla \Q = \S(\Q), 
\end{equation}
with the state vector $\Q = (\rho, \rho v_i, A_{ik}, \rho E)^\transpose$, the nonlinear flux tensor 
$\FF(\Q) = \left (\ff(\Q), \mathbf{g}(\Q) \right )$, with $\ff$ and 
$\mathbf{g}$ the 
fluxes in $x$ and $y$ direction, respectively,  
the non-conservative product $\B(\Q) \cdot \nabla \Q$ containing the curl terms in 
\eqref{eqn.dist}, and the vector of 
potentially stiff algebraic relaxation source terms $\S(\Q)$. As proposed in 
\cite{DumbserCasulli2016,SIMHD} we
\textit{split} the flux tensor $ \FF(\Q) $ into a convective part and a pressure part. However, the 
equations 
for
the distortion $A_{ik}$ as well as its respective contributions to the momentum equation 
and to the total energy conservation law need a special \textit{compatible} and 
structure-preserving discretization using a vertex-based grid staggering. Hence, 
\eqref{eqn.pde} is rewritten as  
\begin{equation}
	\label{eqn.pde.split} 
	\frac{\pd \Q}{\pd t} + \nabla \cdot \left( \FF_c(\Q_c) + \FF_p(\Q) + 
	\FF_v(\Q) 
	\right) + 
	\nabla \G_v(\Q) + \B_v(\Q) \cdot \nabla \Q = \S(\Q), 
\end{equation}
where
\begin{equation} 
	\label{eqn.split.def} 
	\FF_c = \left( \begin{array}{c} \rho v_k \\ \rho v_i v_k \\ 0 \\ \rho v_k ( E_{\rm e} + 
		E_{\rm k} ) \end{array} \right) , 
	\quad 
	\Q_c = \left( \begin{array}{c} \rho  \\ 
		\rho v_i  \\ 0 \\ \rho  ( E_{\rm e} + E_{\rm k} ) \end{array} \right) ,  
	\qquad 
	\FF_p = \left( \begin{array}{c} 0 \\ p \delta_{ik} \\ 0 \\ h \rho v_k \end{array} 
	\right), \qquad 
	\FF_v = \left( \begin{array}{c} 0 \\ -\sigma_{ik}  \\ 0 \\ -v_i \sigma_{ik} 
	\end{array} \right), 
\end{equation}
and
\begin{equation} 
	\G_v(\Q) = \left( \begin{array}{c} 0 \\ 0 \\ A_{im} v_m \\ 0 \end{array} 
	\right),  
	\qquad 
	\B_v(\Q) \cdot \nabla \Q = \left( \begin{array}{c} 0 \\ 0 \\ 
		v_m \left(\frac{\partial A_{ik}}{\partial x_m}-\frac{\partial A_{im}}{\partial x_k}\right) 
		\\ 
		0 \end{array} \right). 
\end{equation} 
Here, $h=E_{\rm i} + p/\rho$ is the specific enthalpy, $\FF_c(\Q_c) $ refers to purely convective 
fluxes 
that will be discretized explicitly 
and  
$\FF_p(\Q) $ are the pressure fluxes that will be discretized implicitly using an edge-based 
staggered grid. The resulting splitting into pressure and convective fluxes is identical to the 
flux-vector splitting scheme of Toro and V\'azquez-Cend\'on forwarded in 
\cite{ToroVazquez}. The  remaining terms $\FF_v(\Q) $,  $\nabla \G_v(\Q)$ and $\B_v(\Q) 
\cdot \nabla \Q$ 
will be carefully discretized in a structure-preserving manner using an explicit scheme on an 
appropriate vertex-based staggered grid. The relaxation source terms $\S(\Q)$  
can become stiff and thus require an implicit discretization on the vertex-based staggered mesh.

The following subsystem  
\begin{equation}
	\label{eqn.pde.ex} 
	\frac{\pd \Q}{\pd t} + \nabla \cdot \left( \FF_c(\Q_c) + \FF_v(\Q) \right) + 
	\nabla \G_v(\Q) + \B_v(\Q) \cdot \nabla \Q = \S(\Q), 
\end{equation}
will be discretized explicitly, apart from the potentially stiff algebraic source terms in $\S$, 
which are discretized implicitly with a simple backward Euler scheme. The discretization method 
presented in the next section will consist in a combination of a classical second order 
MUSCL-Hancock type \cite{toro-book} TVD finite volume scheme for the convective fluxes $\FF_c$, a 
curl-free discretization for the terms $\G_v$ and $\B_v \cdot \nabla \Q$ using compatible gradient 
and curl operators as well as a vertex-based discretization of the terms $\FF_v$. The eigenvalues 
of 
subsystem \eqref{eqn.pde.ex}  in $x$ direction, or more precisely the eigenvalues 
of the matrix $\A_v(\Q) \cdot \mathbf{e}_x$ with $\mathbf{e}_x=(1,0)$ when rewriting 
\eqref{eqn.pde.ex} in quasi-linear form $\partial_t \Q + \mathbf{A}_v(\Q) \cdot \nabla \Q = 
\S(\Q)$, are 
\begin{equation}
	\lambda^{c,v}_{1,2} = u \pm \frac{2}{3} \sqrt{3} c_s, 
	\qquad 
	\lambda^{c,v}_{3,4,5,6} = u \pm c_s, 
	\qquad   
	\lambda^{c,v}_{7,8,\cdots,15} = u. 
	\label{eqn.eval.c}  
\end{equation}
The remaining pressure subsystem, which will be discretized implicitly, reads as follows: 
\begin{equation}
	\label{eqn.pde.im} 
	\frac{\pd \Q}{\pd t} + \nabla \cdot \FF_p(\Q) =0.  
\end{equation}
As already mentioned before, the resulting pressure subsystem is formally identical to the  
Toro-V\'azquez pressure system \cite{ToroVazquez}, hence its eigenvalues in $x$ direction are 
\begin{equation}
	\lambda^p_{1,2} = \frac{1}{2} \left( u \pm \sqrt{u^2 + 4 c_0^2 } \right), 
	\qquad \lambda^p_{3,4,5,\cdots,15} = 0, \qquad 
	\label{eqn.eval.p}  
\end{equation}
with the adiabatic sound speed $c_0$, e.g. for the ideal gas EOS one has, as usual, $c_0^2 = \gamma 
p / \rho$.  

\subsection{Staggered mesh}

To simplify the description of the numerical scheme, we restrict the discussion to two-dimensional 
motion, i.e. we assume that $ \frac{\pd}{\pd x_3} $ vanishes for all fields and thus, we assume a 
two-dimensional physical domain $ \Omega $ spread in $ x_1 = x $ and $ x_2 = y $ and which is 
covered by a set of equidistant and non-overlapping 
Cartesian control volumes $\Omega^{p,q} = [x^{p-\halb},x^{p+\halb}] \times 
[y^{q-\halb},y^{q+\halb}]$ with 
uniform mesh spacings $\Delta x = x^{p+\halb} - x^{p-\halb}$ and $\Delta y = y^{q+\halb} - 
y^{q-\halb}$ 
in $x$ and $y$ direction, respectively, and with $x^{p \pm \halb}=x^p \pm \Delta x / 2$ and 
$y^{q \pm \halb} = y^q \pm \Delta y /2$. Nevertheless, we keep all third components of 
vectors and tensors in the discussion. The 3D extension of the scheme is 
straightforward, defining the normal components of the velocity field on the faces, and keeping 
the discrete distortion field $\Dist$ in the vertices of the primary control volumes. We will 
furthermore use the notation $\mathbf{e}_x = 
(1,0,0)$, 
$\mathbf{e}_y = (0,1,0)$ and $\mathbf{e}_z = (0,0,1)$ for the unit vectors pointing into the 
directions 
of the Cartesian coordinate axes. Also, we use the following notations for the velocity components 
$ u = v_1 $ and $ v = v_2 $.

\begin{figure}[!htbp]
	\begin{center}
		\includegraphics[width=0.4\textwidth]{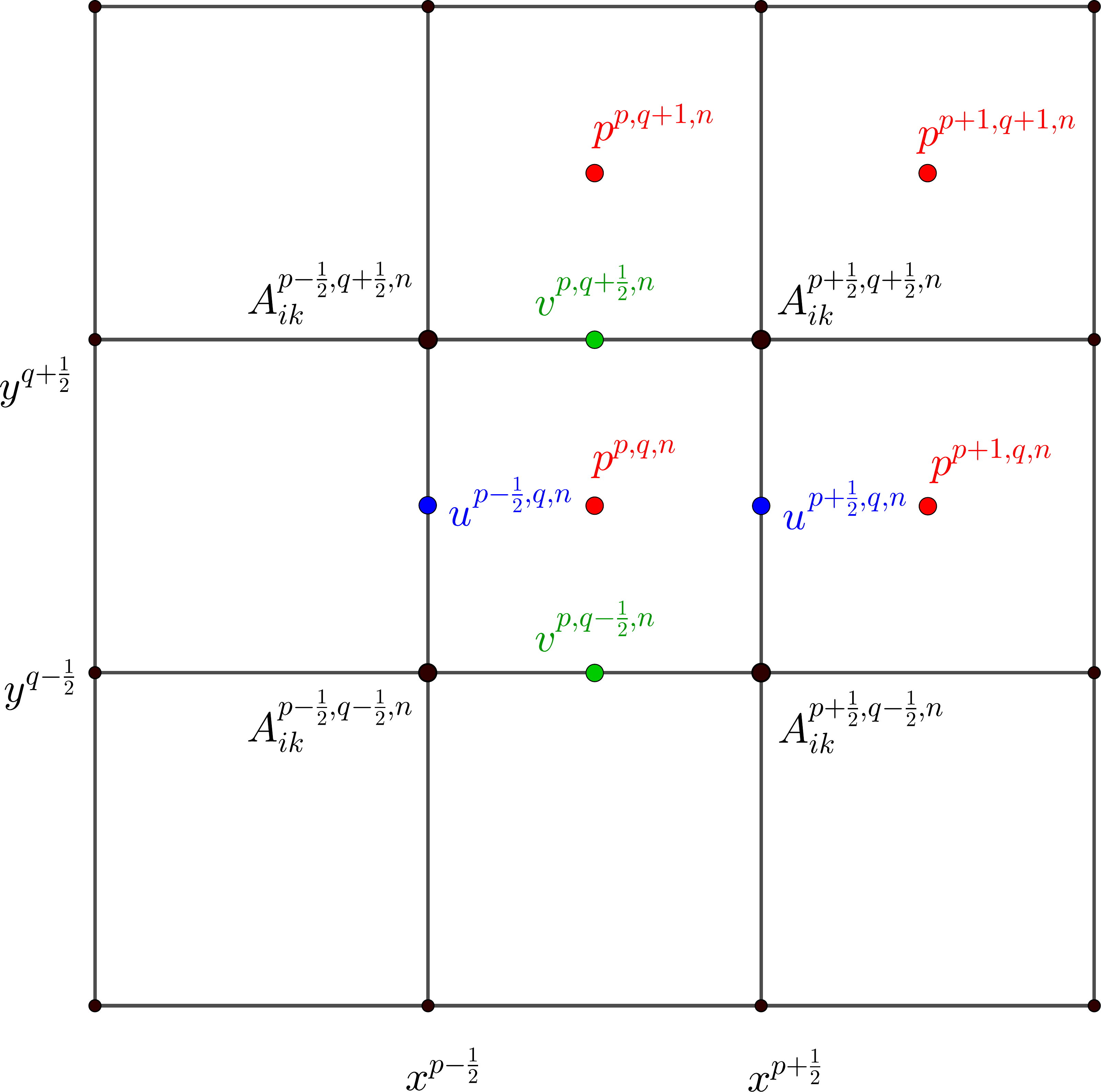}  
		\caption{Staggered mesh configuration with the pressure field $p^{p,q,n}$ defined in the 
		cell barycenters, the velocity field components $u^{p+\halb,q,n}$ and $v^{p,q+\halb,n}$ 
		defined on the edge-based staggered dual grids, respectively, and the distortion field 
		$A_{ik}^{p+\halb,q+\halb,n}$.}  
		\label{fig.staggered}
	\end{center}
\end{figure}

To avoid confusion between tensor indices and discretization indices, throughout this paper we will 
use the \textit{subscripts}  $i,j,k,l,m$ for \textit{tensor indices} and the \textit{superscripts} 
$n,p,q,r,s$ for the \textit{discretization indices} in time 
and space, respectively. 
The discrete spatial coordinates will be denoted by $x^p$ and $y^q$, while the set of discrete 
times will be denoted by $t^n$. For a sketch of the employed staggered grid arrangement of the main 
quantities, see Fig. \ref{fig.staggered}. 

\subsection{Explicit discretization of the nonlinear convective terms and of the corner 
fluxes}\label{sec.explicit}

The semi-implicit scheme used in this paper and proposed in \cite{SIGPR2021} applies an explicit 
discretization 
of the nonlinear convective terms contained in $\FF_c = (\FF_c(\Q),\mathbf{g}_c(\Q))$ 
and of the corner (vertex) fluxes $\FF_v = (\FF_v(\Q),\mathbf{g}_v(\Q))$, 
starting from the known solution $\mathbf{Q}^{p,q,n}$ at time $t^n$. 
The result is a new intermediate state vector $\mathbf{Q}^{p,q,*}$ that is computed via a 
conservative finite volume formulation 
\begin{equation}
	\mathbf{Q}^{p,q,*} = \mathbf{Q}^{p,q,n} 
	- \frac{\Delta t}{\Delta x} \left( \FF_{c,v}^{p+\halb,q} - \FF_{c,v}^{p-\halb,q} 
	\right) 
	- \frac{\Delta t}{\Delta y} \left( \mathbf{g}_{c,v}^{p,q+\halb} - \mathbf{g}_{c,v}^{p,q-\halb} 
	\right), 
	\label{eqn.Qstar} 
\end{equation} 
with the numerical fluxes defined as 
\begin{eqnarray}
	\FF_{c,v}^{p+\halb,q} &=&   \phantom{+} u^{p+\halb,q,n} \halb \left( 
	\mathbf{Q}_{c}\left( \mathbf{Q}^{p+\halb,q,n+\halb}_{-} \right)  + \mathbf{Q}_{c}\left( 
	\mathbf{Q}^{p+\halb,q,n+\halb}_{+} \right) \right) 
	- \halb s^x_{\max} \left( \mathbf{Q}^{p+\halb,q,n+\halb}_{+} - 
	\mathbf{Q}^{p+\halb,q,n+\halb}_{-} \right) \nonumber \\ 
	&& 
	\phantom{u^{p+\halb,q,n}} + \halb \left( \FF_{v}\left( \mathbf{Q}^{p+\halb,q+\halb,n} 
	\right)  + \FF_{v}\left( \mathbf{Q}^{p+\halb,q-\halb,n} \right) \right), 
	\label{eqn.numflux.f} 
\end{eqnarray} 
and 
\begin{eqnarray}
	\mathbf{g}_{c,v}^{p,q+\halb} &=& \phantom{+}  v^{p,q+\halb,n} \halb \left( 
	\mathbf{Q}_{c}\left( \mathbf{Q}^{p,q+\halb,n+\halb}_{-} \right)  + \mathbf{Q}_{c}\left( 
	\mathbf{Q}^{p,q+\halb,n+\halb}_{+} \right) \right) 
	- \halb s^y_{\max} \left( \mathbf{Q}^{p,q+\halb,n+\halb}_{+} - 
	\mathbf{Q}^{p,q+\halb,n+\halb}_{-} \right) \nonumber \\ 
	&& 
	\phantom{ v^{p,q+\halb,n}} + \halb \left( \mathbf{g}_{v}\left( \mathbf{Q}^{p+\halb,q+\halb,n} 
	\right)  + \mathbf{g}_{v}\left( \mathbf{Q}^{p-\halb,q+\halb,n} \right) \right). 
	\label{eqn.numflux.g} 
\end{eqnarray} 
Note that the fluxes above contain the nonlinear convective terms as well as the vertex fluxes 
$\FF_v$ and $\mathbf{g}_v$, which contain the stress tensor $\tensor{\sigma}$. In 
\eqref{eqn.numflux.f} and \eqref{eqn.numflux.g}, the maximum signal 
speeds are computed as 
\begin{eqnarray} 
	s_{\max}^x &=& \max \left( 
	||\boldsymbol{\Lambda}^{c,v}_x(\mathbf{Q}^{p+\halb,q,n+\halb}_{-})||, 
	||\boldsymbol{\Lambda}^{c,v}_x(\mathbf{Q}^{p+\halb,q,n+\halb}_{+})||\right), \nonumber \\ 
	s_{\max}^y &=& \max \left( 
	||\boldsymbol{\Lambda}^{c,v}_y(\mathbf{Q}^{p,q+\halb,n+\halb}_{-})||, 
	||\boldsymbol{\Lambda}^{c,v}_y(\mathbf{Q}^{p,q+\halb,n+\halb}_{+})||\right), 
	\label{eqn.sigspeeds} 
\end{eqnarray} 
with $\boldsymbol{\Lambda}^{c,v}_{k}$ the diagonal matrix of eigenvalues of the explicit subsystem 
\eqref{eqn.pde.ex} in direction $x$ and $y$, respectively. {In \eqref{eqn.sigspeeds} 
	$||\boldsymbol{\Lambda}||$ means the maximum norm of the matrix $\boldsymbol{\Lambda}$, i.e. 
	the 
	maximum absolute value of the matrix.} 
The boundary-extrapolated values are simply computed via a standard MUSCL-Hancock scheme (see 
\cite{toro-book}), as follows: 
\begin{eqnarray}
	\mathbf{Q}^{p+\halb,q,n+\halb}_{-} &=& \mathbf{Q}^{p,q,n}   + \halb \Delta x \, \partial^h_x 
	\mathbf{Q}^{p,q,n}   + \halb \Delta t \partial^h_t \mathbf{Q}^{p,q,n}, \nonumber \\
	\mathbf{Q}^{p+\halb,q,n+\halb}_{+} &=& \mathbf{Q}^{p+1,q,n} - \halb \Delta x \, \partial^h_x 
	\mathbf{Q}^{p+1,q,n} + \halb \Delta t \partial^h_t \mathbf{Q}^{p+1,q,n}, \nonumber  
\end{eqnarray} 
and 
\begin{eqnarray}
	\mathbf{Q}^{p,q+\halb,n+\halb}_{-} &=& \mathbf{Q}^{p,q,n}   + \halb \Delta y \, \partial^h_y 
	\mathbf{Q}^{p,q,n}   + \halb \partial_t \mathbf{Q}^{p,q,n}, \nonumber \\
	\mathbf{Q}^{p,q+\halb,n+\halb}_{+} &=& \mathbf{Q}^{p,q+1,n} - \halb \Delta y \, \partial^h_y 
	\mathbf{Q}^{p,q+1,n} + \halb \partial_t \mathbf{Q}^{p,q+1,n}, \nonumber   
\end{eqnarray} 
with the discrete gradients in space and time computed via  
\begin{eqnarray}
	\partial^h_x \mathbf{Q}^{p,q,n} &=& \textnormal{minmod} \left( \frac{\mathbf{Q}^{p+1,q,n} - 
	\mathbf{Q}^{p,q,n}}{\Delta x}, \frac{\mathbf{Q}^{p,q,n} - \mathbf{Q}^{p-1,q,n}}{\Delta x}  
	\right), \nonumber \\
	\partial^h_y \mathbf{Q}^{p,q,n} &=& \textnormal{minmod} \left( \frac{\mathbf{Q}^{p+1,q,n} - 
		\mathbf{Q}^{p,q,n}}{\Delta y}, \frac{\mathbf{Q}^{p,q,n} - \mathbf{Q}^{p-1,q,n}}{\Delta y}  
		\right), 
	\nonumber 
\end{eqnarray} 
and
\begin{eqnarray}
	%  \partial^h_t \mathbf{Q}^{p,q,n} &=& - \frac{\FF_{c}\left( \mathbf{Q}^{p+\halb,q,n} 
	%\right) - \FF_{c}\left( \mathbf{Q}^{p-\halb,q,n} \right)  }{\Delta x} \nonumber \\    
	%                                  &=& - \frac{\mathbf{g}_{c}\left( \mathbf{Q}^{p,q+\halb,n} 
	%\right) - \mathbf{g}_{c}\left( \mathbf{Q}^{p,q-\halb,n} \right)  }{\Delta x} \nonumber \\    
	\partial^h_t \mathbf{Q}^{p,q,n} &=& - \frac{\FF_{c}\left( \mathbf{Q}^{p,q,n} + \halb 
	\Delta x \, \partial^h_x \mathbf{Q}^{p,q,n}\right) - \FF_{c}\left( \mathbf{Q}^{p,q,n} - 
	\halb \Delta x \, \partial^h_x \mathbf{Q}^{p,q,n}\right)  }{\Delta x} \nonumber \\    
	&& - \frac{ \FF_{v}\left( \mathbf{Q}^{p+\halb,q+\halb,n} \right)  + \FF_{v}\left( 
	\mathbf{Q}^{p+\halb,q-\halb,n} \right) - 
		\FF_{v}\left( \mathbf{Q}^{p-\halb,q+\halb,n} \right)  + \FF_{v}\left( 
		\mathbf{Q}^{p-\halb,q-\halb,n} \right)}{2 \Delta x} \nonumber \\ 
	&&  - \frac{\mathbf{g}_{c}\left( \mathbf{Q}^{p,q,n} + \halb \Delta y \, \partial^h_y 
	\mathbf{Q}^{p,q,n}\right) - \mathbf{g}_{c}\left( \mathbf{Q}^{p,q,n} - \halb \Delta y \, 
	\partial^h_y \mathbf{Q}^{p,q,n}\right)  }{\Delta y} \nonumber \\ 
	&& -\frac{\mathbf{g}_{v}\left( \mathbf{Q}^{p+\halb,q+\halb,n} \right)  + \mathbf{g}_{v}\left( 
	\mathbf{Q}^{p-\halb,q+\halb,n} \right) - 
		\mathbf{g}_{v}\left( \mathbf{Q}^{p+\halb,q-\halb,n} \right)  - \mathbf{g}_{v}\left( 
		\mathbf{Q}^{p-\halb,q-\halb,n} \right)  }{2 \Delta y} 
	. \nonumber  
\end{eqnarray}

\subsection{Discrete divergence, curl and gradient operators}

The main ingredients of the new structure-preserving staggered semi-implicit scheme proposed in this paper are the definitions of
appropriate discrete divergence, gradient and curl operators acting on quantities that are arranged in different and judiciously 
chosen locations on the staggered mesh. The discrete pressure field at time $t^n$ is denoted by $p^{h,n}$ and its degrees of freedom are located in the center of each control volume as $p^{p,q,n}=p(x^p,y^q,t^n)$. Throughout this paper we denote 
with the superscript $h$ the set of all degrees of freedom of the discrete solution and all degrees of freedom generated by a 
discrete operator, in order to ease the notation. The discrete  velocities $v_1^{h,n}$ and $v_2^{h,n}$ are arranged in an edge-based staggered fashion, i.e.  
$u^{p+\halb,q,n}:=v_1^{p+\halb,q,n}=v_1(x^{p+\halb},y^q,t^n)$ and 
$v^{p,q+\halb,n}:=v_2^{p,q+\halb,n}=v_2(x^p,y^{q+\halb},t^n)$. 
The discrete distortion field $\Dist^{h,n}$ is defined on 
the \textit{vertices} of each spatial control volume as  
$A_{ik}^{p+\halb,q+\halb,n} = A_{ik}(x^{p+\halb},y^{q+\halb},t^n)$. For clarity, 
see again Fig.\,\ref{fig.staggered}.  

\begin{figure}[!htbp]
	\begin{minipage}[c]{0.5\textwidth} 
		\includegraphics[width=0.67\textwidth]{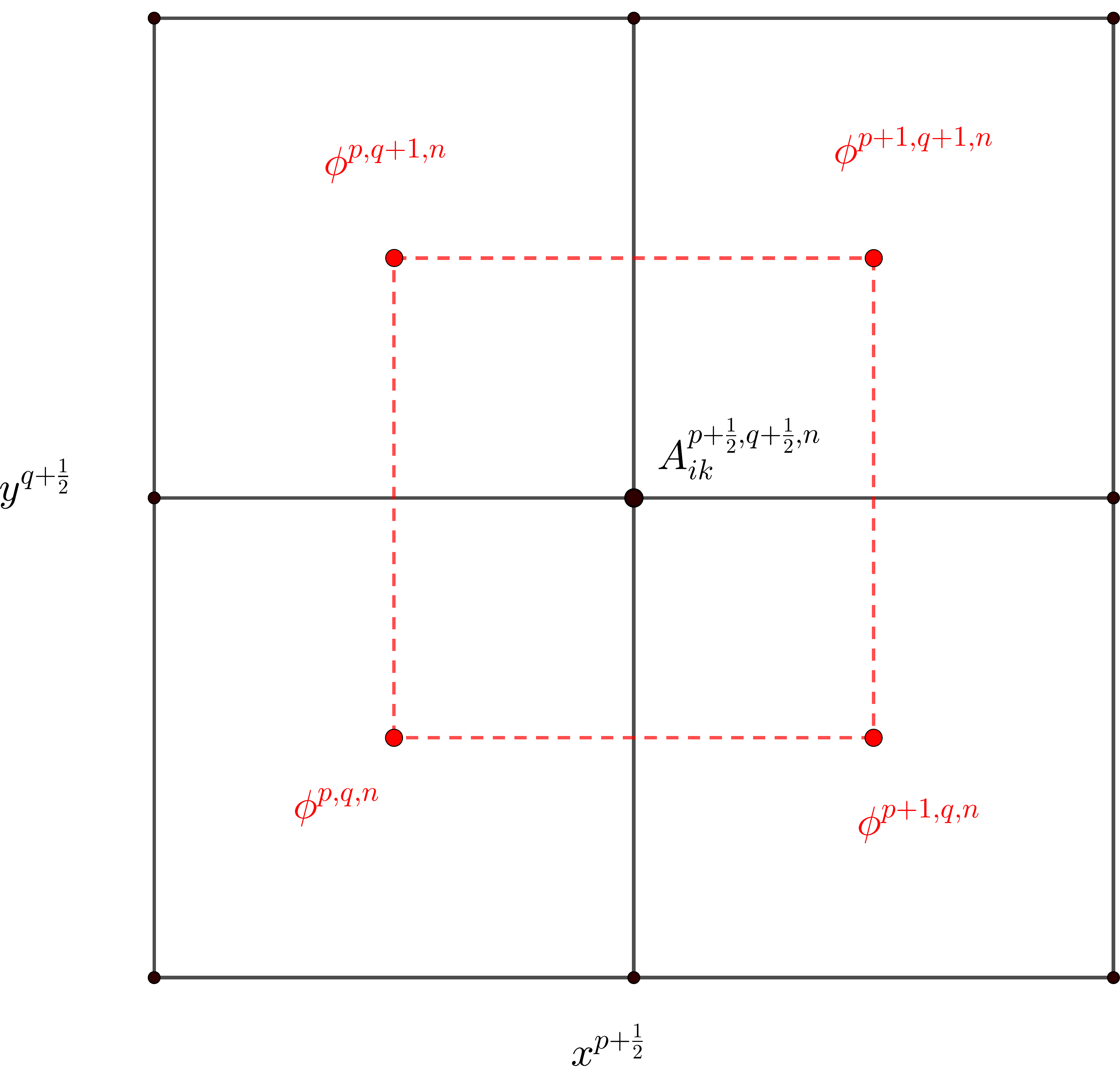} 
	\end{minipage} 
	\begin{minipage}[c]{0.5\textwidth} 
		\includegraphics[width=0.65\textwidth]{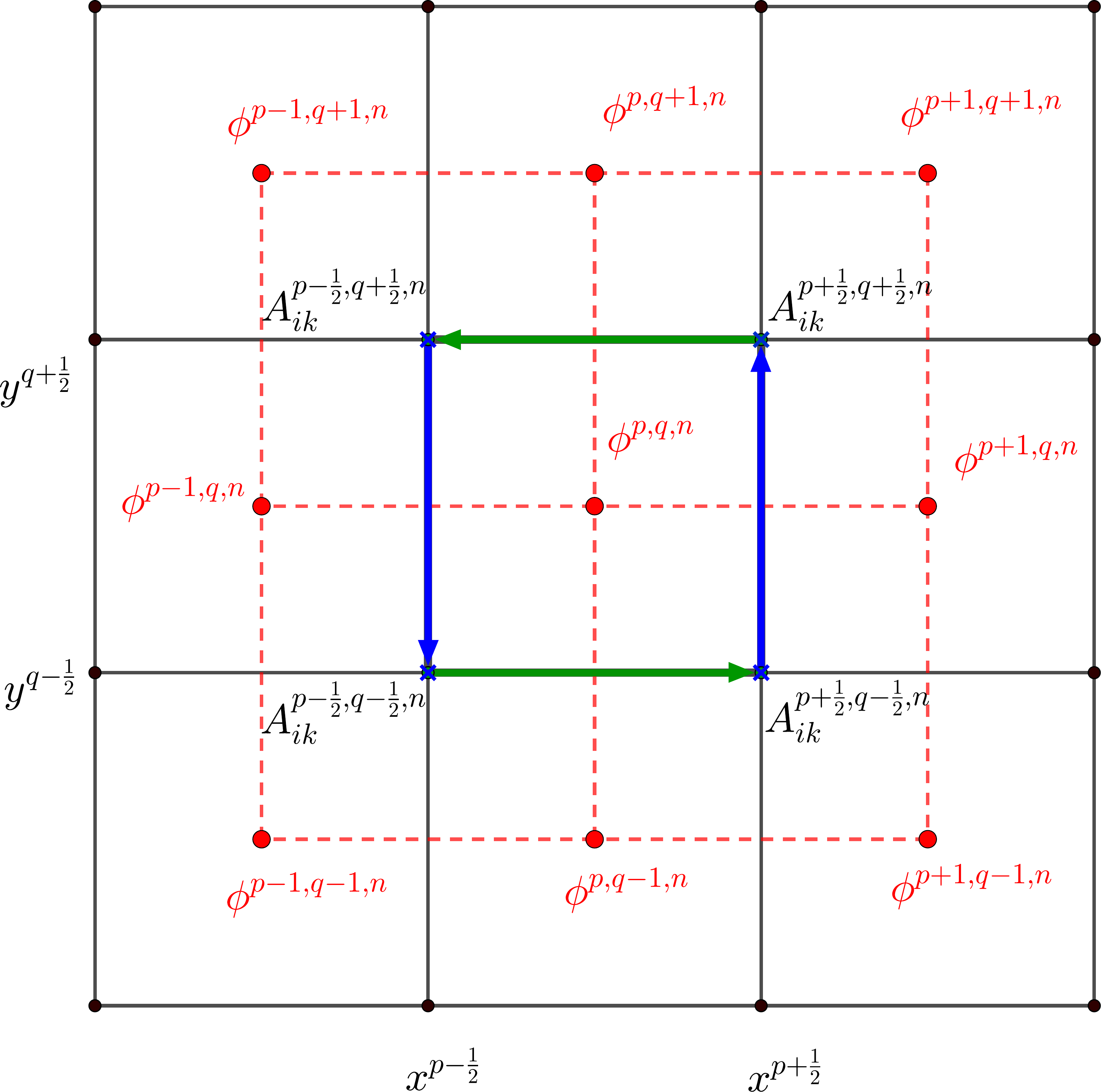} 
	\end{minipage}    
	%\end{tabular} 	
	\caption{Left: stencil of the discrete gradient operator, which computes the corner gradient of a scalar field defined in the cell barycenter. Right: stencil of the discrete curl operator, which defines the curl inside the cell barycenter using the vector field components defined in the corners of the primary control volume. In the
		right panel we also show the total 9-point stencil that is needed for the discrete identity $\nabla^h \times \nabla^h \phi^{h,n} = 0$.}  
	\label{fig.grad.curl}
\end{figure}

The \textit{discrete divergence operator}, $\nabla^h \cdot$, acting on a discrete vector field 
$\vv^{h,n}$ is abbreviated 
by $\nabla^h \cdot \vv^{h,n}$ and its degrees of freedom are given by  
\begin{equation}
\nabla^{p,q} \cdot \vv^{h,n}  =  
\frac{u^{p + \halb,q,n} - u^{p - \halb,q,n}}{\Delta x} + 
\frac{v^{p,q + \halb,n} - v^{p,q - \halb,n}}{\Delta y}, 
\label{eqn.div} 
\end{equation}
i.e. it is based on the \textit{edge-based} staggered values of the field $\vv^{h,n}$. It defines  
a discrete divergence on the control volume $\Omega^{p,q}$ via the Gauss theorem, 
\begin{equation}
\nabla^{p,q} \cdot \vv^{h,n} = \frac{1}{\Delta x \Delta y} \int \limits_{\Omega^{p,q}} \nabla \cdot 
\vv d \xx = \frac{1}{\Delta x \Delta y} \int \limits_{\partial \Omega^{p,q}} \vv \cdot 
\nn \, dS,  
\label{eqn.gauss}
\end{equation}
based on the mid-point rule for the computation of the integrals along each edge of $\Omega^{p,q}$. In \eqref{eqn.gauss} the
outward pointing unit normal vector to the boundary $\partial \Omega^{p,q}$ of $\Omega^{p,q}$ is 
denoted by $\nn$. 
In a similar manner, the $z$ component of the \textit{discrete curl}, $\nabla^h \times $, of a 
discrete vector field 
$\vv^{h,n}$
is denoted by $\left( \nabla^h \times \vv^{h,n} \right) \cdot \mathbf{e}_z$ and its degrees of 
freedom are naturally defined as
\begin{eqnarray}
\left( \nabla^{p,q} \times \vv^{h,n} \right) \cdot \mathbf{e}_z &=& \varepsilon_{3jk} 
\partial_j^{p,q} v_k^{h,n} \nonumber \\ 
&=&  
\halb \left (\frac{v^{p + \halb,q +\halb,n} - v^{p - \halb,q + \halb,n}}{\Delta x} + 
\frac{v^{p + \halb,q -\halb,n}  - v^{p - \halb, q -\halb,n} }{\Delta x} \right ) 
\nonumber \\ &-&  
\halb \left (\frac{u^{p + \halb,q +\halb,n} - u^{p + \halb,q - \halb,n}}{\Delta y} 
+ \frac{u^{p - \halb,q +\halb,n} - u^{p - \halb, q -\halb,n} }{\Delta y} \right ), 
\label{eqn.rot} 
\end{eqnarray}
making use of the \textit{vertex-based} staggered values of the field $\vv^{h,n}$, see the right 
panel in Fig. \ref{fig.grad.curl}. In Eqn. \eqref{eqn.rot} the symbol $\varepsilon_{ijk}$ is again 
the usual Levi-Civita tensor. Eqn. \,\eqref{eqn.rot} defines a discrete
curl on the control volume $\Omega^{p,q}$ via the Stokes theorem
\begin{equation}
\left( \nabla^h \times \vv^{h,n} \right) \cdot \mathbf{e}_z = \frac{1}{\Delta x \Delta y} \int 
\limits_{\Omega^{p,q}} 
\left( \nabla \times  \vv \right) \cdot \mathbf{e}_z \, d\xx = \frac{1}{\Delta x \Delta y} 
\int \limits_{\partial \Omega^{p,q}} \vv \cdot \bm{t} \, dS,  
\label{eqn.stokes}
\end{equation}
based on the trapezoidal 
rule for the computation of the integrals along each edge of $\Omega^{p,q}$ (here, $ \bm{t} $ 
stands 
for the tangent vector to the face of the control volume). Since the distortion 
field $\Dist$ transforms as a vector and not as a rank 2 tensor ($\Dist$ is a triad and thus a set
of three vectors), the degrees of freedom of the $z$-component of the discrete curl of 
$\Dist^{h,n}$ simply read  
\begin{eqnarray}
\left (\left( \nabla^{p,q} \times \Dist^{\!h,n} \right) \cdot \mathbf{e}_z \right )_i &=&  
\varepsilon_{3jk} 
\partial_j^{p,q} A_{ik}^{h,n}  \nonumber \\ 
&=&  
\halb \left (\frac{A_{i2}^{p + \halb,q +\halb,n} - A_{i2}^{p - \halb,q + \halb,n}}{\Delta x} + 
\frac{A_{i2}^{p + \halb,q -\halb,n} - A_{i2}^{p - \halb, q -\halb,n} }{\Delta x} \right ) 
\nonumber \\ &-&  
\halb \left (\frac{A_{i1}^{p + \halb,q +\halb,n} - A_{i1}^{p + \halb,q - \halb,n}}{\Delta y} + 
\frac{A_{i1}^{p - \halb,q +\halb,n} - A_{i1}^{p - \halb, q -\halb,n} }{\Delta y} \right ). 
\label{eqn.rotA} 
\end{eqnarray}
Last but not least, we need to define a discrete gradient operator that is compatible with the discrete curl,
so that the continuous identity
\begin{equation}
\nabla \times \nabla \phi = 0
\label{eqn.rotgrad} 
\end{equation}
also holds on the discrete level. If we define a scalar field in the barycenters of the control volumes $\Omega^{p,q}$ as
$\phi^{p,q,n}=\phi(x^p,y^q,t^n)$ then the corner gradient generates a natural discrete gradient operator $\nabla^{h}$ 
of the discrete scalar field $\phi^{h,n}$  that defines a discrete gradient in all vertices of the mesh. 
The corresponding degrees of freedom generated by $\nabla^{h} \phi^{h,n}$ read (see the left panel 
of Fig. \ref{fig.grad.curl})
\begin{equation}
\label{discr.grad}
\nabla^{p+\halb,q+\halb}  \phi^{h,n} = \partial_k^{p+\halb,q+\halb} \phi^{h,n} = \left( 
\begin{array}{c}  
\halb \left( \frac{\phi^{p + 1, q + 1,n} - \phi^{p, q + 1,n}}{\Delta x} 
+ \frac{\phi^{p + 1, q,n} - \phi^{p, q,n} }{\Delta x} \right) \\
\halb \left( \frac{\phi^{p + 1, q + 1,n} - \phi^{p + 1, q,n}}{\Delta y} + \frac{\phi^{p, q + 
		1,n} 
	- \phi^{p,q,n} }{\Delta 
	y} \right)  \\ 
\displaystyle 0
\end{array} \right).
%	= 
%	\left( \begin{array}{c}  
%	\halb \left( \PD_1^{p + \halb, q + 1}\phi^{h,n} + \PD_1^{p + \halb, q}\phi^{h,n} \right) \\
%	\halb \left( \PD_2^{p + 1, q + \halb}\phi^{h,n} + \PD_2^{p, q + \halb}\phi^{h,n} \right) \\ 
%	\displaystyle 0
%	\end{array} \right), 
\end{equation}
%Here, 
%\begin{equation}\label{FD}
%	\PD_1^{p+\halb,q} \phi^{h,n} = \frac{\phi^{p+1,q,n} - \phi^{p,q,n}}{\Delta x}, 
%	\qquad
%	\PD_2^{p,q+\halb} \phi^{h,n} = \frac{\phi^{p,q+1,n} - \phi^{p,q,n}}{\Delta y}.
%\end{equation}

It is then straightforward to verify that an immediate consequence of \eqref{eqn.rot} and 
\eqref{discr.grad} is 
\begin{equation}
\nabla^h \times \nabla^h \phi^{h,n} = 0, 
\label{eqn.discrete.curl} 
\end{equation}
i.e. one obtains a discrete analogue of \eqref{eqn.rotgrad}. 
We furthermore define the following averaging operators from the three different staggered meshes 
to the cell barycenter $(x^p,y^q)$: 
\begin{eqnarray}
A_{ik}^{p,q,n} &=& \frac{1}{4} \left( A_{ik}^{p-\halb,q-\halb,n} + A_{ik}^{p+\halb,q-\halb,n} +  
A_{ik}^{p-\halb,q+\halb,n} + A_{ik}^{p+\halb,q+\halb,n} \right), 
\nonumber \\ 
u^{p,q,n} &=& \frac{1}{2} \left( u^{p-\halb,q,n} + u^{p+\halb,q,n} \right),     
\nonumber \\                                 								   
v^{p,q,n} &=& \frac{1}{2} \left( v^{p,q-\halb,n} + v^{p,q+\halb,n} 
\right).                                   								   
\end{eqnarray}

%Finally, we note that definition \eqref{discr.grad} of the discrete gradient fulfills the following 
%discrete product rule 
%\begin{equation}\label{discr.prod.rule.x}
%\pd_1^{p+\halb,q+\halb}(\phi^{h,n}\psi^{h,n}) =
%\left( 
%\phi^{p+\halb,*}\PD_1^{p+\halb,*}\psi^{h,n} 
%\right)^{*,q+\halb}
%+
%\left(
%\psi^{p+\halb,*}\PD_1^{p+\halb,*}\phi^{h,n}
%\right)^{*,q+\halb}.
%\end{equation}
%\begin{equation}\label{discr.prod.rule.y}
%\pd_2^{p+\halb,q+\halb}(\phi^{h,n}\psi^{h,n}) =
%\left( 
%\phi^{*,q+\halb}\PD_2^{*,q+\halb}\psi^{h,n} 
%\right)^{p+\halb,*}
%+
%\left(
%\psi^{*,q+\halb}\PD_2^{*,p+\halb}\phi^{h,n}
%\right)^{p+\halb,*}.
%\end{equation}
%for arbitrary scalar functions $ \phi $ and $ \psi $ stored at the same grid, see 
%Appendix\,\ref{sec.appendix.producd.rule}. 
%Here, the asterisk in the superscripts mean that this indexes is taken from the original grid.
%\textcolor{red}{[If however, the functions are stored on 
%different grids like $ J_k $ and $ v_k $ in $ \pd_k(J_m v_m) $ then the validity of this product 
%rule is not clear...]}

\subsection{Explicit, curl-free compatible discretization of the distortion field} 

The key ingredient of the numerical method \cite{SIGPR2021} is the proper discretization of the 
thermal impulse equation and of the PDE for the distortion field. We employ the following 
compatible discretization 
\begin{eqnarray}
	A_{ik}^{p+\halb,q+\halb,n+1} &=&   A_{ik}^{p+\halb,q+\halb,n} - \Delta t \, 
	\partial^{p+\halb,q+\halb}_k \left( A_{im}^{h,n} v_m^{h,n} \right)  \nonumber \\ 
	&-& \Delta t \frac{1}{4} \sum \limits_{r=0}^{1} \sum \limits_{s=0}^{1} v_m^{p+r,q+s,n} \left( 
	\partial_m^{p+r,q+s} A_{ik}^{h,n} - \partial_k^{p+r,q+s} A_{im}^{h,n} \right) \nonumber \\
	&-& \Delta t \frac{\left| \det(\Dist)^{p+\halb,q+\halb,n+1} \right|^{\frac{5}{3}}}{\tau}  
	A_{ij}^{p+\halb,q+\halb,n+1} {\dev{G}_{jk}}^{p+\halb,q+\halb,n+1}. 
	\label{eqn.Ah}
\end{eqnarray}
It is easy to check that in the homogeneous case (when $\tau \to \infty$ and therefore the 
algebraic source term vanishes)  
for an initially curl-free vector field $\Dist^{h,n}$ that satisfies 
$\nabla^{h} \times \Dist^{h,n} = 0$ also $\nabla^{h} \times \Dist^{h,n+1} = 0$ holds. To see this, 
one needs  
to apply the discrete curl operator $\nabla^h \times$ to \eqref{eqn.Ah}.  One realizes 
that the second row of \eqref{eqn.Ah}, which contains the discrete curl of $\Dist^{h,n}$ 
vanishes immediately, 
due to $\nabla^{h} \times \Dist^{h,n} = 0$. The third row vanishes because $\tau \to \infty$. 
The curl of the first term on the right hand side in the first row of \eqref{eqn.Ah} is zero 
because of 
$\nabla^{h} \times \Dist^{h,n} = 0$ and the curl of the second term is zero because of 
$\nabla^h \times \nabla^h \phi^{h,n} = 0$, with the auxiliary scalar field $\phi^{h,n} = 
A_{im}^{h,n} v_m^{h,n} $, whose 
degrees of freedom are computed as $\phi^{p,q,n} = A_{im}^{p,q,n} v_m^{p,q,n} $ after averaging of 
the velocity vector and 
the distortion into the barycenters of the control volumes $\Omega^{p,q}$.
{Note that on uniform grids this average is second order accurate.}

\subsection{Implicit solution of the pressure equation} \label{sec.pressure.system}

Up to now, the contribution of the pressure to the momentum and to the total energy conservation 
laws, i.e. the terms contained in the pressure fluxes $\FF_p$,  
have been excluded. The discrete momentum equations including the pressure terms read (recall that 
$ u = v_1 $ and $ v = v_2 $)
\begin{eqnarray}
\label{eqn.rhou2d} 
(\rho u)^{p+\halb,q,n+1} &=& (\rho u)^{p+\halb,q,*} - \frac{\Delta t}{\Delta x} \left( 
p^{p+1,q,n+1} - p^{p,q,n+1} \right),  \nonumber \\ 
(\rho v)^{p,q+\halb,n+1} &=& (\rho v)^{p,q+\halb,*} - \frac{\Delta t}{\Delta y} \left( 
p^{p,q+1,n+1} - p^{p,q,n+1} \right),  
\end{eqnarray} 
where pressure is taken \textit{implicitly}, while all nonlinear convective terms and the vertex 
fluxes have already been discretized \textit{explicitly} via the operators $(\rho u)^{p+\halb,q,*} 
$ and 
$(\rho v)^{p,q+\halb,*}$ given in \eqref{eqn.Qstar} and after averaging of the obtained 
quantities back to the 
edge-based staggered dual grid. {Recalling that $ E_{\rm i}(\rho,p) $, $ E_{\rm e}(A_{ik}) 
	$, and $ E_{\rm k}(v_k) $ are 
	three contributions to the specific total energy $ E $ given by \eqref{eqn.EOS},} a 
preliminary form of the discrete total energy equation reads 
\begin{eqnarray}
\label{eqn.rhoE2d.prelim} 
\rho E_{\rm i}\left( p^{p,q,n+1} \right) + \rho E_{\rm e}^{p,q,n+1} + \rho \tilde{E_{\rm 
k}}^{p,q,n+1}  
&=& \rho E^{p,q,*}  
\nonumber \\      
&-& \frac{\Delta t}{\Delta x} \left(  \tilde{h}^{p+\halb,q,n+1} (\rho u)^{p+\halb,q,n+1} - 
\tilde{h}^{p-\halb,q,n+1} (\rho u)^{p-\halb,q,n+1} \right)\phantom{.}     \nonumber \\
&-& \frac{\Delta t}{\Delta y} \left(  \tilde{h}^{p,q+\halb,n+1} (\rho v)^{p,q+\halb,n+1} - 
\tilde{h}^{p,q-\halb,n+1} (\rho v)^{p,q-\halb,n+1} \right). %\nonumber \\  
\end{eqnarray}
Here, we have used the abbreviation $\rho E_{\rm i}\left( p^{p,q,n+1} \right) = \rho^{p,q,n+1} 
E_{\rm i}\left( p^{p,q,n+1}, \rho^{p,q,n+1} \right)$.   
Inserting the discrete momentum equations \eqref{eqn.rhou2d} into the discrete energy equation \eqref{eqn.rhoE2d.prelim} and making tilde symbols explicit via a simple Picard iteration 
(using the lower index $r$ in the following), as suggested in \cite{DumbserCasulli2016,SIMHD}, leads to the following  discrete wave equation for the unknown pressure:  
\begin{eqnarray}
\label{eqn.p2d} 
\rho^{p,q,n+1} E_{\rm i}\left( p_{r+1}^{p,q,n+1}, \rho^{p,q,n+1} \right) & & \nonumber \\
- \frac{\Delta t^2}{\Delta x^2} \left(  {{h}_{r}^{p+\halb,q,n+1}} \left( p_{r+1}^{p+1,j,n+1}-p_{r+1}^{p,q,n+1} \right) 
- {{h}_{r}^{p-\halb,q,n+1}} \left( p_{r+1}^{p,q,n+1}-p_{r+1}^{p-1,q,n+1} \right) \right) & & \nonumber \\  
- \frac{\Delta t^2}{\Delta y^2}\left(  {{h}_{r}^{p,q+\halb,n+1}} \left( p_{r+1}^{p,q+1,n+1}-p_{r+1}^{p,q,n+1} \right) 
- {{h}_{r}^{p,q-\halb,n+1}} \left( p_{r+1}^{p,q,n+1}-p_{r+1}^{p,q-1,n+1} \right) \right)  
& = & b_{r}^{p,q,n}, % \nonumber \\ 
\end{eqnarray} 
with the known right hand side 
\begin{eqnarray}
b_r^{p,q,n}  &=& \rho E_{}^{p,q,*} - \rho E_{\rm e}^{p,q,n+1} - \rho E_{{\rm k},r}^{p,q,n+1} 
\nonumber \\
&-& \frac{\Delta t}{\Delta x} \left( 
{h}_{r}^{p+\halb,q,n+1} (\rho u)^{p+\halb,q,*} - 
{h}_{r}^{p-\halb,q,n+1} (\rho u)^{p-\halb,q,*} \right) \phantom{.} \nonumber \\ 
&-& \frac{\Delta t}{\Delta y} \left( {h}_{r}^{p,q+\halb,n+1} (\rho v)^{p,q+\halb,*} - 
{h}_{r}^{p,q-\halb,n+1} (\rho v)^{p,q-\halb,*} \right). 
\label{eqn.rhs.2d} 
\end{eqnarray} 
The density at the new time $\rho_{}^{p,q,n+1} = \rho_{}^{p,q,*} $ is already known from \eqref{eqn.Qstar}, and also the energy 
contribution $\rho E_{\rm e}^{p,q,n+1}$ of the distortion field $\Dist^{h,n+1}$ is already known, 
after averaging onto the main grid of the 
staggered field components that have been evolved in the vertices
via the compatible discretization \eqref{eqn.Ah} and \eqref{eqn.Ah}.  
The final system for the pressure \eqref{eqn.p2d} forms a \textit{mildly nonlinear system} of the form 
\begin{equation}
\rho \mathbf{E}_{\rm i} \left( \mathbf{p}_{r+1}^{n+1} \right) + \mathbf{M}_r \cdot 
\mathbf{p}_{r+1}^{n+1} = \mathbf{b}_r^{n} 
\label{eqn.nonlinear} 
\end{equation}
as in \cite{DumbserCasulli2016}, with a linear part contained in $\mathbf{M}_r$ that is symmetric and at least positive semi-definite. {In one space dimension 
	$\mathbf{M}_r$ is tridiagonal, while it is pentadiagonal in two space dimensions. } Hence, with the usual  assumptions on the nonlinearity detailed in \cite{CasulliZanolli2012}, it can be efficiently solved with 
the nested Newton method of Casulli and Zanolli \cite{CasulliZanolli2010,CasulliZanolli2012}. 
{As already stated in \cite{DumbserCasulli2016} for equations of state whose dependence of the internal energy density is \textit{linear} in the pressure, the system \eqref{eqn.nonlinear} reduces to a \textit{linear} one. Only for more complex cubic or tabulated equations of state, \eqref{eqn.nonlinear} becomes nonlinear.} 

We further note that in the incompressible 
limit (i.e. when the Mach number tends to zero, ${\rm Ma} \to 0$), following the asymptotic 
analysis performed in 
\cite{KlaMaj,KlaMaj82,Klein2001,Munz2003}, 
the pressure tends to a constant and the contribution of the kinetic energy $\rho E_{\rm k}$  
can be neglected w.r.t. $\rho E_{\rm i}$. Therefore, in the incompressible limit the system 
\eqref{eqn.p2d} tends to the usual pressure Poisson equation of incompressible flow solvers. 
In each Picard iteration, after the solution of the pressure system \eqref{eqn.p2d}, the enthalpies at the interfaces are  recomputed and the momentum is updated  by 
\begin{eqnarray}
\label{eqn.rhou2d.pic} 
(\rho u)_{r+1}^{p+\halb,q,n+1} &=& (\rho u)^{p+\halb,q,*} - \frac{\Delta t}{\Delta x} \left( 
p_{r+1}^{p+1,q,n+1} - p_{r+1}^{p,q,n+1} \right),  \\
(\rho v)_{r+1}^{p,q+\halb,n+1} &=& (\rho v)^{p,q+\halb,*} - \frac{\Delta t}{\Delta y} \left( 
p_{r+1}^{p,q+1,n+1} - p_{r+1}^{p,q,n+1} \right),  
\end{eqnarray} 
from which the new kinetic energy density $(\rho E)_{{\rm k},r+1}^{p,q,n+1}$ can be computed after 
averaging the momentum onto the main grid.  
At the end of the Picard iterations, the total energy is updated as 
\begin{eqnarray}
\label{eqn.rhoE2d} 
(\rho E)_{}^{p,q,n+1} &=& (\rho E)_{}^{p,q,*}  
- \frac{\Delta t}{\Delta x} \left( {h}^{p+\halb,q,n+1} (\rho u)^{p+\halb,q,n+1} - 
{h}_{}^{p-\halb,q,n+1} (\rho u)^{p-\halb,q,n+1} \right) \nonumber \\
& & \phantom{(\rho E)_{}^{p,q,*}} 
- \frac{\Delta t}{\Delta y} \left( {h}_{}^{p,q+\halb,n+1} (\rho v)^{p,q+\halb,n+1} - 
{h}_{}^{p,q-\halb,n+1} (\rho v)^{p,q-\halb,n+1} \right),   
% \nonumber \\ 
\end{eqnarray} 
while the final momentum is averaged back onto the main grid. This completes the description of our 
new curl-free semi-implicit finite volume scheme for the GPR model of continuum mechanics in the 
two-dimensional case recently proposed in \cite{SIGPR2021}. {Due to the \textit{operator splitting} 
between the explicit subsystem and the implicit pressure system and due to the fully implicit 
treatment of the algebraic source terms (backward Euler), the scheme proposed in this paper is 
globally only first order accurate in time. To obtain higher order in time, an IMEX Runge-Kutta 
scheme should be used, see \cite{PareschiRusso2000,PareschiRusso2005,BDLTV2020,BDT2021}.}

\section{Numerical results}\label{sec.results}

Throughout this section, we use the notations $ u = v_1 $, $ v = v_2 $ for the velocity components, 
and $ 
x = x_1 $, $ y = x_2 $ for the coordinates. In all the presented test cases, the CFL number for the 
explicit step of SISPFV scheme was set 
to~$ 0.95 $ and was chosen based on the shear sound speed $ \csh $ which was set to $ 10 $ or 
higher and thus, greater then the flow velocity which was typically of the order of $ 1 $.

\subsection{Couette flow}\label{sec.Couette}

\begin{figure}[!htbp]
	%	\captionsetup{format=plain}
	\begin{center}
		\includegraphics[draft=false,trim=0 0 0 0,clip, scale=0.5]{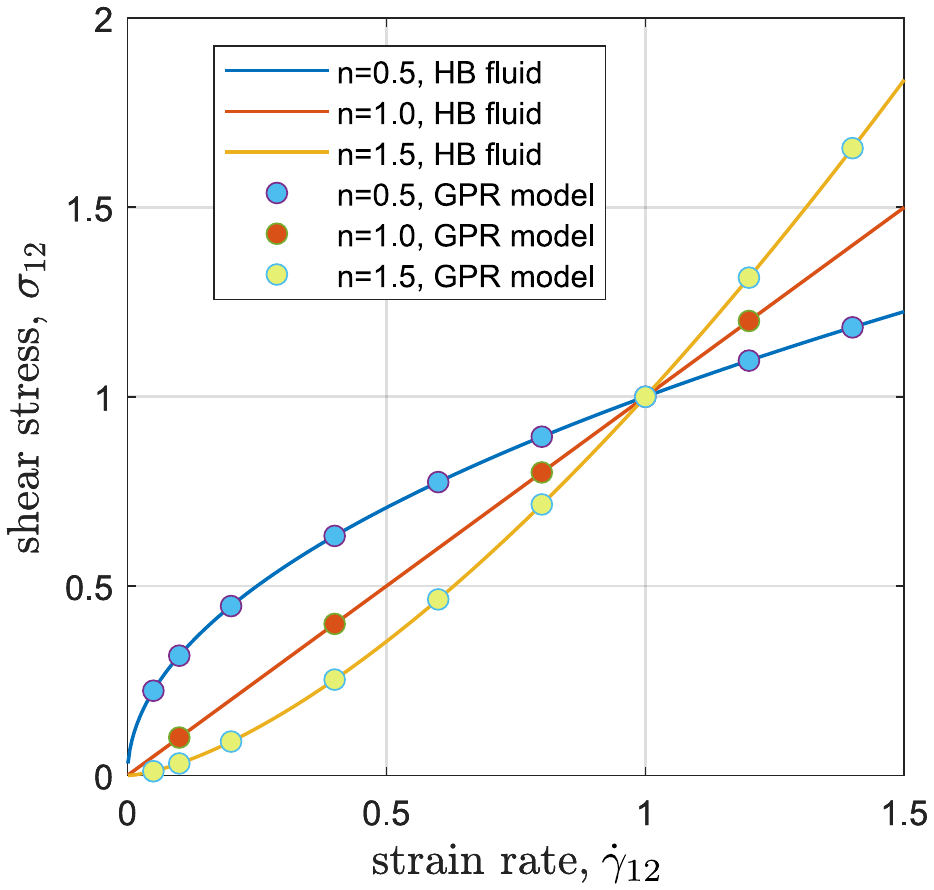}
		\hspace{1cm}
		\includegraphics[draft=false,trim=0 2 0 0,clip, scale=0.515]{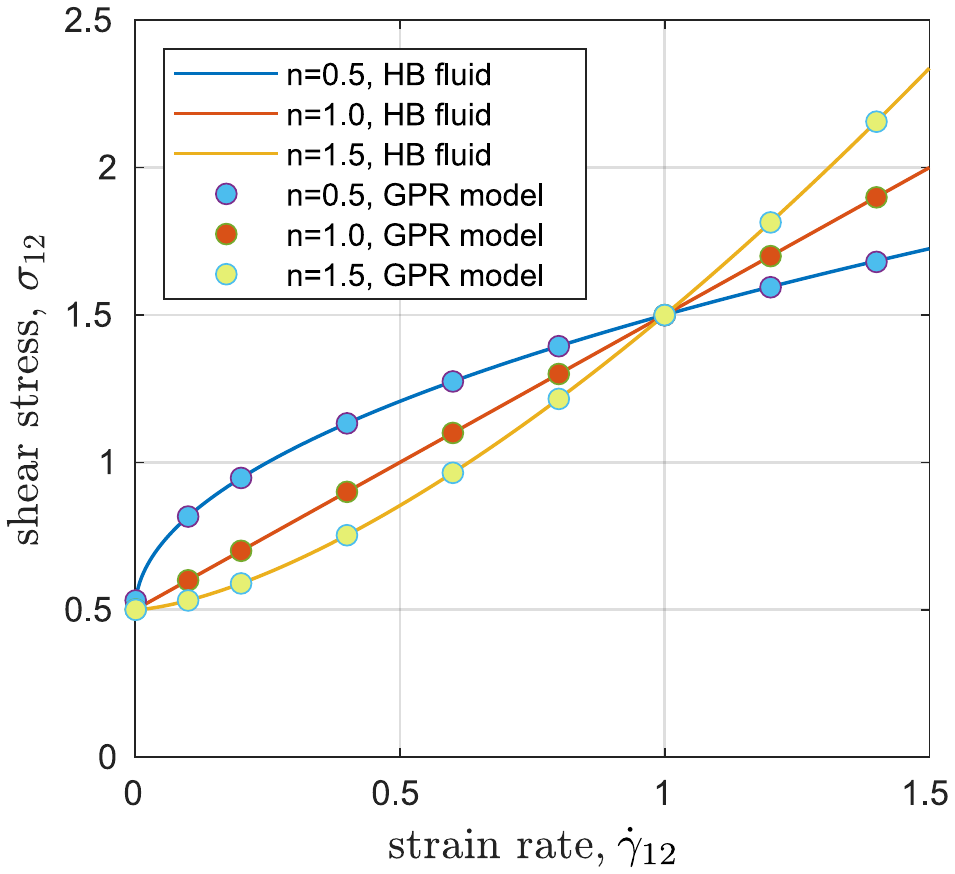}
		\caption{Couette flow. Comparison of the numerical solution to the GPR model obtained with  
		the SPSIFV scheme (symbols) versus the analytical solution of the HB model for a range of 
		$ \dot{\gamma}_{12}$, and for $ \sigmaY = 0 $ (left, no 
			yield 
			stress) and $ \sigmaY = 0.5 $ (right). Other material parameters are  $ \kappa = 1.0 $, 
			$ n = 0.5 $, $ n = 1.0 $, and $ n = 1.5 $. The 
			parameters of 
			the 
			GPR model are $ \csh = 10 $, $ \taus = 10^{10} $.
		}  
		\label{fig:Couette}
	\end{center}
\end{figure}

In the first example, we verify the model in the standard Couette flow problem with ($ \sigmaY >0 
$) and without ($ \sigmaY = 0 $) the yield stress. The two-dimensional 
computational domain $ (x,y) \in [0,1]\times[0,1]  $ has periodic boundary conditions in $ x 
$-direction, and no-slip conditions at $ y=0 $ and $ y=1 $. Additionally, the boundary $ y=1 $ is 
moving in the positive direction at a constant velocity. We solve system \eqref{eqn.GPR} with the 
SISPFV scheme \cite{SIGPR2021} on a grid composed of $ 4\times100 $ elements until time $ t=10 $ 
for a range of velocities of the moving boundary. At time $ t = 10 $, the velocity $ u(y) $ has 
already reach 
a constant slope while the stress $ \sigma_{12}(y) $ is constant (symbols in 
Fig.\,\ref{fig:Couette}). 
The initial conditions are set to $ \rho = 1 $, $ \Dist = \II $, $ \vv = 0 $. The material 
parameters 
of the GPR model are $ \csh = 10 $, while the relaxation time $ \tau $ is defined from the 
Herschel-Bulkley viscosity which has the parameters $ \sigmaY =0 $ or $ \sigmaY = 0.5 $, $ \kappa = 
1 $, $ n=0.5 ,1.0 $, or $ 1.5 $. The relaxation time in the solid state is taken as $ \taus = 
10^{10} $, while the fluid state relaxation time is computed from \eqref{eqn.tauf}.
In Fig.\,\ref{fig:Couette}, one can notice that the yield-stress and power-law rheology of the HB 
model is well captured by the 
GPR model.

We also use the Couette flow example to demonstrate a peculiar feature of the distortion field. 
Namely, its intrinsic rotational dynamics. Recall that the distortion $ \Dist $ can be viewed as a 
local frame field (local basis triad) consisting of three linearly independent vectors. As has been 
noticed earlier \cite{HPR2016,DPRZ2016}, the relaxation process in $ \Dist $ governed by the 
right-hand side in \eqref{eqn.dist} acts in such a way that the tangential stresses tend to reduce 
but also the basis triad $ \Dist $ rotates permanently even in the case when the velocity field is 
stationary. In the current formulation, the distortion spin does not affect the stress tensor $ 
\tensor{\sigma} $, e.g. see \cite{GodRom2003}, and does not violate the angular momentum 
conservation because the stress $ \tensor{\sigma} $ stays always symmetric. An extension of the 
model towards accounting for the distortion spin and associating it to, for example, unresolved 
small-scale eddies in turbulent flows is discussed in \cite{PRD-Torsion2019} in the context of the 
Riemann-Cartan geometry. 

\begin{figure}[!htbp]
	%	\captionsetup{format=plain}
	\begin{center}
		\includegraphics[draft=false,trim=0 0 0 0,clip, 
		scale=0.5]{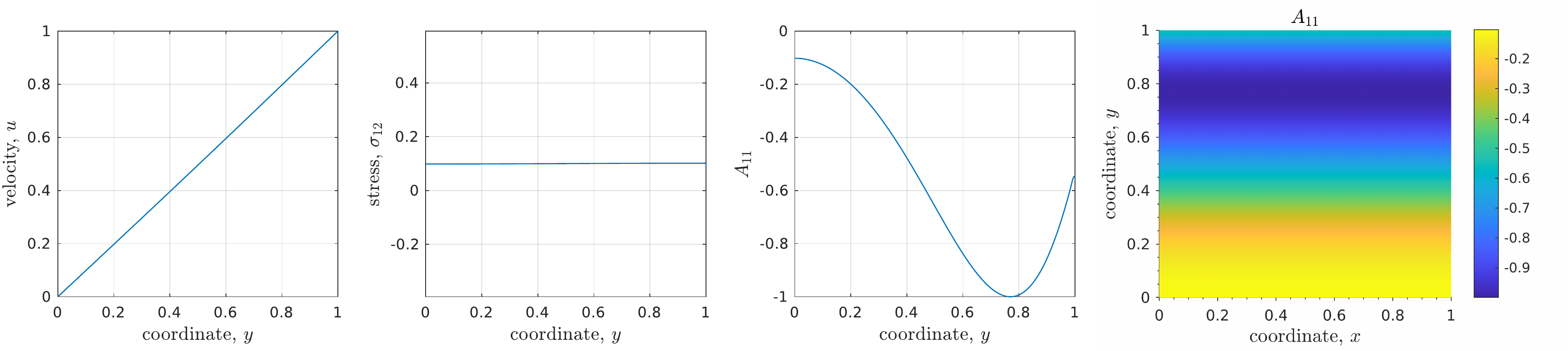}
		\\
		\includegraphics[draft=false,trim=0 0 0 0,clip, 
		scale=0.5]{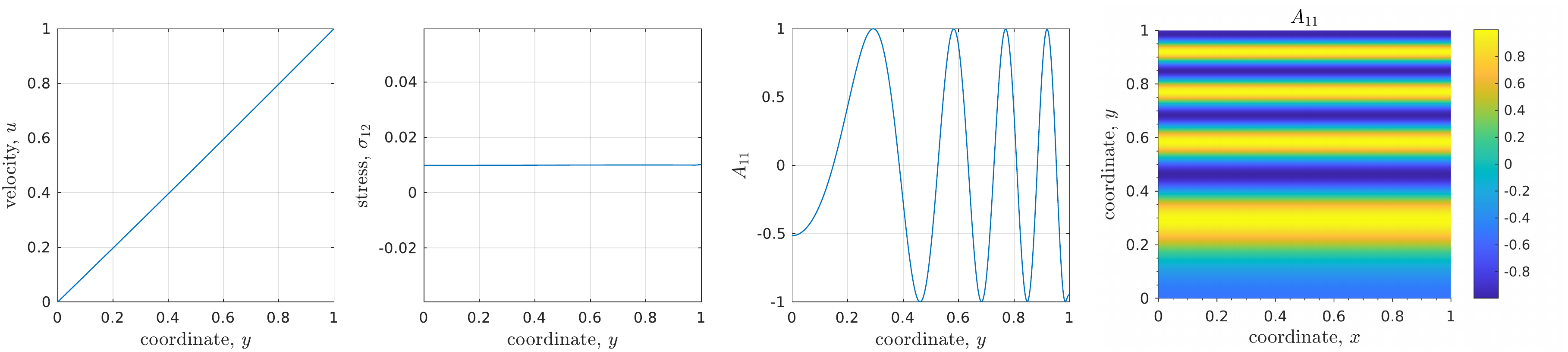}
		\\
		\includegraphics[draft=false,trim=0 0 0 0,clip, 
		scale=0.5]{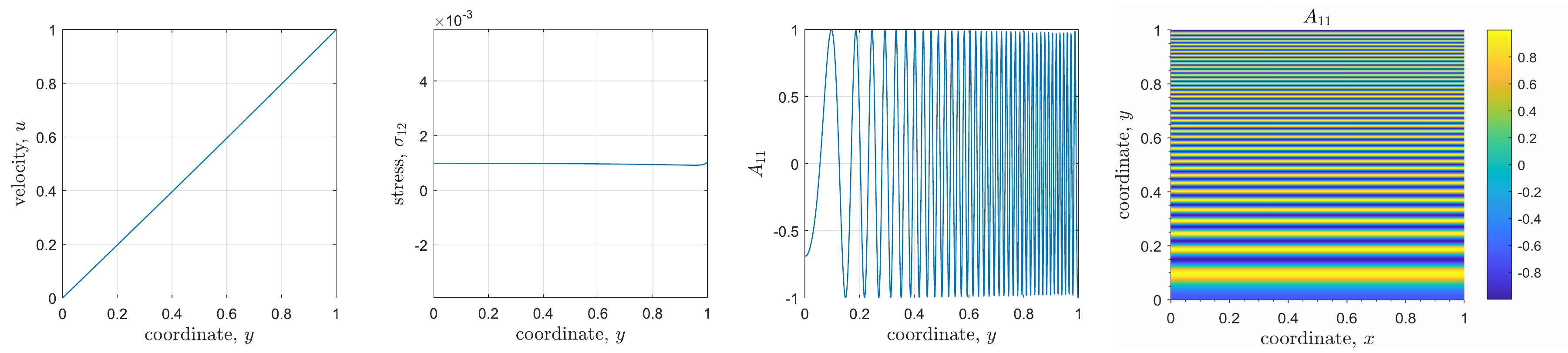}
		\caption{Couette flow of a Newtonian fluid obtained with our model for the Newtonian 
		viscosity $ \eta = 10^{-1} $ (top row, final time $ t = 5 $), $ \eta = 10^{-2} $ (middle 
		row, final 
		time $ t = 50 $), and $ \eta = 10^{-3} $ (bottom, final time $ t = 500 $). Cuts of the 
		velocity $ u $, 
		stress $ \sigma_{12} $, 
		and $ A_{11} $ component of the distortion, as well as a snapshot of $ A_{11} $ (right 
		column). 
		The third and fourth columns demonstrate 
		different orientations of the distortion field of the fluid layers due to the intrinsic 
		rotations. The higher the Reynolds number the higher the orientational heterogeneity in the 
		distortion field.
		}  
		\label{fig:Couette.A}
	\end{center}
\end{figure}

Fig.\,\ref{fig:Couette.A} shows strong heterogeneity in the distortion field due to the intrinsic 
rotations despite the velocity and the stress are homogeneous and steady. In this regards, one may 
recall the polar decomposition for the distortion matrix $ \Dist = \bm{R}\sqrt{\GG} $, with $ 
\bm{R} $ being an orthonormal matrix, and the fact that the stress $ \tensor{\sigma} $ 
\eqref{eqn.stress} does not depend on the rotation $ \bm{R} $, see \cite{GodRom2003}. To 
demonstrate this feature of the model, we
carried out three simulations of the Couette flow in the domain $ (x,y) \in [0,1]\times[0,1] $ with 
the GPR model and SISPFV scheme of a 
Newtonian fluid with $ \eta = 10^{-1} $, 
$ \eta = 10^{-2} $, and $ \eta = 10^{-3} $, see Fig.\,\ref{fig:Couette.A}. The numerical solutions 
are shown at 
times approximately right after the 
steady state is 
reached, $ t=5 $ for $ \eta = 10^{-1} $,  $ t=50 $ for $ \eta=10^{-2} $, and $ t = 500 $ for $ \eta 
= 10^{-3} $. Initially, the fluid is at rest 
with the parameters $ \rho =1 $, $ \vv = 0 $, $ \Dist = \II $ and $ \csh = 10 $. The velocity of 
the moving 
boundary ($ y = 1 $) is set 
to 1. A mesh of $ 4\times400 $ elements was used for $ \eta = 10^{-1} $ and $ 10^{-2} $, while to 
resolve the strong heterogeneity of $ \Dist $ at $ \eta = 10^{-3} $ we were need to use a quite 
fine mesh of $ 4 \times 1600 $ elements. A slight deviation from the constant value in $ 
\sigma_{12} $ is visible for the case $ \eta = 10^{-3} $ (bottom) which is due to the fact that at 
this mesh resolution and with the second-order accuracy of our SISPFV scheme we are still not able 
to resolve high heterogeneity of the distortion field in the interval $ y \in [0.8,1] $ of the 
most intense heterogeneity of $ \Dist $. Remark that the actual 
deformation of the fluid elements which is stored in $ \GG $ are of the order of $ 10^{-3} $ while 
the elements of the rotation matrix $ \bm{R} $ are of the order of $ 1 $. Therefore, one may 
conclude that for high Reynolds number, the lack of the resolution of rotational peculiarities of 
the 
distortion field might result in errors in the stress tensor.

\subsection{Plane Hagen-Poiseuille flow}

In the second numerical example, we test the GPR model in the plane Hagen-Poiseuille flow. The 
two-dimensional 
computational domain $ (x,y) \in [0,1]\times[0,1]  $ has periodic boundary conditions in $ x 
$-direction, and no-slip conditions at $ y=0 $ and $ y=1 $. We solve system \eqref{eqn.GPR} with 
the SISPFV scheme on a grid composed of $ 4\times100 $ elements. The pressure drop 
$ \Delta p $ is set to 0.1. 
In the initial conditions, we set the material velocity equal to the analytical solution 
\eqref{eqn.HB.Pois}, and other state variables were set as $ \rho = 1 $, $ \Dist = \II $. The final 
integration time is $ t = 1 $. The analytical solution to the HB 
model is given by (e.g. see \cite{Bleyer2015})
\begin{equation}\label{eqn.HB.Pois}
	u(y)=\left\{
	\begin{array}{ll}
		\frac1m f \left( \left (\frac{y_0}{h} \right )^m - \left (\frac{y_0-y}{h}\right )^m 
		\right),& y \leq 
		y_0,
		\\[2mm]
		\frac1m f \left ( \frac{y_0}{h}\right )^m ,& y_0 \leq y \leq h-y_0,
		\\[2mm]
		\frac1m f \left( \left (\frac{y_0}{h} \right )^m - \left (\frac{y-(h-y_0)}{h}\right 
		)^m\right),&
		y > h - y_0,
	\end{array} \right.
\end{equation}
where $ f = \frac{h}{V} \left( \frac{\Delta p \, h}{\kappa}\right)^{1/n} $, $ m = 1 + 1/n $,
$ y_0 = h(1/2 - \Bi/f^n)$, $ h $ is the channel width, $ V $ is a velocity scale, and $ \Bi $ is 
the generalized Bingham number which for the HB model can be defined as 
 \begin{equation}\label{BiNumber.gen}
 	\Bi = \frac{\sigmaY}{\kappa}\left (\frac{h}{V}\right )^n 2^{1-n},
 \end{equation}
and which quantifies the effect of yield stress versus power-law, e.g. \cite{Sverdrup2018}. In the 
examples of this section, we set $ h=1.0 $ (dimension in the $ y $-direction) and we set the 
velocity scale $ V=1 $ for $ n=1 $ and $ V=2.03 $ for $ n\neq1 $.

\begin{figure}[!htbp]
	\begin{center}
		\includegraphics[draft=false,trim=0 0 0 0,clip,scale=0.6]{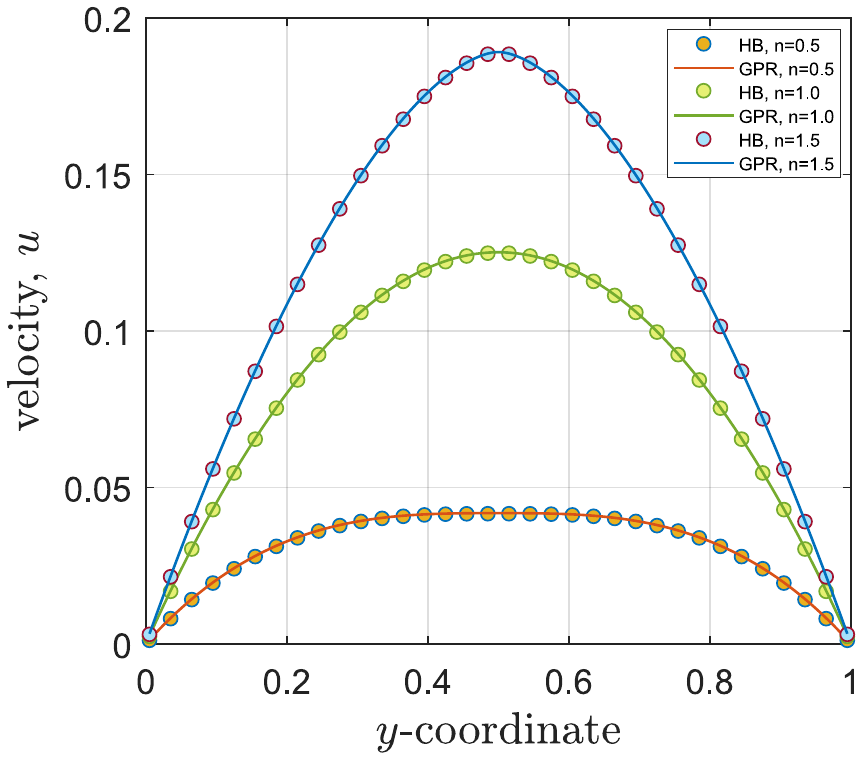}
		\hspace{0.5cm}
		\includegraphics[draft=false,trim=0 0 0 0,clip,scale=0.26]{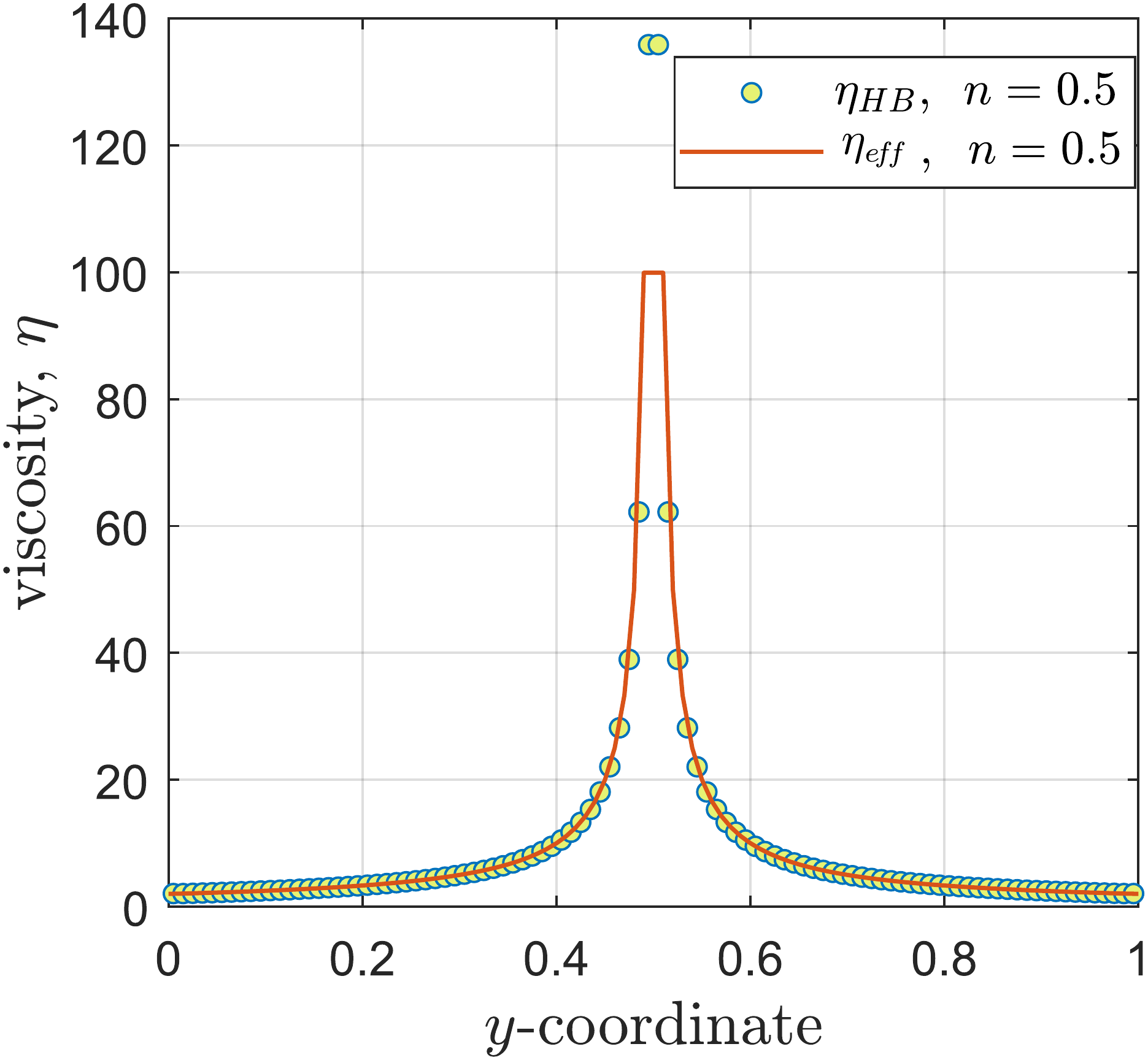}
		\hspace{0.5cm}
		\includegraphics[draft=false,trim=0 0 0 0,clip,scale=0.26]{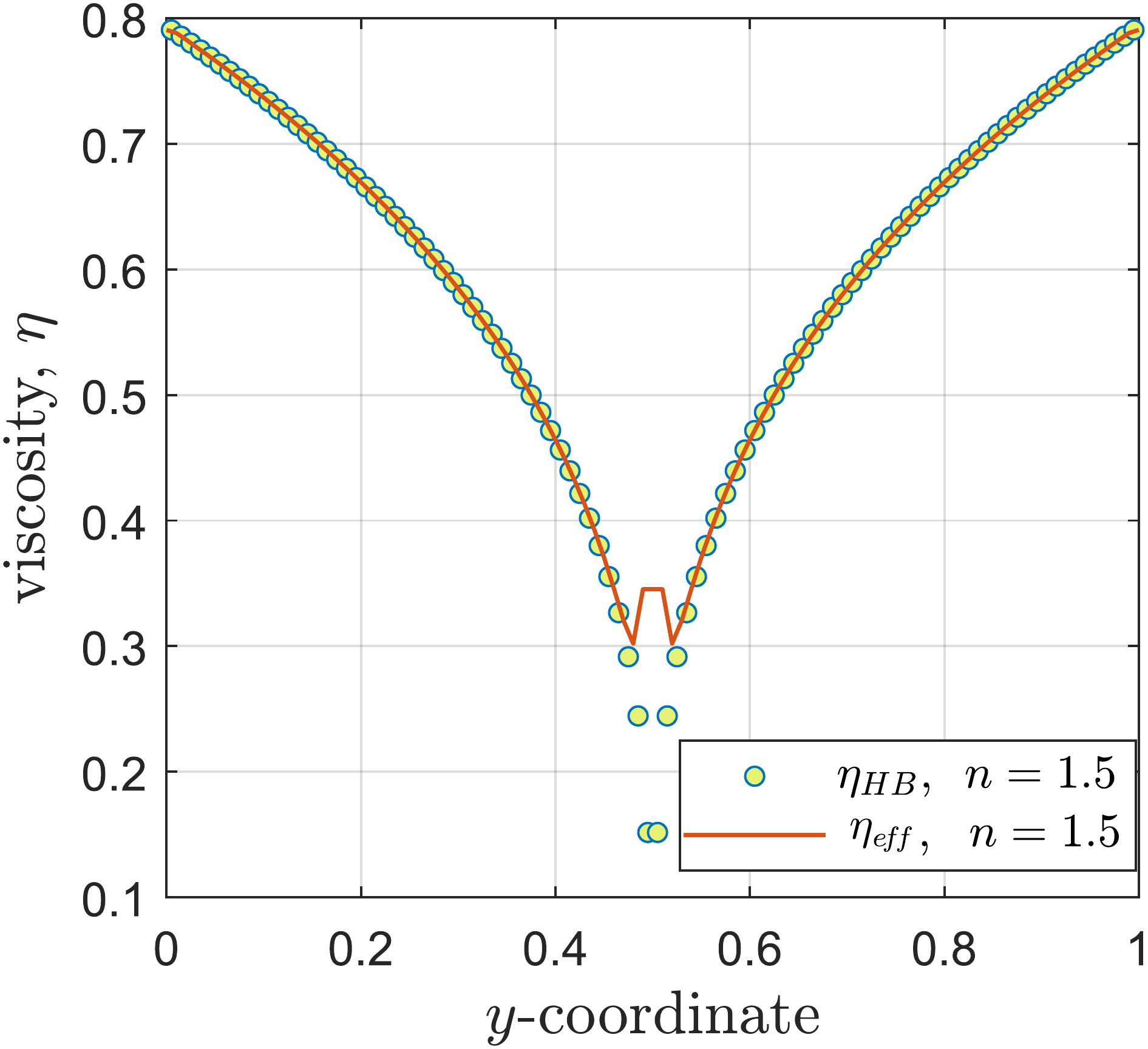}
		\caption{Hagen-Poiseuille flow. Comparison of the numerical solution to the GPR model  
			obtained with the SISPFV scheme against the analytical solution 
			to the Herschel-Bulkley model ($ \sigmaY = 0 $) for the power-law exponents $ n=0.5 $, 
			$ 
			n = 1.0 $, and $ n = 1.5 $ (left subfigure, from bottom to top). The middle and right 
			subfigures show the 
			effective viscosity of the GPR model $ \etaeff = \frac16 \rho \tau \csh^2  $ (lines) 
			and the HB viscosity (symbols) computed from the analytical solution 
			\eqref{eqn.HB.Pois}. 
		}  
		\label{fig:Pois-HB-GPR}
	\end{center}
\end{figure}

Fig.\,\ref{fig:Pois-HB-GPR} depicts the numerical solution to the GPR model and the analytical 
solution \eqref{eqn.HB.Pois} to the HB model for a range of power-law exponents $ n = 0.5 $, $ 1.0 
$, and $ 1.5 $ and zero yield-stress $ \sigmaY = 0 $ ($ \Bi = 0 $). The consistency index was taken 
as $ \kappa = 
1.0 $.  The shear sound speed in the GPR model was taken as $ \csh = 10 $. The relaxation time in 
the solid state is taken as $ \taus = 
10^{20} $, while the fluid state relaxation time is computed from \eqref{eqn.tauf}. Overall, a good 
agreement between the 
solutions is achieved.

\begin{figure}[!htbp]
	\begin{center}
		\includegraphics[draft=false,trim=0 0 0 0,clip,scale=0.26]{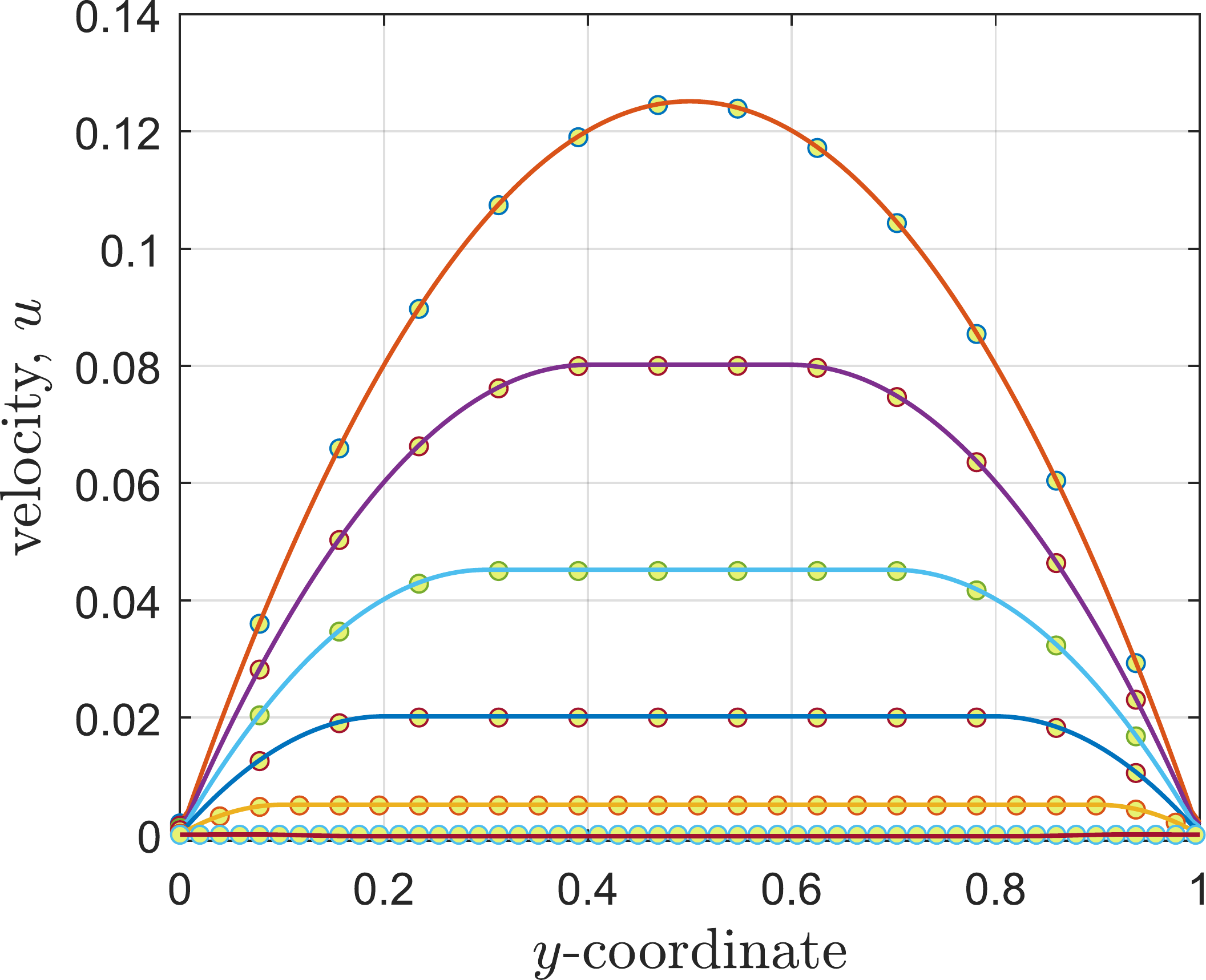}
		\hspace{0.1cm}
		\includegraphics[draft=false,trim=0 0 0 0,clip,scale=0.26]{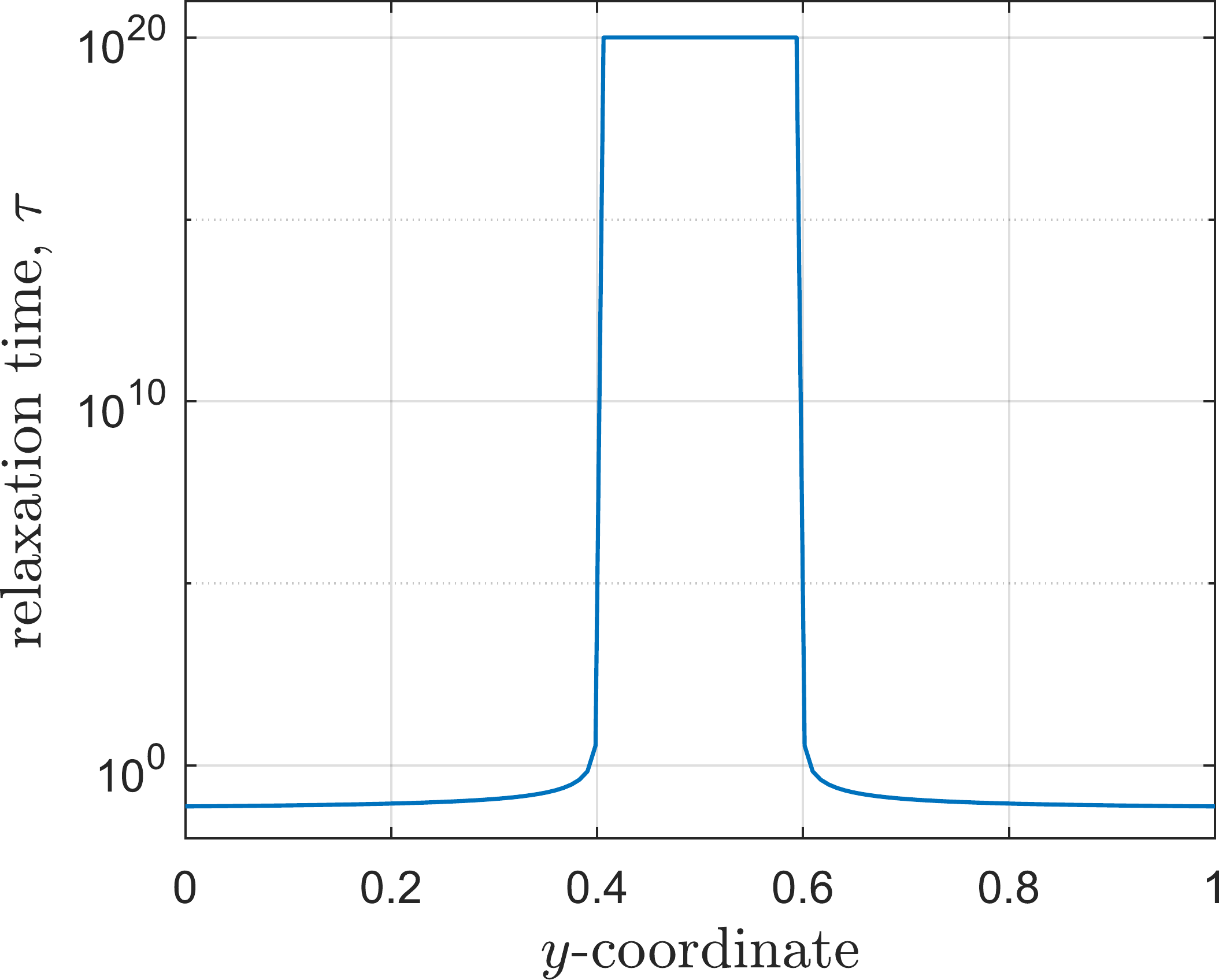}
		\hspace{0.1cm}
		\includegraphics[draft=false,trim=0 0 0 0,clip,scale=0.26]{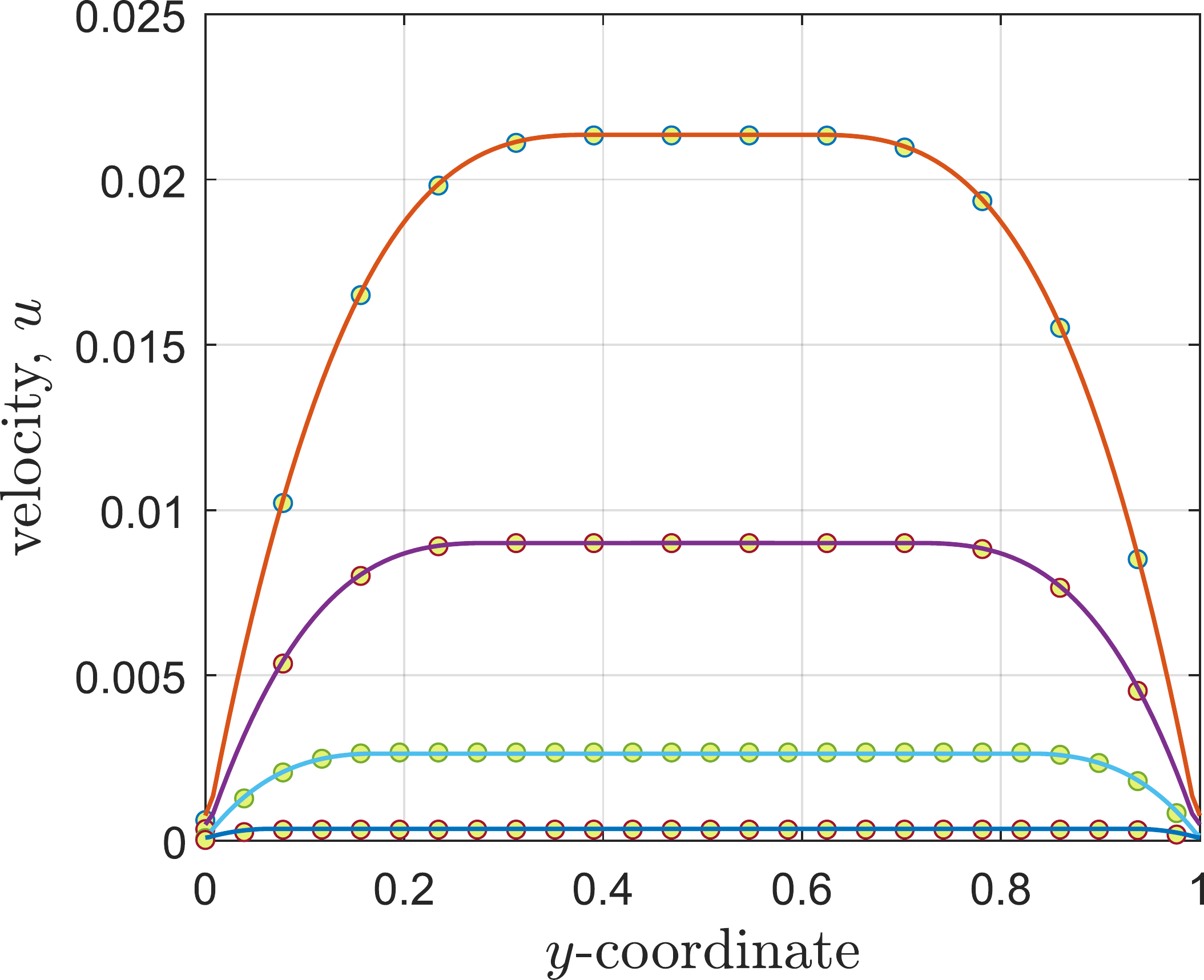}
		\caption{Hagen–Poiseuille flow of Herschel-Bulkley fluid for $ n=1.0 $ (left) and $ n=0.5 $ 
		(right). Comparison of the 
		numerical
		solution $ u(y) $ to the GPR model (lines) against the analytical solution 
		\eqref{eqn.HB.Pois} to the HB model 
		(symbols)
		for the 
		range of $ \Bi $ 
			numbers: $ \Bi =  
			0 $, $ 0.1 $, $ 0.2 $, $ 0.3 $, $ 0.4 $, and $ 0.5 $ (from top to bottom) for $ n=1.0 $ 
			and $ \Bi = 0.1 $, $ 0.2 $, $ 0.3 $, and $ 0.4 $ for $ n=0.5 $. The middle subfigure 
			shows a typical behavior of the relaxation time ($ n=1.0 $, $ \Bi=0.1 $).
		}  
		\label{fig:Pois-GPR-HB-n1}
	\end{center}
\end{figure}

Fig.\,\ref{fig:Pois-GPR-HB-n1} (left subfigure) shows the numerical solution to the GPR model for a 
range Bingham 
numbers $ \Bi = 0.0 $, $ 0.1 $, $ 0.2 $, $ 0.3 $, $ 0.4 $, and $ 0.5 $ and power-law index $ n=1 $. 
A good agreement between 
the GPR solution and the solution to the HB model is achieved. The middle subfigure of 
Fig.\,\ref{fig:Pois-GPR-HB-n1} 
also 
shows a typical behavior of the relaxation time $ \tau(\Dist) $. The numerical solution for the 
same range of Bingham numbers but the power-law index $ n=0.5 $ is shown in 
Fig.\,\ref{fig:Pois-GPR-HB-n1} (right subfigure). Also, a good agreement between the solution of 
our 
hyperbolic model and the 
parabolic model is achieved.

\subsection{Lid-driven cavity flow}

In the third example, we consider another classical benchmark problem for the numerical solution of 
incompressible Navier-Stokes equations, the lid-driven cavity problem. As previously, we compute 
the numerical
solution to the GPR model with the SISPFV scheme and it with the numerical solutions 
\cite{Sverdrup2018,Syrakos2014} to the Herschel-Bulkley model 
for the both 
cases $ \sigmaY = 0 
$ and $ \sigmaY > 0 $.

\subsubsection{Power-law fluid}
For the computational setup in this section, the computational domain is given by $ [0, 1] \times 
[0, 1]$. The initial
condition is simply given by $ \rho=1 $, $ \vv=0 $, and $ \Dist = \II $. The shear sound speed was 
set to $\csh = 10 $. The lid velocity ($ y = 1 $) was set to 1 and no-slip boundary condition is 
imposed on the other boundaries. The parameters of the Herschel-Bulkley model 
are $ \kappa = 10^{-2} $, $ n = 0.5, 1.0, 1.5 $, and also in this section, $ \sigmaY = 0 $. 
The Reynolds number estimated based on the consistency index $ \kappa $ is $ \Re = 100 $.
The 
computational domain was discretized with $ 256\times256 $ elements for $ n=1.0 $ and $ n=1.5 $
but we had to use a very fine mesh of $ 1600\times1600 $ elements in the most difficult for us case 
$ n=0.5 $. This is conditioned by the fact that for $ n=0.5 $, locally, the relaxation time drops 
down to $ 10^{-5} $ and below which results in a very stiff source term in \eqref{eqn.dist}. To 
efficiently
deal with such a stiffness in the source term it is necessary to have an asymptotic preserving 
scheme while we recall that our SISPFV scheme is
only quasi asymptotic-preserving property, see Section\,\ref{sec.scheme} and \cite{SIGPR2021}. 
\begin{figure}[!htbp]
	\begin{center}
		\begin{tabular}{c} 
			\includegraphics[draft=false,width=0.75\textwidth]{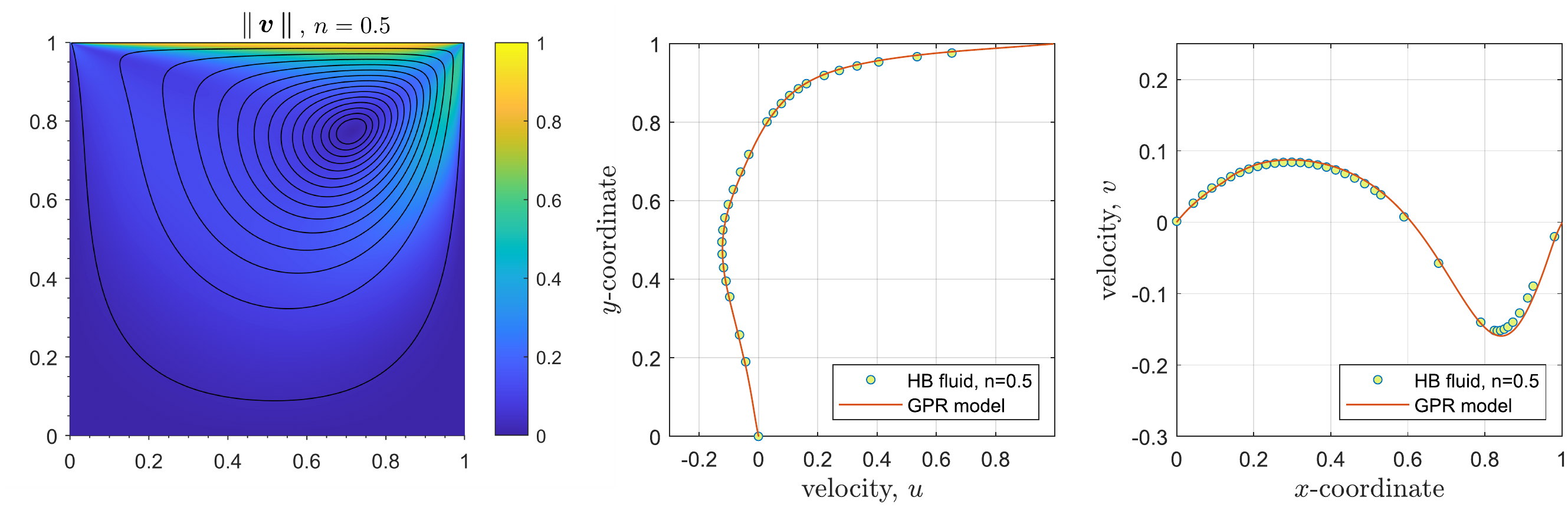}
			\\
			\includegraphics[draft=false,width=0.75\textwidth]{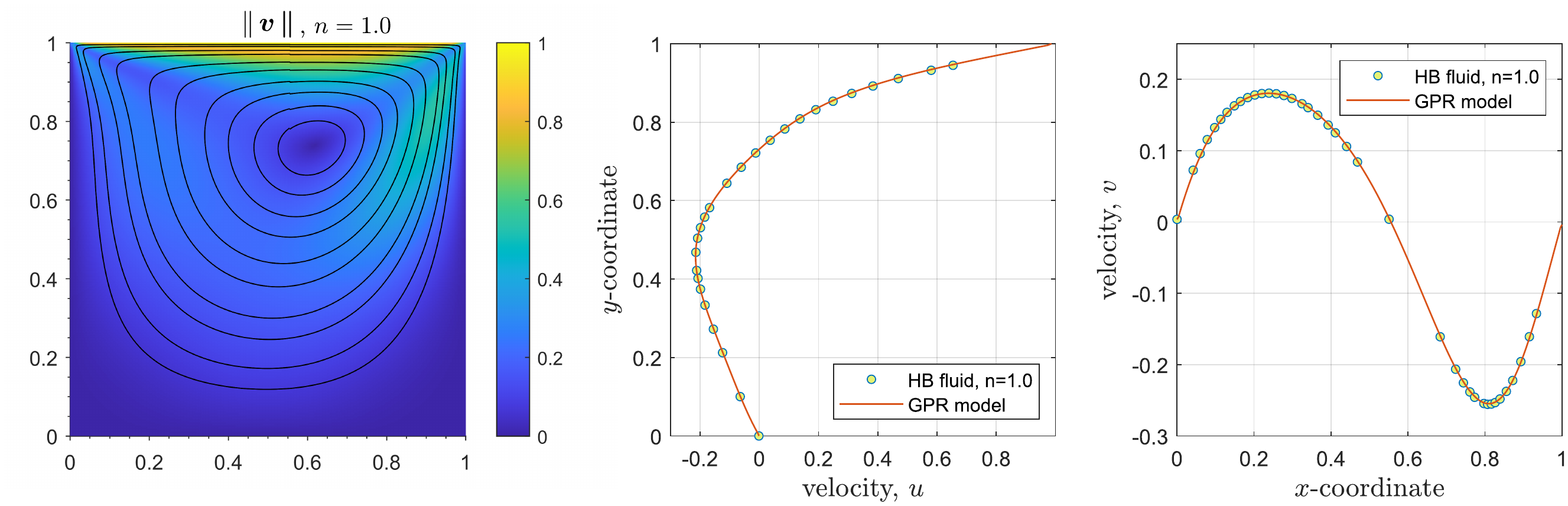}
			\\
			\includegraphics[draft=false,width=0.75\textwidth]{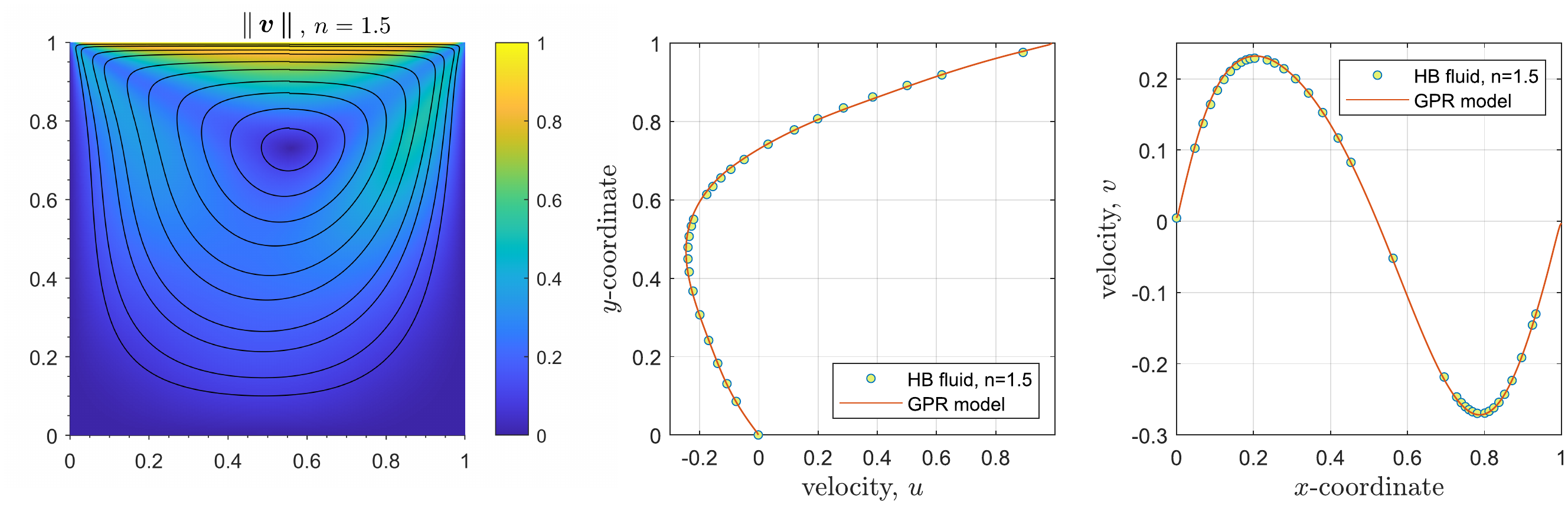}
			\\
		\end{tabular} 
		\caption{Lid-driven cavity flow, for the GPR model with the power-law 
			exponents $ n = 0.5 $ 
			(top row, $ t=30 $), $ n = 
			1.0 $ (middle row, $ t=20 $), and $ n =1.5 $ (bottom row, $ t=20 $) and $ \csh = 10 $. 
			Middle and right column 
			shows the cuts of 
			the velocity field in $ x = 0.5 $ and $ y = 0.5 $ (red lines) and reference numerical 
			solution (symbols)
			from 
			\cite{Sverdrup2018} obtained with the Herschel-Bulkley model for $ \kappa = 10^{-2} $, 
			zero 
			yield stress, and the same power-law exponents $ n = 0.5 $, $ 1.0 $, and $ 1.5 $.
		}  
		\label{fig:Cavity-GPR-HB}
	\end{center}
\end{figure}

Fig.\,\ref{fig:Cavity-GPR-HB} shows the snapshots of $ \Vert\vv\Vert $ alongside with the 
streamlines and the cuts $ u(y) $ and $ v(x) $ of the velocity 
field at $ x = 0.5 $ and $ y = 0.5 $, accordingly. Overall, one can notice a good agreement 
between the numerical solution obtained with the SISPFV scheme for the first-order hyperbolic GPR 
model and the reference 
solution \cite{Sverdrup2018}
to the parabolic Navier-Stokes equations with the Herschel-Bulkley viscosity.

\begin{figure}[!htbp]
	\begin{center}
		\begin{tabular}{rrr}
			\includegraphics[draft=false,scale=0.25]{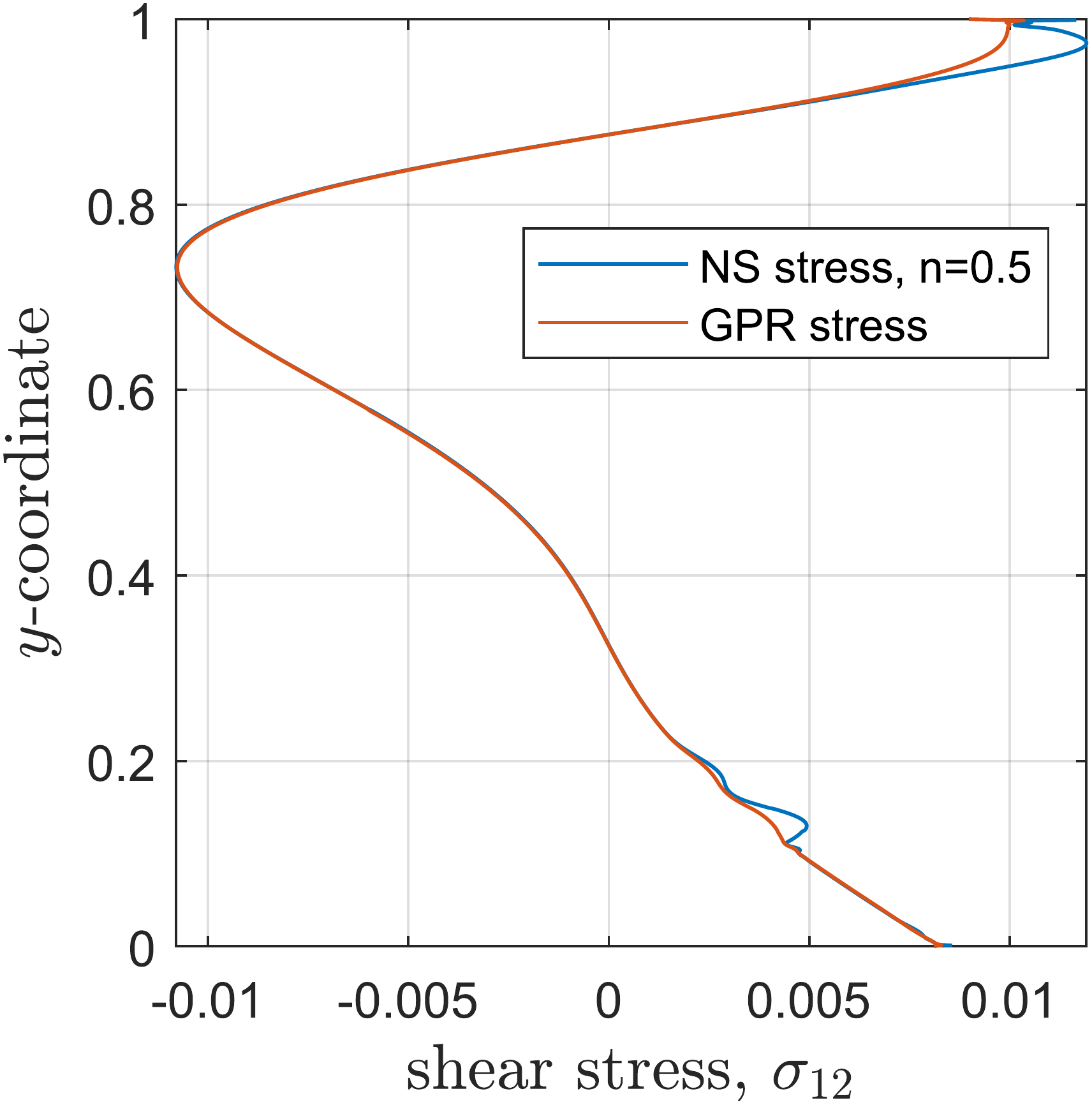}
			&
			\includegraphics[draft=false,scale=0.25]{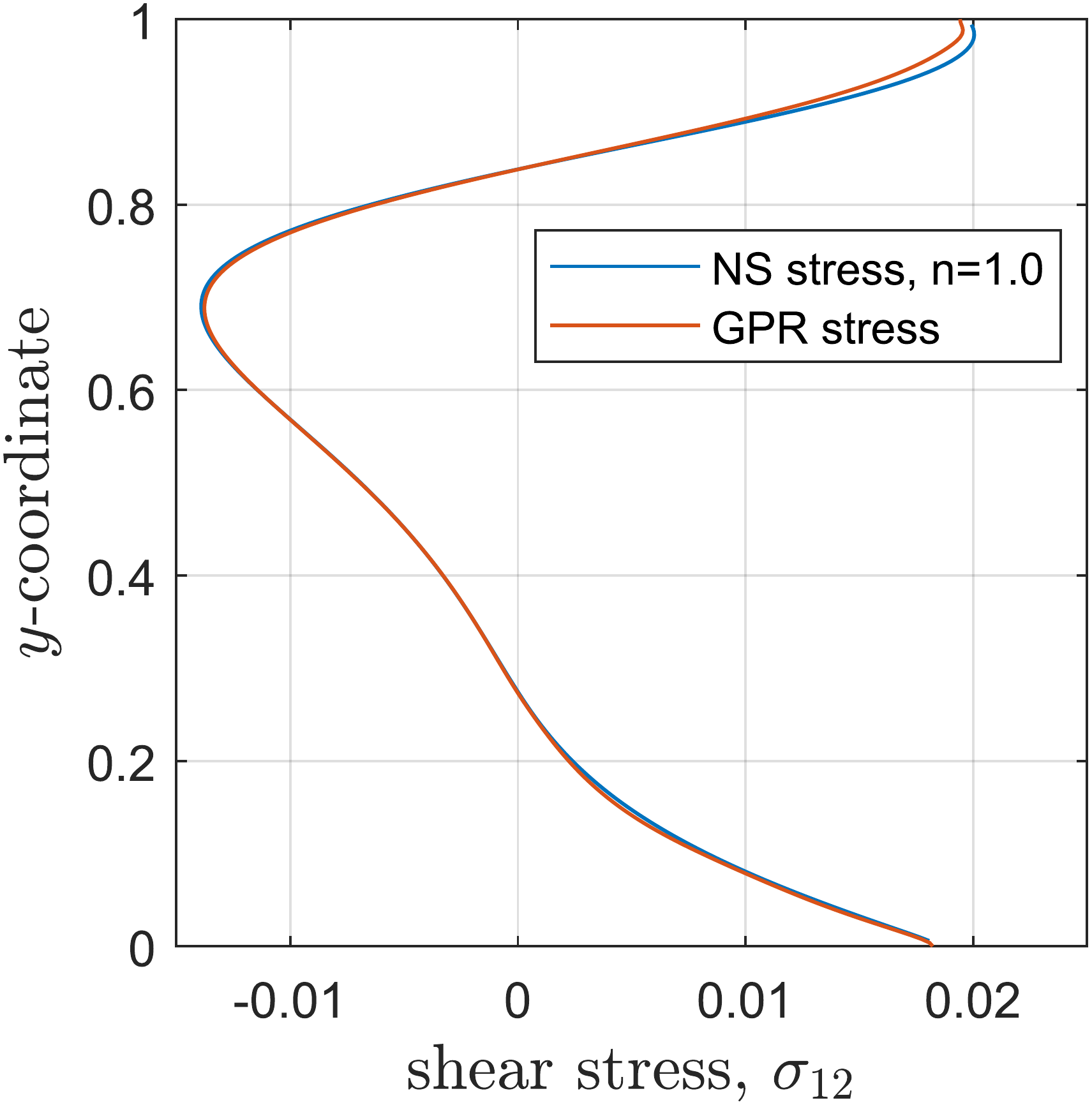}
			&
			\includegraphics[draft=false,scale=0.25]{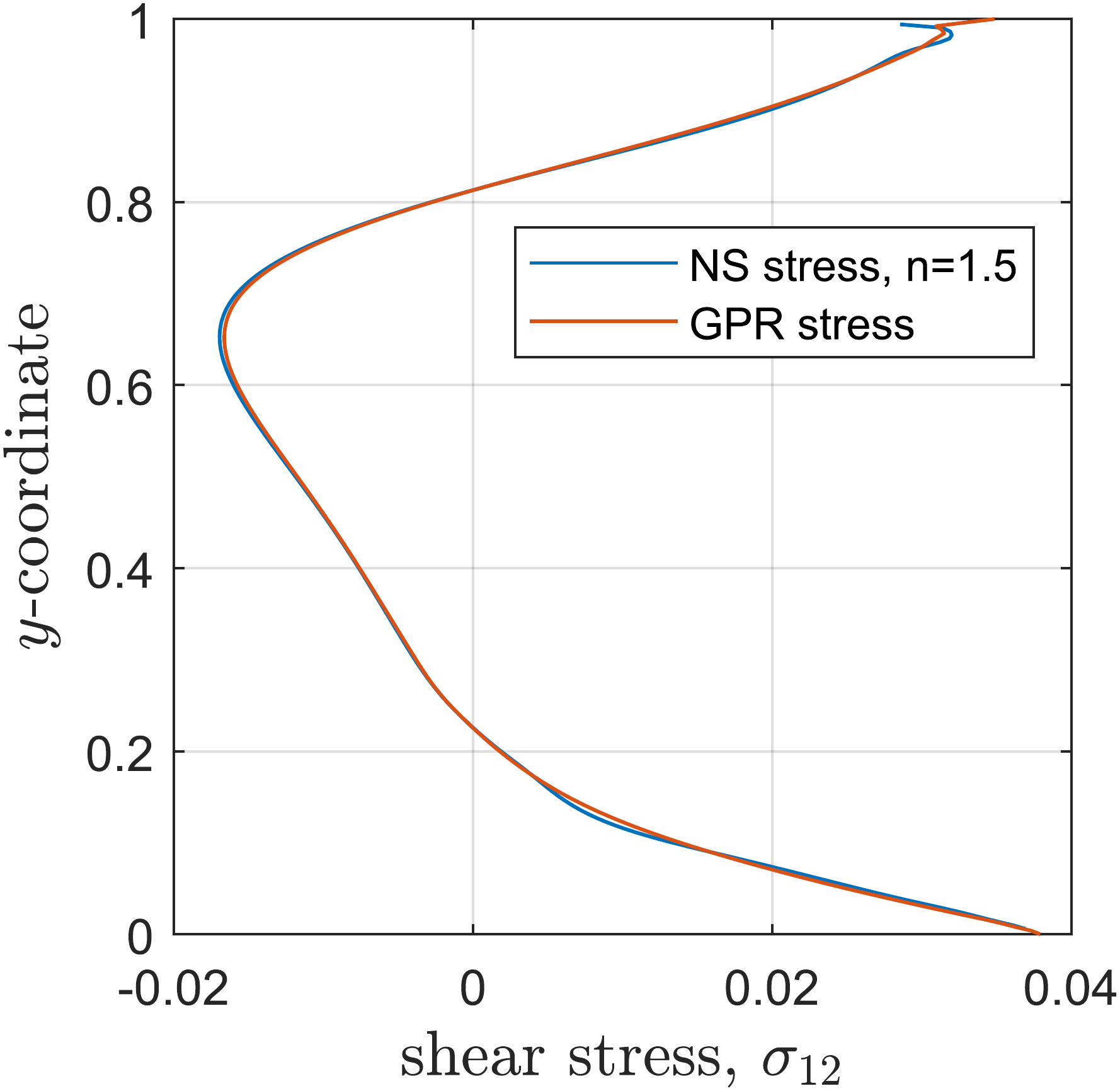} \\
			\includegraphics[draft=false,scale=0.25]{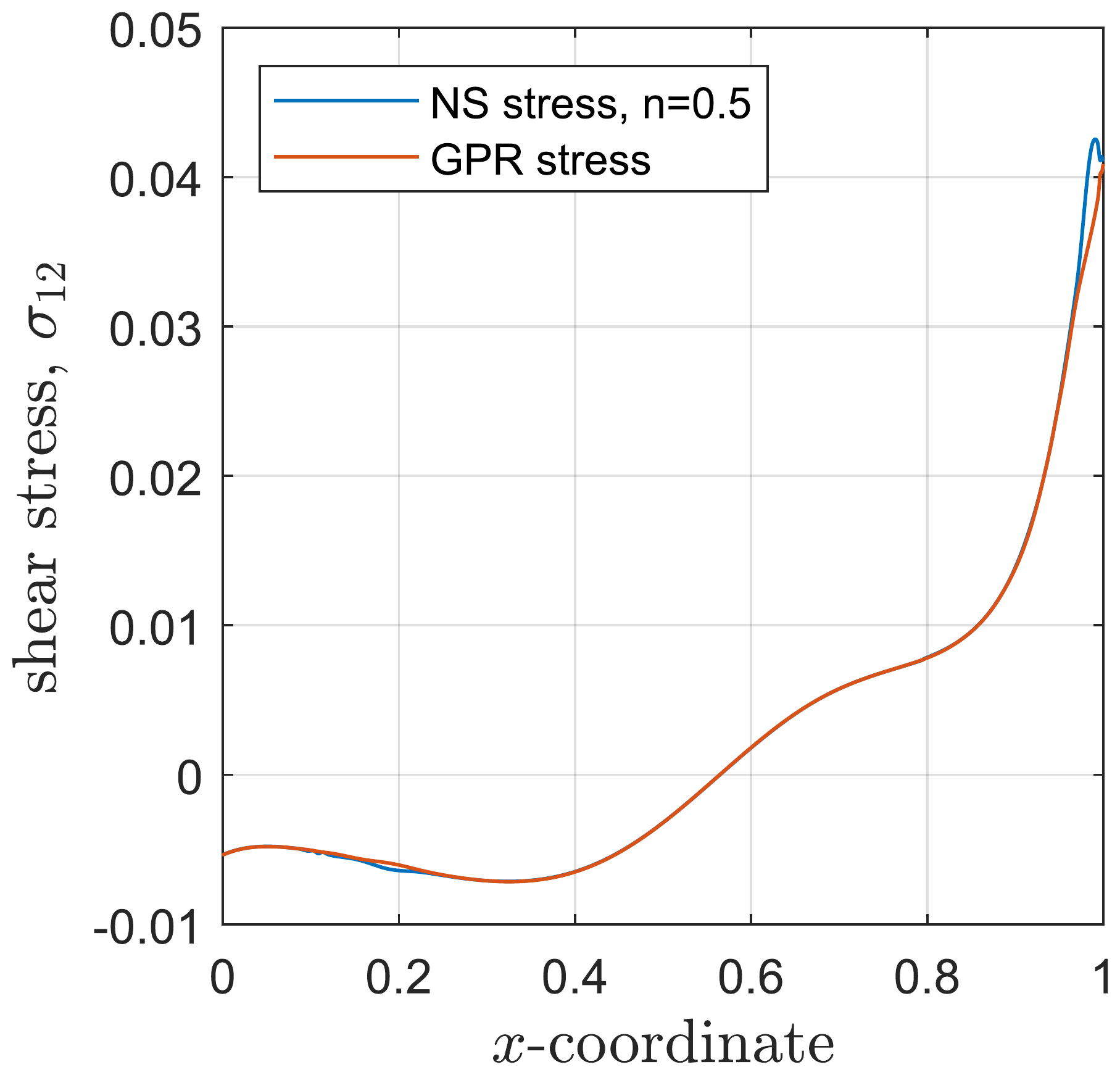}	
			&
			\includegraphics[draft=false,scale=0.25]{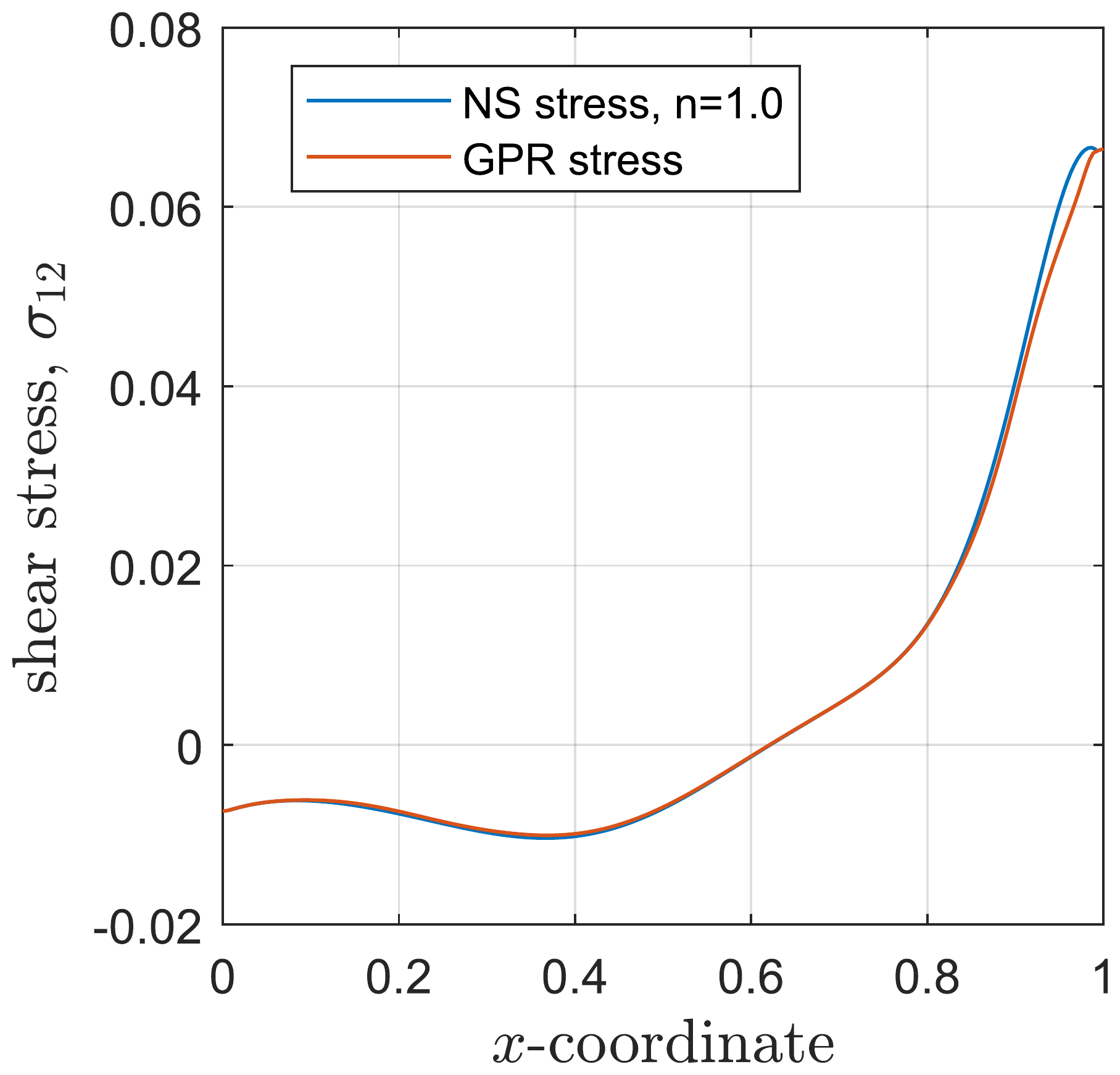}	
			&
			\includegraphics[draft=false,scale=0.25]{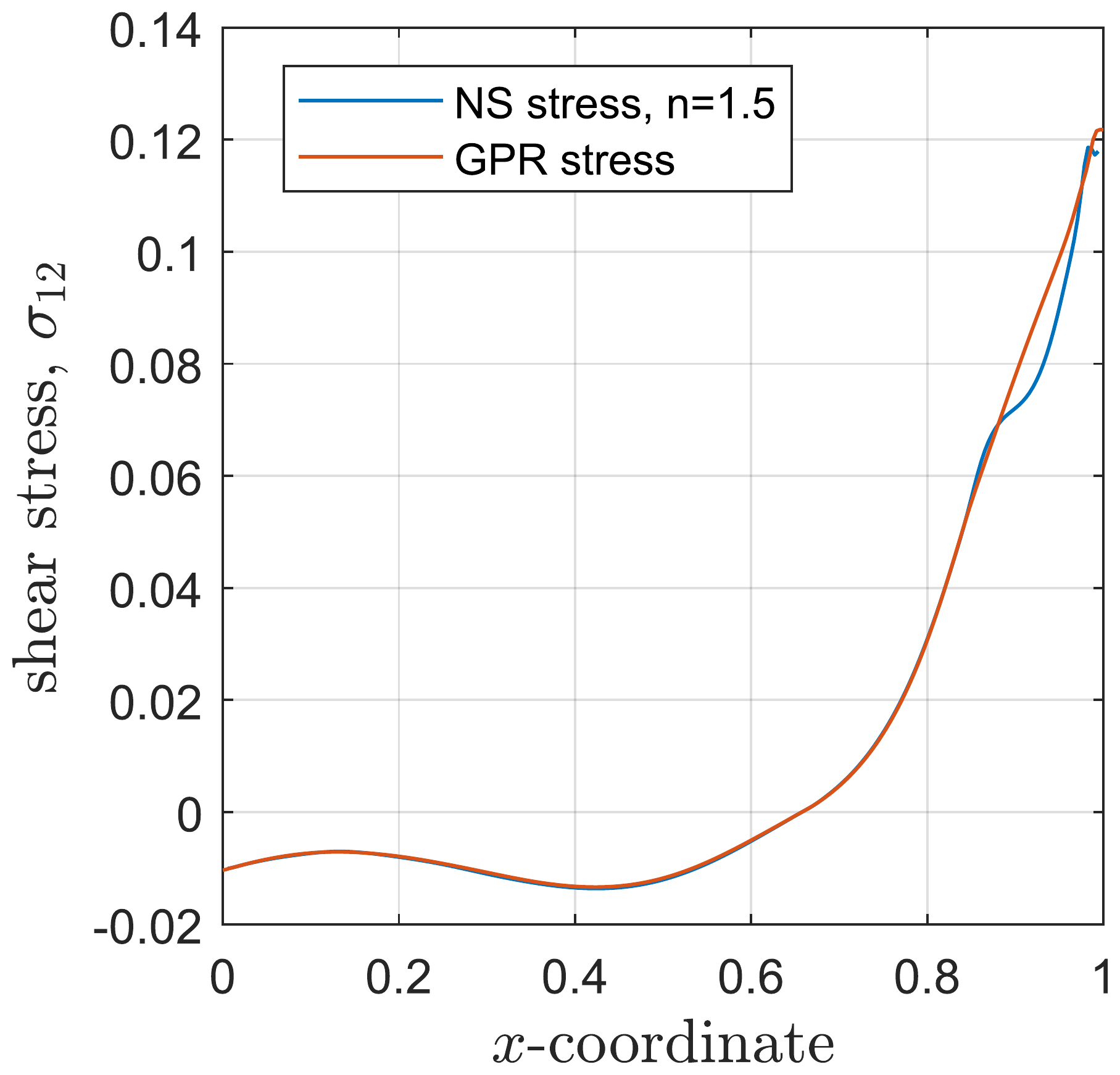}
		\end{tabular}
		\caption{Lid-driven cavity flow. Comparison of the 
		Navier-Stokes stress $ \tensor{\sigma}_{\ms{NS}} = 
		\eta(\dot{\gamma}) \dotgam $ (computed from the velocity gradient of the 
		GPR solution as a 
		post-processing) and the stress of the GPR model computed from the distortion field 
		$ \tensor{\sigma}=-\rho \Dist^\transpose E_{\Dist} $. Despite an overall good agreement being 
		achieved some deviations are also visible which are due to the fact that our scheme is not 
		perfectly asymptotic preserving but only quasi-asymptotic preserving.
		}  
		\label{fig:Cavity-GPR-HB-eta-T12}
	\end{center}
\end{figure}

Because the velocity fields depicted in Fig.\,\ref{fig:Cavity-GPR-HB} agree well with the reference 
solution of the Navier-Stokes equations with the Herschel-Bulkley viscosity, we can use this 
velocity data 
in order to compute the Navier-Stokes stress tensor \eqref{eqn.Newton.law} and compare it with the 
stress 
tensor of the GPR model computed from the distortion field according to \eqref{eqn.stress}. Note 
that the Navier-Stokes $ \tensor{\sigma}_{\mathsmaller{NS}} $ \eqref{eqn.Newton.law} and the stress 
of the GPR model $ 
\tensor{\sigma} $ \eqref{eqn.stress} are of different nature. The Navier-Stokes 
stress is of dissipative nature and is non-local (depends on the spatial gradient of the state 
variables) while $ \tensor{\sigma} $ is of elastic nature (non-dissipative) and computed locally 
(no space derivatives of state variables). 
Nevertheless, one can 
see in Fig.\,\ref{fig:Cavity-GPR-HB-eta-T12} that the two agree reasonably well.

\begin{figure}[!htbp]
	\begin{center}
			\includegraphics[draft=false,width=0.3\textwidth]{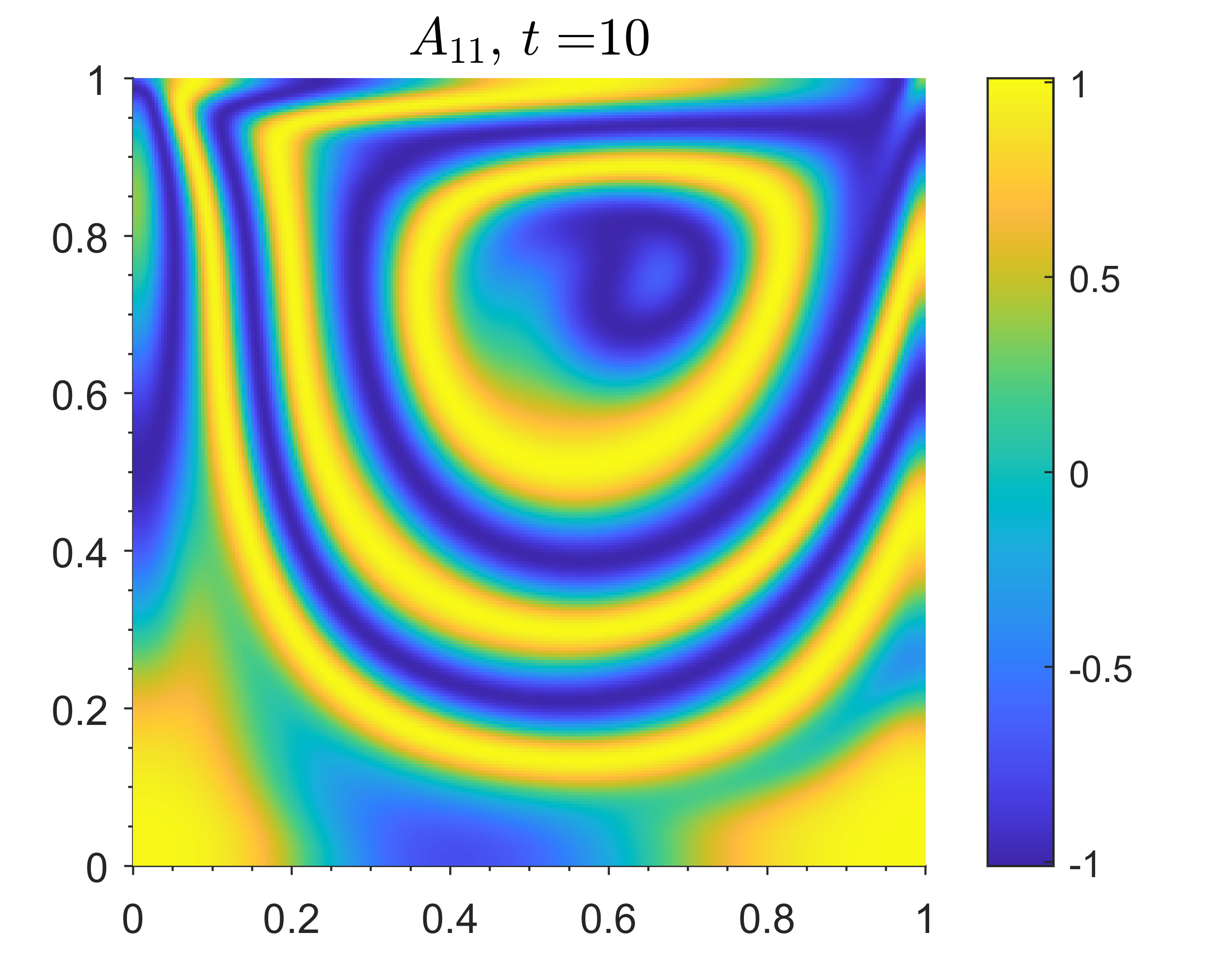} 
			\includegraphics[draft=false,width=0.3\textwidth]{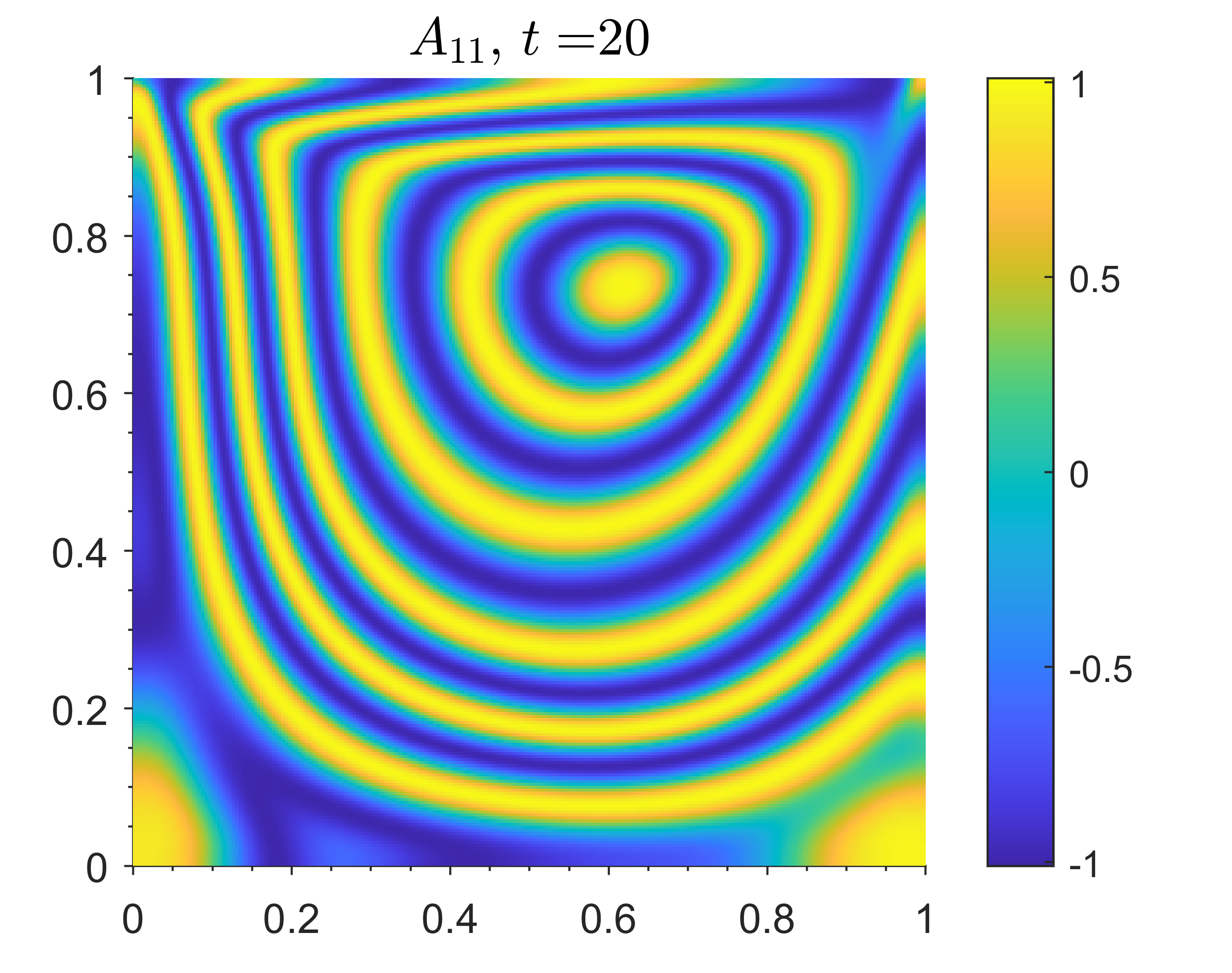}
			\includegraphics[draft=false,width=0.3\textwidth]{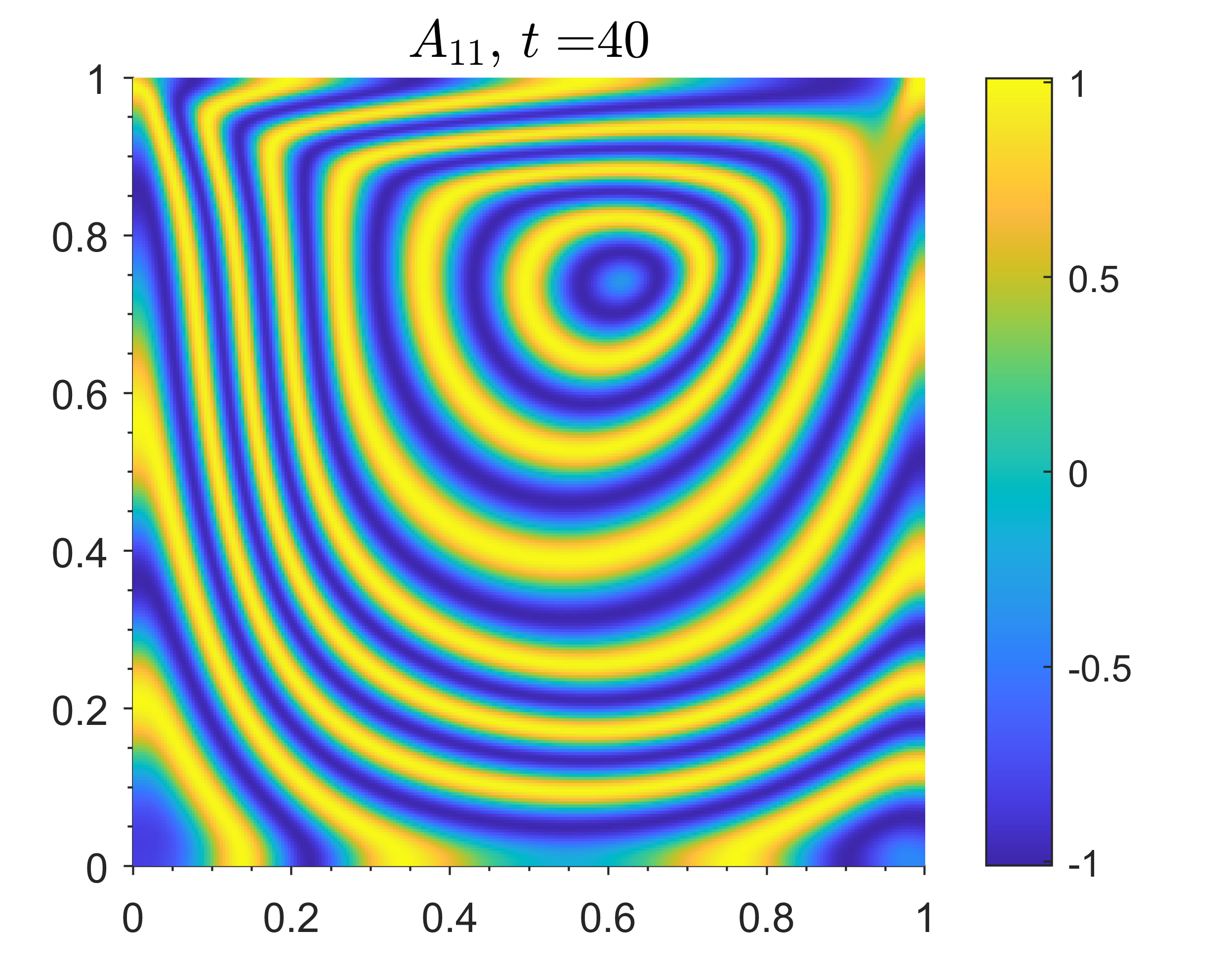}
			\\
			\includegraphics[draft=false,width=0.3\textwidth]{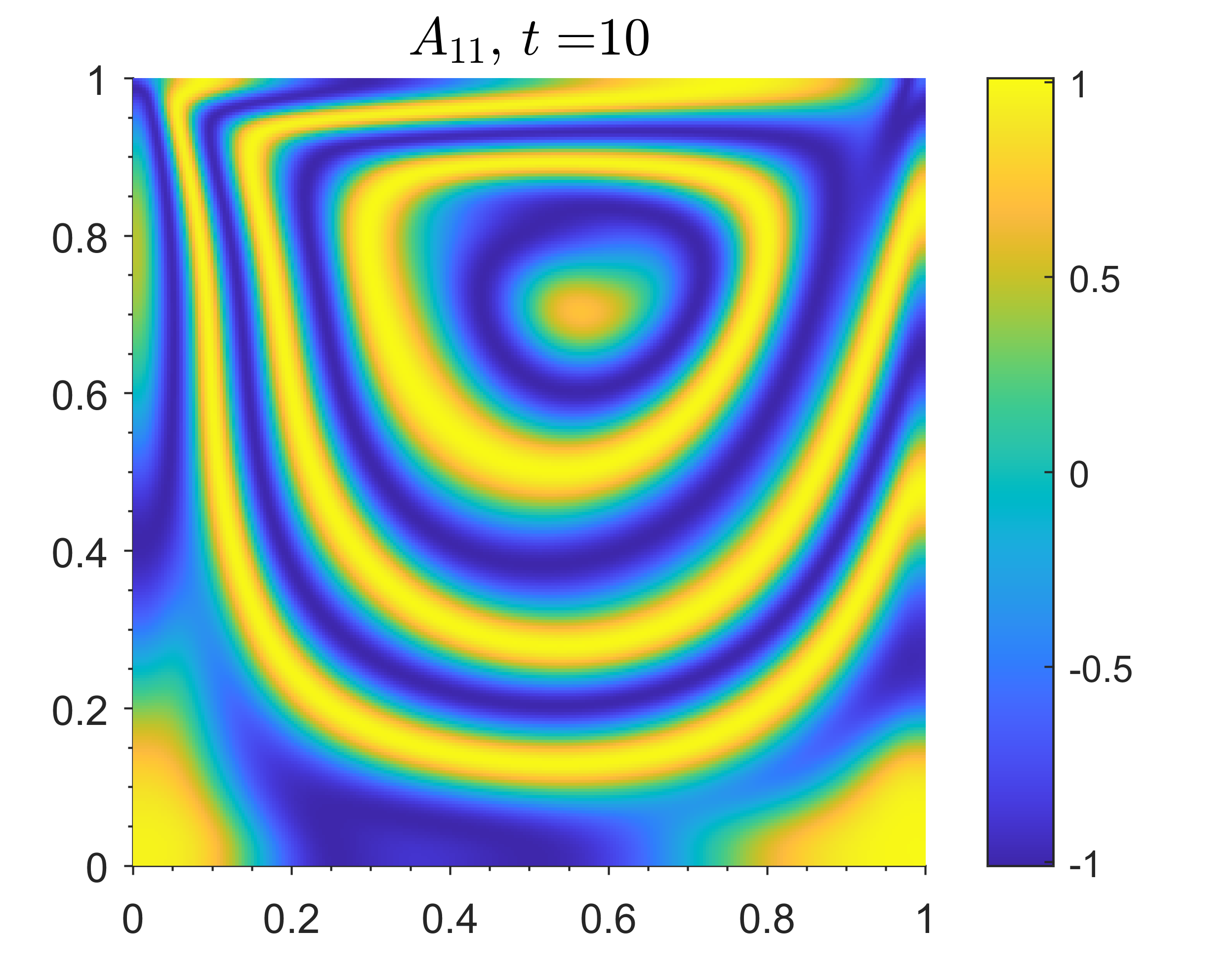}
			\includegraphics[draft=false,width=0.3\textwidth]{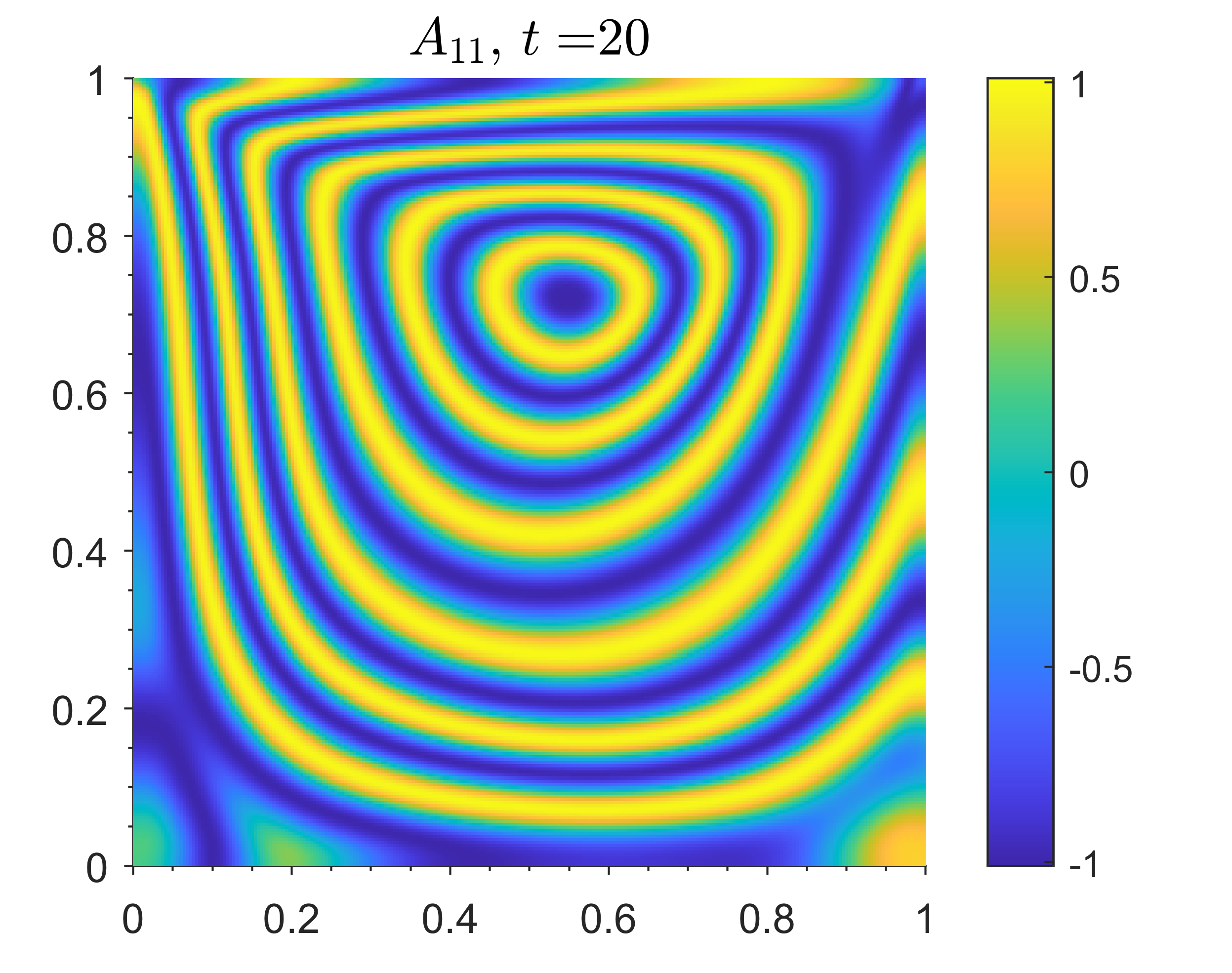}
			\includegraphics[draft=false,width=0.3\textwidth]{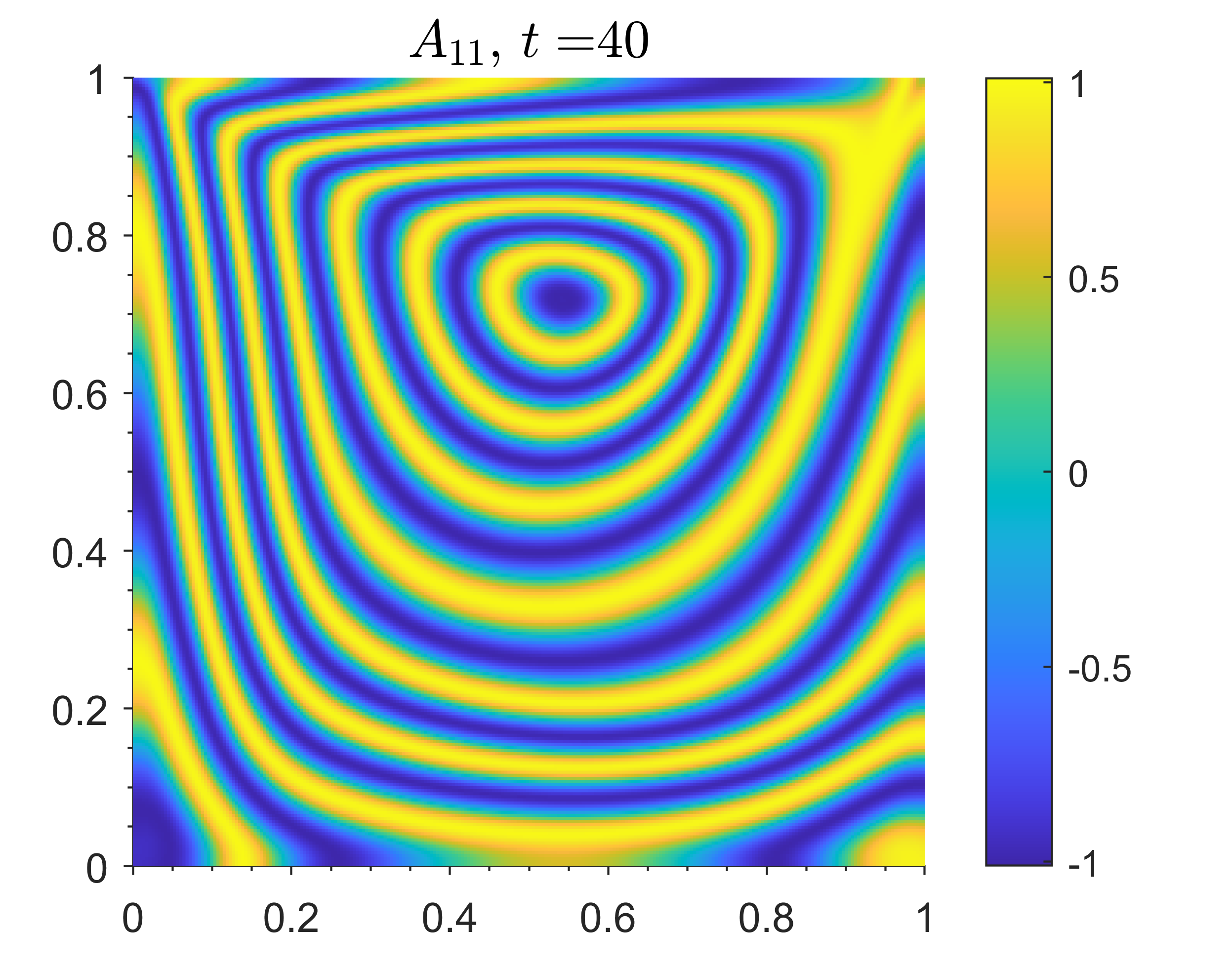}
			\\
		\caption{Snapshots of $ A_{11} $ for the lid-driven cavity flow of the GPR model with the 
		power-law exponents $ n = 1.0 $ 
			(top row) and $ n =1.5 $ (bottom row) at times $ t = 10 $, $ 20 $, and $ 40 $. 
		}  
		\label{fig:Cavity-GPR-A11}
	\end{center}
\end{figure}

\begin{figure}[!htbp]
	\begin{center}
		\includegraphics[draft=false,width=0.3\textwidth]{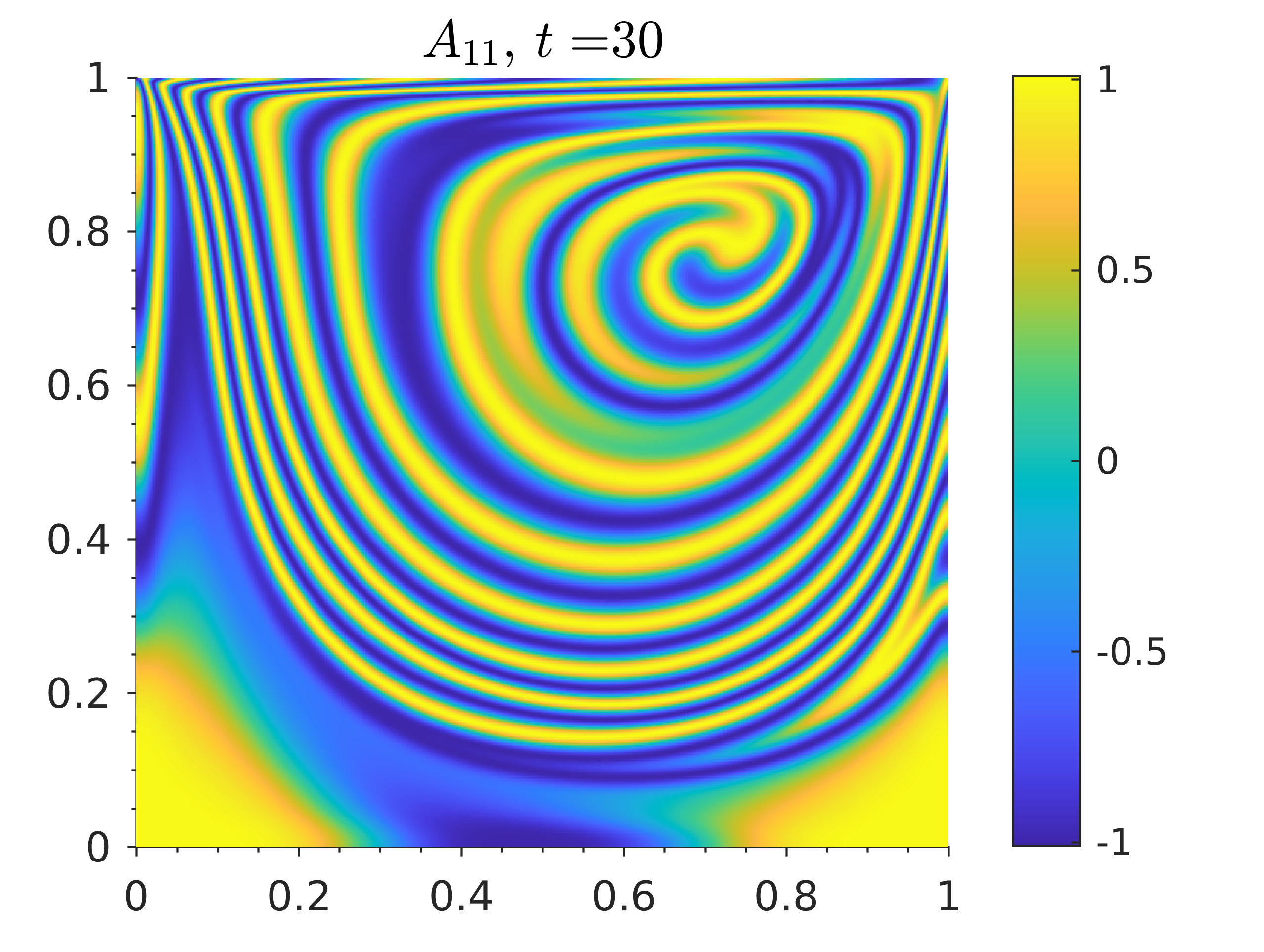} 
		\includegraphics[draft=false,width=0.3\textwidth]{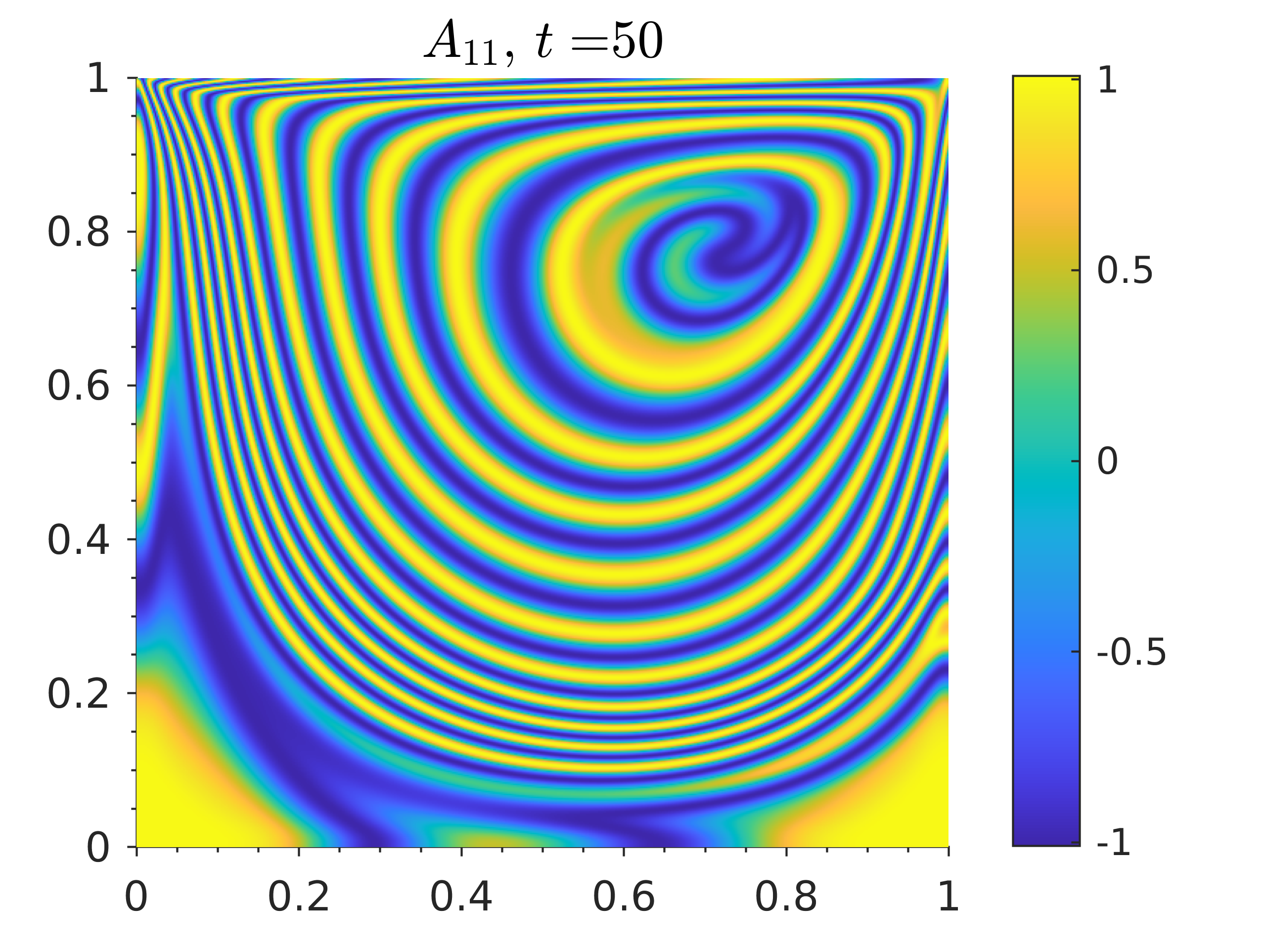}
		\caption{Snapshots of $ A_{11} $ for the lid-driven cavity flow of the GPR model with the 
			power-law exponent $ n = 0.5 $ at times $ t = 30 $ and $ 50 $. 
		}  
		\label{fig:Cavity-GPR-A11-n05}
	\end{center}
\end{figure}

Finally, Fig.\,\ref{fig:Cavity-GPR-A11} and Fig.\,\ref{fig:Cavity-GPR-A11-n05} show non-trivial 
dynamics 
of the distortion field. We plot only $ A_{11} $ component but the others have a similar 
distribution. The real deformations of the material elements are tiny and contained in the 
metric tensor $ \GG = \Dist^\transpose \Dist $ and they are hidden beyond the rotations $ \bm{R} $ 
(see the discussion to Fig.\,\ref{fig:Couette.A} in Section\,\ref{sec.Couette})  
whose elements vary between $ -1 $ and $ 1 $. Also, note that the 
dynamics of $ \Dist $ is never steady even if other quantities (density, velocity) reach the 
steady-state. Thus, the colors in Fig.\,\ref{fig:Cavity-GPR-A11} and 
Fig.\,\ref{fig:Cavity-GPR-A11-n05} 
keep evolving in time while preserving the overall pattern.

\subsection{Viscoplastic fluids}

In the last numerical example, we run the simulation of the lid-driven cavity flow of our model in 
the elastoviscoplastic regime, i.e. with 
$ \sigmaY > 0 $. In this example, we only consider $ n=1.0 $ (i.e. the Herschel-Bulkley viscosity 
reduces to the Bingham viscosity).  First, we run the model at Bingham 
numbers $ \Bi=1 $, $ 10 $, and $ 100 $ at low Reynolds number $ \Re = 1 $ and then at the same 
Bingham numbers but at Reynolds $ \Re=100 $. In all the simulation of this section, the yield 
stress $ \sigmaY $ is computed from \eqref{BiNumber.gen} with $ h = 1 $, $ V=2.03 $. The lid 
velocity is set to $ 1 $ while the Reynolds number is controlled via setting $ \kappa=1 $ ($ \Re=1 
$) or $ \kappa = 10^{-2} $ ($ \Re=100 $). Initially, the material is at rest with the parameters $ 
\rho = 1 $, $ \vv =0  $, $ \Dist = \II $, while the simulation lasts until the velocity reaches the
steady state.

Recall that the ``\emph{softness}'' of the solid state in our model is characterized by the shear 
sound speed $ \csh $ and because in the reference solution of the Bingham model we shall compare 
with, the solid 
state is treated as infinitely rigid we were need to take also a higher shear sound $ \csh $ than 
in 
the 
previous examples. We chose $ \csh = 50 $ for all the simulations presented below.

Fig.\,\ref{fig:Cavity-GPR-Bi-Re1} depicts the results at Reynolds number $ Re = 1$ and $ \Bi = 1 $, 
$ 10 $, and $ 100 $. We plot the snapshots of $ \sigma/\sigmaY $ which can be used to distinguish 
the solidified ($ \sigma/\sigmaY < 1 $) and fluid ($ \sigma/\sigmaY > 1 $) regions. One can also 
use for that purpose the relaxation time $ \tau $ also depicted in 
Fig.\,\ref{fig:Cavity-GPR-Bi-Re1}. We can notice overall a good agreement with the reference 
solution reproduced from \cite{Syrakos2014} and obtained with a steady state finite volume solver 
for the Bingham model. To compute the solution, a mesh of $ 256^2 $ cells was used for $ \Bi = 1 $ 
and $ \Bi = 10 $, while the a mesh of $ 512^2 $ cells was used in the case of $ \Bi = 100 $.

\begin{figure}[!htbp]
	\begin{center}
			\includegraphics[draft=false,scale=0.5]{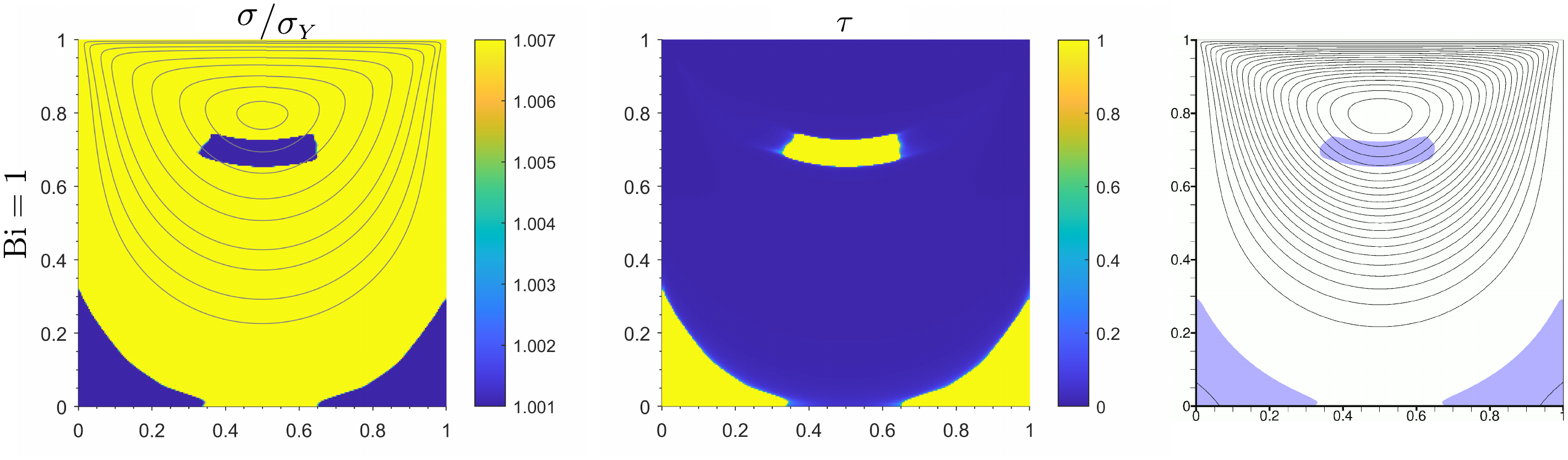}\\
			\includegraphics[draft=false,scale=0.5]{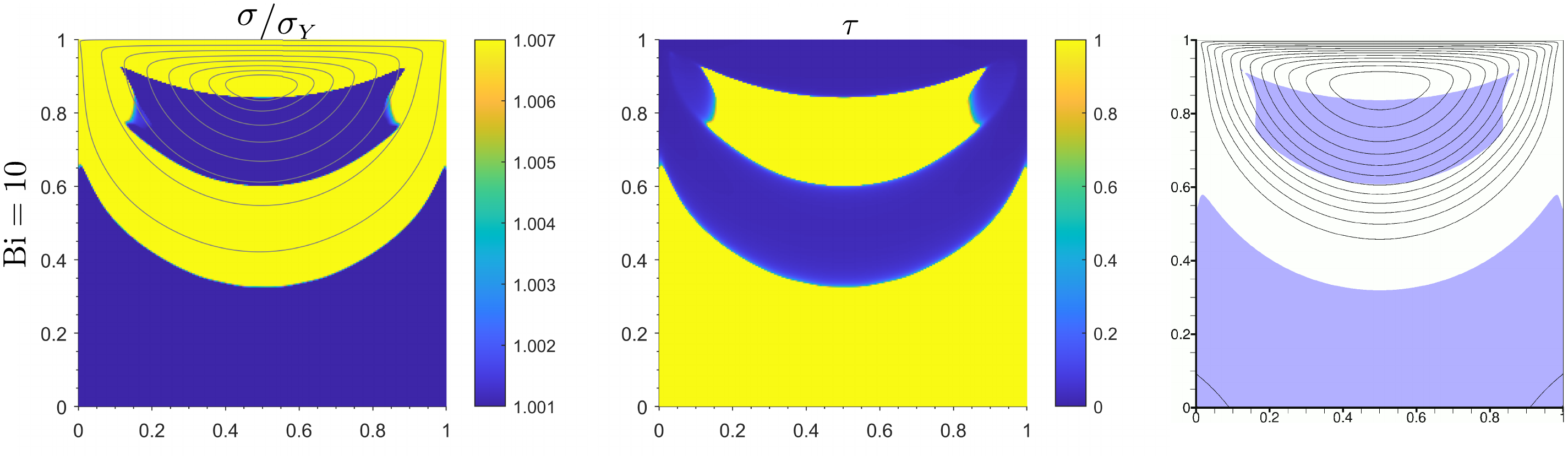}\\
			\includegraphics[draft=false,scale=0.5]{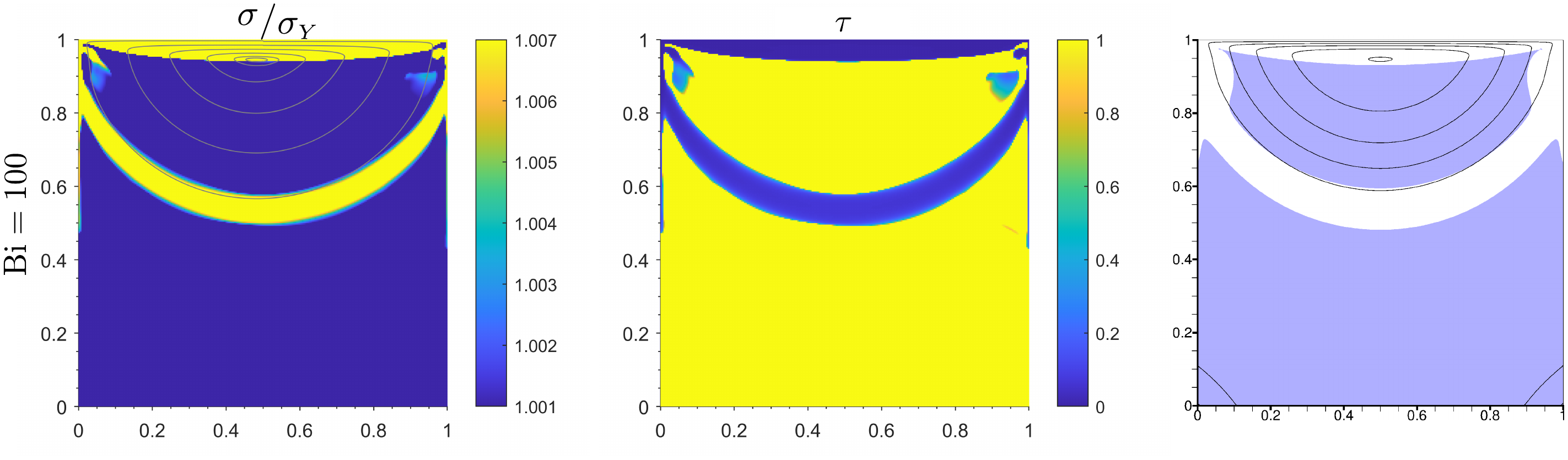}
		\caption{Lid-driven cavity flow of GPR model in the elastoviscoplastic regime at low 
		Reynolds number $ 
		\Re = 1 $ ($ \kappa = 1 $) for $ \Bi = 1 $ (top row), $ \Bi = 10 $ (middle row), and $ \Bi 
		= 100 $ (bottom 
		row). The power-low exponent is set to $ n = 1.0 $. The equivalent stress (left column) can 
		be used to visualize the unyielded regions (blue, $ \sigma / \sigma_Y >1 $). Second columns 
		shows the relaxation time with the yellow color denoting the unyielded regions (for 
		visualization, we color all cells with $ \tau >1 $ in yellow). The 
		reference 
		solution of the Bingham model (reproduced from \cite{Syrakos2014}) is shown in the right 
		column. The shaded regions denote the unyielded regions $ \sigma > \sigma_Y $. 
		}  
		\label{fig:Cavity-GPR-Bi-Re1}
	\end{center}
\end{figure}

\begin{figure}[!htbp]
	\begin{center}
		\begin{tabular}{c} 
			\includegraphics[draft=false,width=0.75\textwidth]{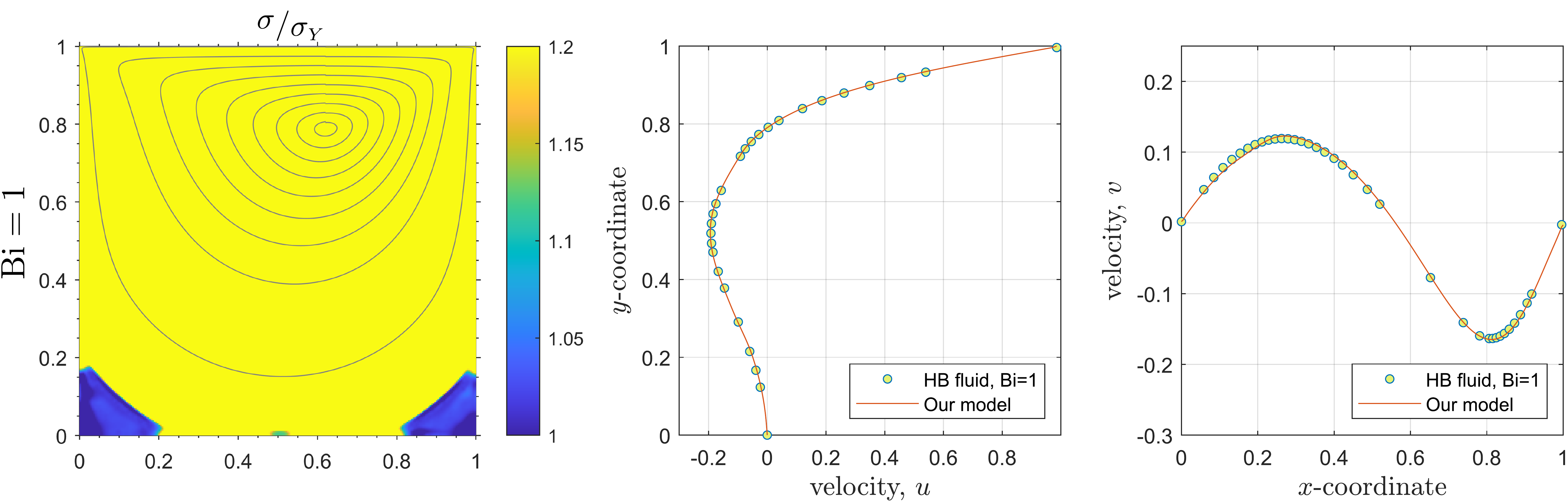}
			\\
			\includegraphics[draft=false,width=0.75\textwidth]{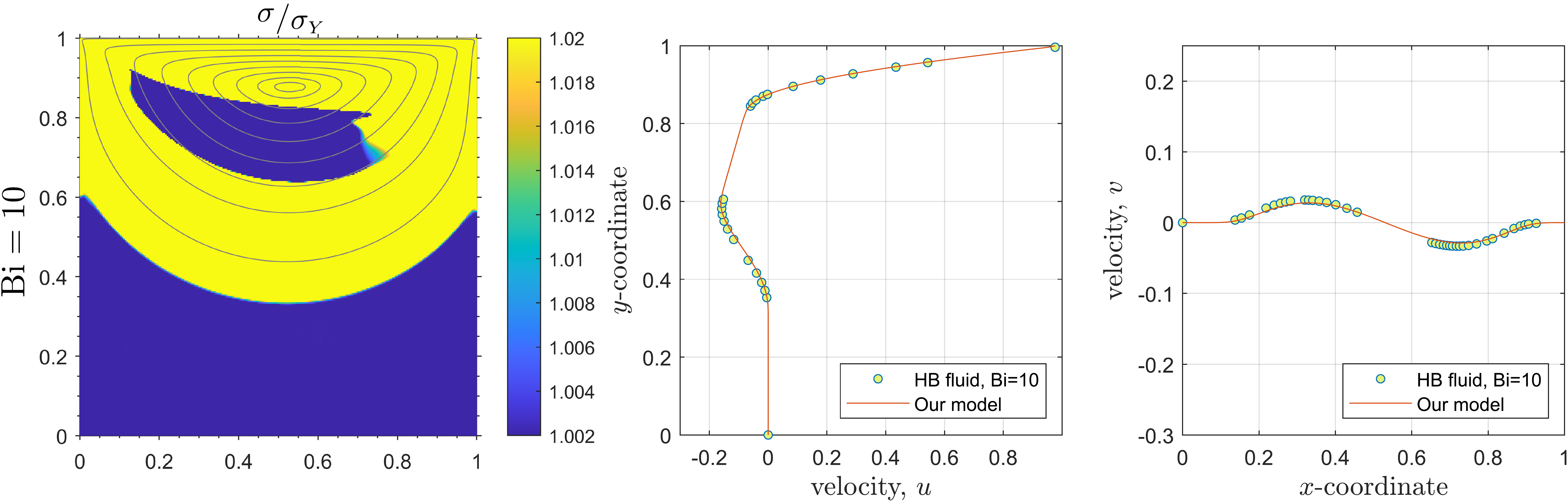}
			\\
			\includegraphics[draft=false,width=0.75\textwidth]{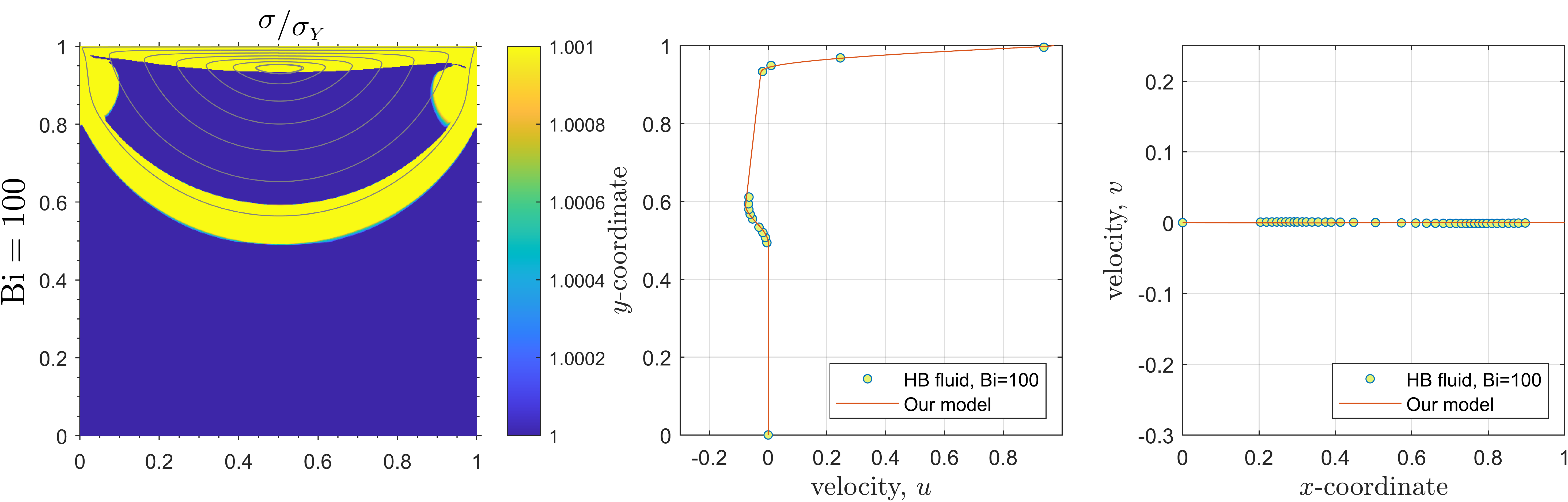}
%			
%\includegraphics[draft=false,width=0.75\textwidth]{Cavity-nHB1.5-xy-cuts-streams}
%			\\
		\end{tabular} 
		\caption{Lid-driven cavity flow of the GPR model in the elastoviscoplastic regime at $ 
		Re=100 $ ($ \kappa=10^{-2} $) and different Bingham numbers: $ \Bi=1 $ (top row, final time 
		$ t=10 $), $ \Bi=10 
		$ (middle 
		row, final time $ t=1 $), and $ 
		\Bi=100 $ (bottom row, final time $ t=0.1 $). Equivalent stress $ 
		\sigma=\Vert\tensor{\sigma}\Vert $ and 
		streamlines (left column), and $ x=0.5 $ (middle column) and $ y=0.5 $ (right column) cuts 
		of the velocity field compared against the solution to the Bingham model (data has been 
		extracted from \cite{Sverdrup2018}). In the left column, the blue color correspond to the 
		solidified regions.
		}  
		\label{fig:Cavity-GPR-Bi-Re100}
	\end{center}
\end{figure}

\begin{figure}[!htbp]
	\begin{center}
			\includegraphics[draft=false,scale=0.5]{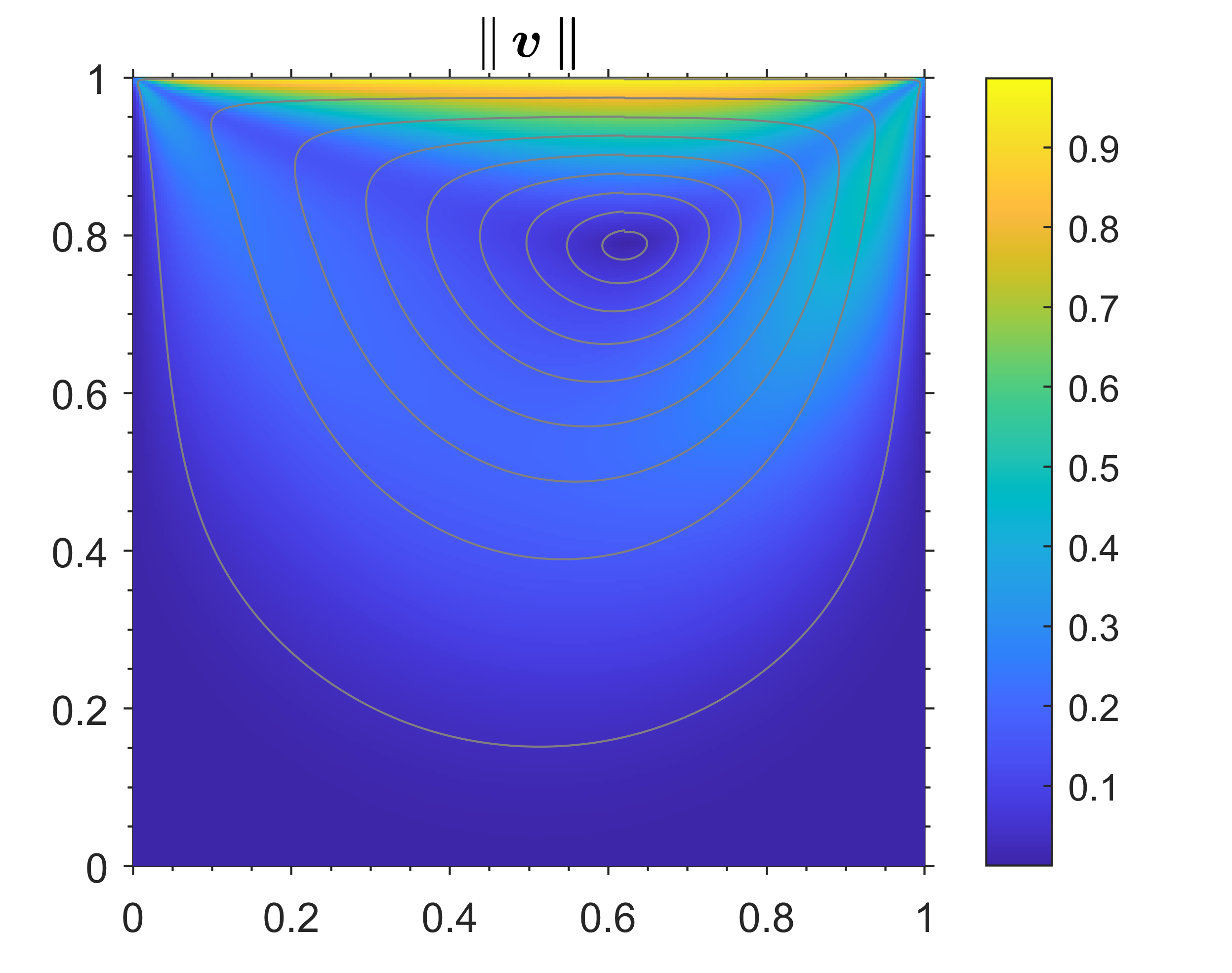}
			\includegraphics[draft=false,scale=0.5]{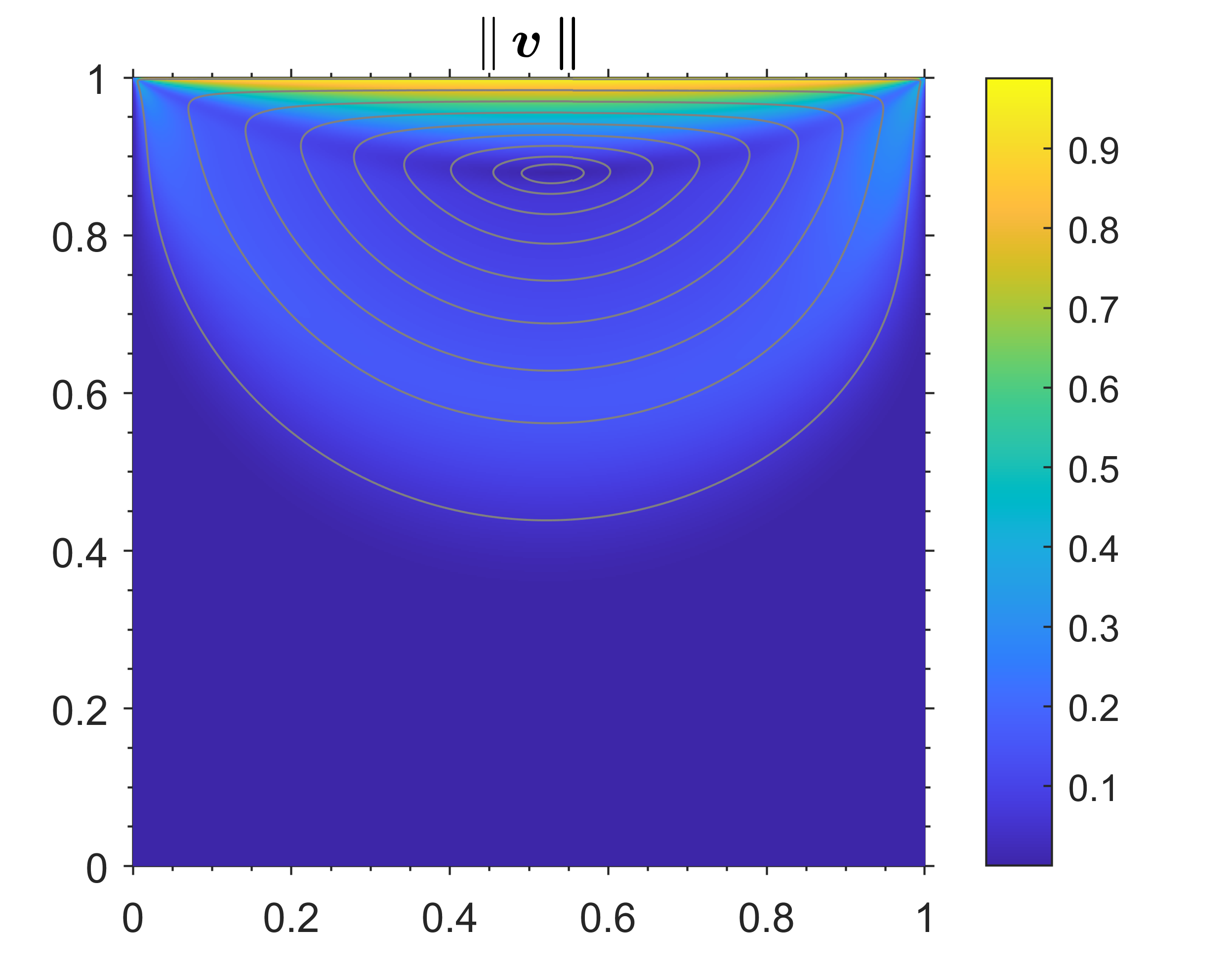}
			\includegraphics[draft=false,scale=0.5]{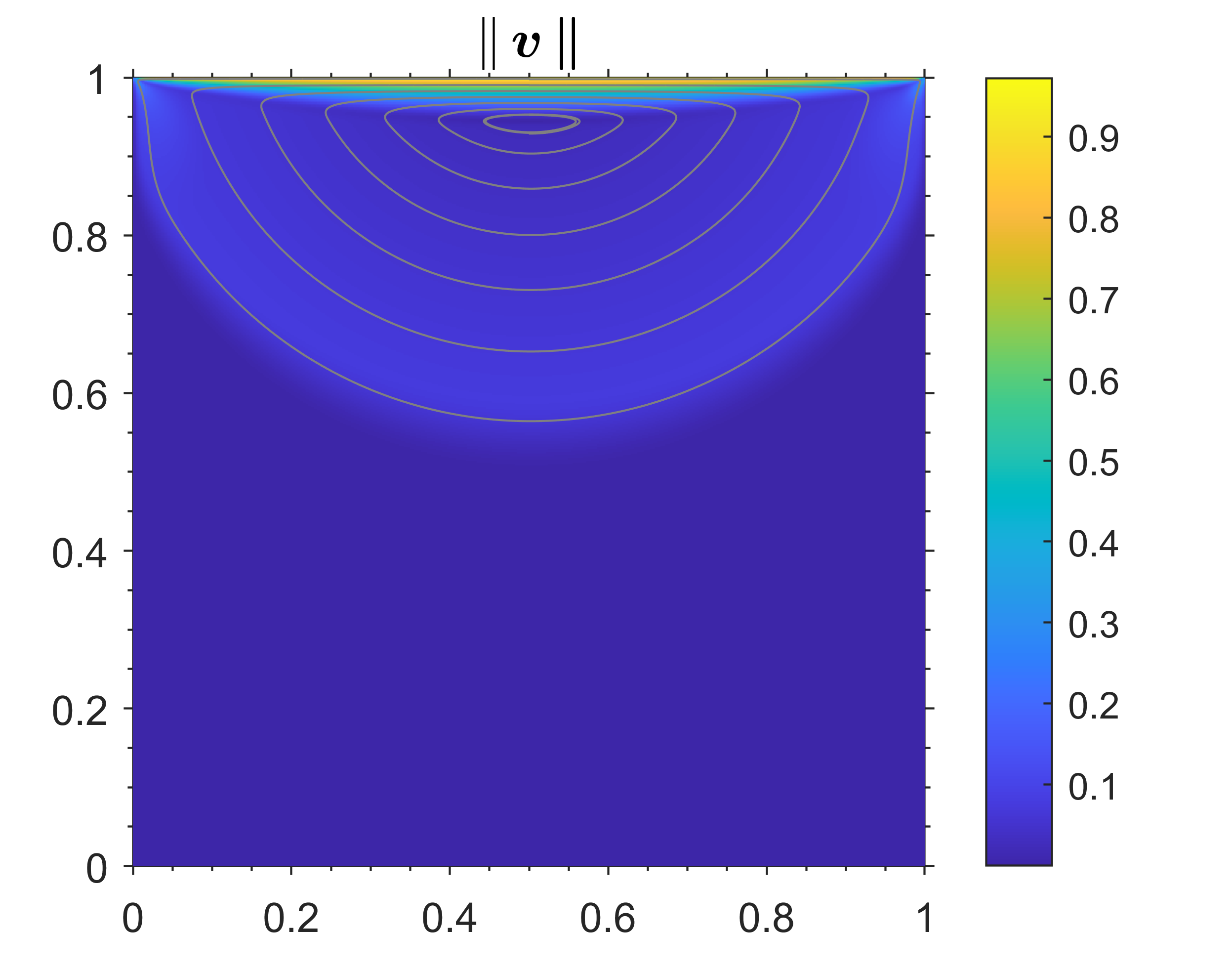}
			%			
			%\includegraphics[draft=false,width=0.75\textwidth]{Cavity-nHB1.5-xy-cuts-streams}
			%			\\
		\caption{Lid-driven cavity flow of the GPR model in the elastoviscoplastic regime with $ 
		n=1 $ at $ 
			Re=100 $ and different $ \Bi $ numbers: $ \Bi=1 $ (left), $ \Bi=10 $ (middle), 
			and $ 
			\Bi=100 $ (right). The norm of the velocity field and streamlines showing the absence 
			of 
			the flow in the solidified region. 
		}  
		\label{fig:Cavity-GPR-Bi-Re100-vel}
	\end{center}
\end{figure}

Finally, Fig.\,\ref{fig:Cavity-GPR-Bi-Re100} and Fig.\,\ref{fig:Cavity-GPR-Bi-Re100-vel} show the 
solution to the GPR model at $ \Re = 100 $. This time, we also have the possibility of quantitatively 
comparing the solution to the hyperbolic GPR model with the solution to the parabolic Navier-Stokes 
equations with the HB rheology, see the velocity cuts in Fig.\,\ref{fig:Cavity-GPR-Bi-Re100}. One 
can 
notice a good agreement with the reference solution. Also, we plot the snapshots of the velocity 
field in Fig.\,\ref{fig:Cavity-GPR-Bi-Re100-vel} in order to demonstrate that there is no flow 
in the solidified regions attached to the bottom of the cavity (the plug). The same computational 
meshes of $ 256^2 $ (for $ \Bi=1 $, $ 10 $) and $ 512^2 $ (for $ \Bi=100 $) were used for the case 
of $ \Re=100 $. 

\section{Conclusion and outlook} 

In this paper, we have studied the ability of the unified first-order hyperbolic model of 
continuum fluid and solid mechanics \eqref{eqn.GPR} to describe flows of non-Newtonian and, in 
particular, of viscoplastic fluids (or yield-stress fluids) \cite{FrigaardReview2014}. Recently, 
it was already shown in \cite{Jackson2019a} that the model can deal with non-Newtonian fluids with 
the 
power-law viscosity (without a yield stress), while here, for the first time, the model was used to 
describe the stress-driven solid-fluid transformations in the presence of the yield stress. Via 
direct comparison of the numerical solution of our model against analytical or numerical solutions 
to the Navier-Stokes equations with the Herschel-Bulkley viscosity on several standard benchmark 
test cases (Couette, Hagen-Poiseuille, and lid-driven cavity flows) it was demonstrated that the 
solution to the parabolic Navier-Stokes equations can be well approximated with the solution of our 
first-order hyperbolic model \eqref{eqn.GPR}.

Note that, previously, the yield-stress fluids were modeled either with a pure viscous classical 
approach relying on the Navier-Stokes equations (Bingham-type models) and thus ignoring the 
elasticity of the unyielded 
state, or with more advanced elastoviscoplastic approaches which take into account the elasticity 
of the unyielded 
material \cite{Saramito2007,Saramito2009,Benito2008,Fusi2007}. The latter, rely on the idea of 
additive splitting of the total Cauchy stress $ \tensor{\sigma} $ into two parts $ 
\tensor{\sigma}(\doteps,\tensor{\varepsilon}) 
= 
\tensor{\sigma}_{\ms{NS}}(\doteps) 
+ \tensor{\sigma}_{\ms{E}}(\tensor{\varepsilon}) $ with $ 
\tensor{\sigma}_{\ms{NS}}(\doteps) $ being the 
conventional 
Navier-Stokes stress defined as the function of the strain rate $ \doteps $, and $ 
\tensor{\sigma}_{\ms{E}} $ being the  
elastic stress which is a function of the strain $ \tensor{\varepsilon} $. The two stresses $ 
\tensor{\sigma}_{\ms{NS}} $ and $ \tensor{\sigma}_{\ms{E}} $ are conceptually different. Namely, 
the 
Navier-Stokes stress being the non-local stress (i.e. it involves the first gradient of the state 
vector) and fully dissipative in nature, while the elastic stress $ \tensor{\sigma}_{\ms{E}} $ 
being local and 
reversible. This makes it unclear the description of a 
general solid-fluid transformation (melting/solidification) with such an approach. 

On the other hand, in our theory, the stress tensor, in both fluid and solid state, is computed 
in a unified manner $ \tensor{\sigma} = -\rho \Dist^\transpose E_\Dist $ based \emph{only} on the 
strain 
measure, the distortion field $ 
\Dist $, while the use of the strain rate $ \doteps $ is not required at all. This, we believe, is 
one of the main novelties of our theory in comparison with the conventional 
aforementioned approaches, which should make our approach attractive for modeling of processes 
involving general solid-fluid transformations including thermally-driven melting and 
solidification. Note that the heat conduction, being ignored in this study, can be also described 
by the first-order hyperbolic equations with relaxation-type source terms similar to 
\eqref{eqn.dist}, 
see \cite{RmenskiMalyshev1987,SIGPR2021,DPRZ2016}. 

Because all the benchmark tests considered in this paper involve incompressible flows, we have 
employed our new Semi-Implicit Structure Preserving Finite Volume (SISPFV) scheme proposed in 
\cite{SIGPR2021} which treats the 
pressure part of the model in an implicit manner and thus, allows to run the model in the low-Mach 
number regime ($ {\rm Ma}\to 0 $). Further development of the scheme will concern an implicit 
treatment of the elastic part of the model which will allow to eliminate the shear sound speed from 
the stability condition of the numerical scheme resulting in a bigger time step. Also, in order to 
efficiently run the model in the 
stiff 
relaxation limit ($ \tau  \ll 1 $), we plan to incorporate an IMEX 
Runge-Kutta-type time-integration in 
order to improve the accuracy of the current SISPFV scheme in time. Furthermore, for better 
tracking of the yield surface in flows of viscoplastic fluids, or propagation of general 
solidification fronts, 
one could 
use Arbitrary Lagrangian-Eulerian (ALE) schemes \cite{Boscheri2013,Boscheri2014,Boscheri2015} or 
even more general schemes on \emph{non-conforming}
meshes \cite{Gaburro2020a,Gaburro2020,Busto2020}.

Further applications of the model will concern the modeling melting and solidification processes in 
additive manufacturing \cite{Francois2017,King2017,Khairallah2016}. For this purposes we shall use 
our first-order hyperbolic model for heat conduction 
\cite{RmenskiMalyshev1987,DPRZ2016,SIGPR2021,SHTC-GENERIC-CMAT} 
and 
hyperbolic surface tension model advanced by Gavrilyuk~\textit{et al}  in \cite{Berry2008a} and 
further 
developed by Schmidmayer~\textit{et al} \cite{Schmidmayer2017} and Chiocchetti~\textit{et al}
\cite{Chiocchetti2020} in conjunction with a simple diffuse interface approach 
\cite{Tavelli2019,Kemm2020,Busto2020,Cracks2020,Cracks2021}
for tracking the free boundaries.
%=============================================================================
%==========    A C K N O W L E D G M E N T S
\section*{Acknowledgments}

The research presented in this paper was partially funded by the European Union's Horizon 2020 
Research and Innovation  Programme under the project \textit{ExaHyPE}, grant no. 671698 (call 
FETHPC-1-2014). 
The authors also acknowledge funding from the Istituto Nazionale di Alta Matematica (INdAM) through 
the GNCS group and the program \textit{Young Researchers Funding 2018} via the research project 
\textit{Semi-implicit structure-preserving schemes for continuum mechanics}. 
E.R. acknowledges the financial support from the state contract of 
the Sobolev Institute of Mathematics (project no. 0314-2019-0012).  
Results by E.R. obtained in Sec.2 were done under the support of the Russian Science Foundation grant 19-77-20004.
M.D., I.P., and S.C. acknowledge the financial support received from the Italian Ministry of 
Education, 
University and Research (MIUR) in the 
frame of the Departments of Excellence Initiative 2018--2022 attributed to DICAM of the University 
of Trento (grant L. 232/2016). 
S.C. acknowledges the financial support received by the Deutsche Forschungsgemeinschaft (DFG) under 
the project \textit{Droplet Interaction Technologies (DROPIT)}, grant no. GRK 2160/1. 
W.B., M.D. and I.P. also received financial support in the frame of 
the PRIN Project 2017 \textit{Innovative numerical methods for evolutionary partial differential 
equations and applications}. M.D. and I.P. have also received funding from the University of Trento 
via the \textit{Strategic Initiative Modeling and Simulation} and via the \textit{UniTN Starting 
Grant initiative}.
I.P. acknowledges a partial support by ANR-11-LABX-0040-CIMI within the program ANR-11-IDEX-0002-02.

The authors are grateful to the Leibniz Rechenzentrum (LRZ) for awarding access to the 
SuperMUC supercomputer based in Munich, Germany.

%\section*{References}
\bibliographystyle{plainurl}
\bibliography{library}

\end{document}